\documentclass[a4paper,11pt]{article}
\pdfoutput=1

\usepackage{jheppub}

\usepackage[T1]{fontenc}
\usepackage[latin9]{inputenc}
\usepackage{amsmath}
\usepackage{amssymb}
\usepackage{esint}
\usepackage{lmodern}
\usepackage{slashed}
\usepackage{dsdshorthand}
\usepackage{tikz}

\usepackage{mathdots}
\usetikzlibrary{arrows.meta}

\usepackage{tabularx}
\usetikzlibrary{arrows}
\newtheorem*{theorem*}{Theorem}



\title{Lining up a Positive Semi-Definite Six-Point Bootstrap}

\author[a]{Ant\'onio Antunes,}
\author[a]{Sebastian Harris,}
\author[a,b]{Apratim Kaviraj,}
\author[a,c]{Volker Schomerus}

\preprint{DESY-23-220}
\affiliation[a]{Deutsches  Elektronen-Synchrotron  DESY,  Notkestr.   85,  22607  Hamburg,  Germany}
\affiliation[b]{Department of Physics, Indian Institute of Technology - Kanpur, Kanpur 208016,  India}
\affiliation[c]{II. Institut f\"ur Theor. Physik, Universit\"at Hamburg, Luruper Chaussee 149, D-22761 Hamburg}

\emailAdd{antonio.antunes@desy.de}
\emailAdd{sebastian.harris@desy.de}
\emailAdd{akaviraj@iitk.ac.in}
\emailAdd{volker.schomerus@desy.de}

\abstract{In this work, we initiate a positive semi-definite numerical bootstrap program for multi-point correlators.
Considering six-point functions of operators on a line, we reformulate the crossing symmetry
equation for a pair of comb-channel expansions as a semi-definite programming problem. We provide two
alternative formulations of this problem. At least one of them turns out to be amenable to numerical
implementation. Through a combination of analytical and numerical techniques, we obtain rigorous
bounds on CFT data in the triple-twist channel for several examples.}

\begin{document}
\addtolength{\abovedisplayskip}{0pt}
\addtolength{\belowdisplayskip}{1pt}
\addtolength{\baselineskip}{3pt}
\maketitle
\flushbottom

\section{Introduction}
The numerical conformal bootstrap \cite{Rattazzi:2008pe} has by now reached the status of a mature method
in the non-perturbative study of conformal field theories \cite{Poland:2018epd}. In its standard incarnation,
one extracts the information contained in the crossing equation for the four-point functions of a few low
lying operators by acting with a finite basis of functionals and rewriting the problem as a semi-definite
program (SDP), taking advantage of reflection positivity of the correlators \cite{Poland:2011ey,Kos:2014bka}.
This has led, for example, to the highest precision determination of the critical exponents of the 3d Ising
model, by establishing a small allowed island in the space of scaling dimensions \cite{ElShowk:2012ht,
El-Showk:2014dwa,Kos:2014bka,Kos:2016ysd}. While many other relevant examples have been addressed
successfully within the numerical conformal bootstrap, the treatment of the 3d Ising model remains
a milestone for modern conformal bootstrap. This makes it a good stage to discuss the prospects of
a higher-point bootstrap.

Having seen how far the island of the 3d Ising model has shrunk by considering a small number of four-point
correlators, one may certainly wonder whether it might be possible to end up with just a single point in the
space of scaling dimensions by pushing numerical precision while remaining with a finite (though possibly
increased) number of correlators. While this is not rigorously excluded, there is little supportive
evidence for such a scenario. On the contrary, it seems more likely not to be the case. Given that the
full set of CFT data is only probed if the four-point functions of all operators are taken into
account, the exact solution of any CFT appears to be a genuinely infinite problem.

This is the moment in which higher-point functions enter the scene, since even a single such
function contains the dynamical information of infinitely many four-point functions. In fact, the
OPE of two scalars in 3d already contains operators in all possible parity-even representations of the conformal
group. This is because in 3d the most general Lorentz representation is a symmetric traceless tensor
(STT).\footnote{There can also be odd parity scalars and STTs obtained from contraction with the
Levi-Civita tensor. In the subsequent discussion, we restrict ourselves to the parity-even sector. The parity odd sector emerges from the scalar-STT OPE, and the discussion can easily be generalised to describe this sector as well.} It follows that five- and six-point functions of a single scalar field
actually suffice to recover all the OPE coefficients of the theory, at least in the absence of
internal symmetries. For five points, the STT-STT-scalar data can be extracted, and, with a
pairwise OPE (the so-called snowflake expansion, briefly reviewed in Section~\ref{sec:sixptcomb}),
the six-point function gives us access to STT-STT-STT data.

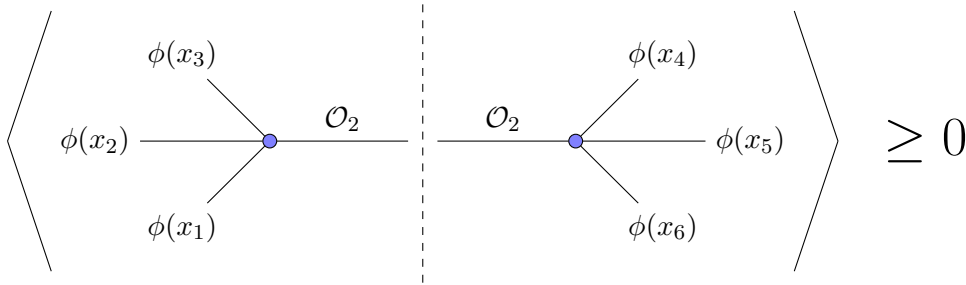
\begin{figure}[ht]
    \centering
    \begin{tikzpicture}[scale=1.15]
        \node (E1) at (-0.5,-1) {$\phi(x_1)$};
        \node (E2) at (-1.5,0) {$\phi(x_2)$};
        \node (E3) at (-0.5,1) {$\phi(x_3)$};
        \node (E4) at (5,1) {$\phi(x_4)$};
        \node (E5) at (6,0) {$\phi(x_5)$};
        \node (E6) at (5,-1) {$\phi(x_6)$};

        \node[shape=circle,fill=blue!50,draw=black, scale = 0.5] (V1) at (0.5,0) {};
       \node[shape=circle,fill=blue!50,draw=black, scale = 0.5] (V4) at (4,0) {};

        \node (Z1) at (2.2,0){};
        \node (Z2) at (2.3,0){};

        \draw (E1) -- (V1);
        \draw (E2) -- (V1);
        \draw (E3) -- (V1);
        \draw (E4) -- (V4);
        \draw (E5) -- (V4);
        \draw (E6) -- (V4);

        \path[-, draw] (V1) edge node[above] {$\mathcal{O}_2$} (Z1);

        \path[-, draw] (Z2) edge node[above] {$\mathcal{O}_2$} (V4);

        \node (Vert1) at (2.25,-1.75){};
        \node (Vert2) at (2.25,1.75){};
        \draw[dashed] (Vert1) -- (Vert2);

        \coordinate (Lu) at (-2,1.5);
        \coordinate (Lc) at (-2.5,0);
        \coordinate (Ld) at (-2,-1.5);
        \draw (Lu) -- (Lc);
        \draw (Lc) -- (Ld);

        \coordinate (Ru) at (6.5,1.5);
        \coordinate (Rc) at (7,0);
        \coordinate (Rd) at (6.5,-1.5);
        \draw (Ru) -- (Rc);
        \draw (Rc) -- (Rd);

        \node (Ge0) at (8,0) {\huge $\ge 0$};

\end{tikzpicture}
    \caption{ Schematic representation of reflection positivity of the six-point function in the comb-channel expansion.
	The sets $\{x_1,x_2,x_3\}$ and $\{x_4,x_5,x_6\}$ of insertion points are assumed to be related by a reflection. 	\label{fig:reflection}
	}
\end{figure}

There is one problem with the snowflake OPE decomposition, however, namely that it is incompatible
with reflection positivity. This implies that it cannot be analysed with the rigorous machinery of
SDPs. While truncated crossing equations have been recently used to obtain some estimates in the
five-point case \cite{Poland:2023vpn}, a positive semi-definite formulation would bring us back to
the world of rigorous error bars which makes the standard numerical bootstrap so
popular.\footnote{On the other hand, analytical methods, which work at large spin, have already
produced some initial results in the five- and six-point case \cite{Bercini:2020msp,
Antunes:2021kmm,Kaviraj:2022wbw}.} Fortunately, the six-point function provides us with a
positive opportunity: Taking OPEs sequentially splits the correlator into two halves related
by reflection, the so-called comb-channel expansion, reviewed in Section~\ref{sec:sixptcomb},
manifesting STT-STT-scalar data, in a way that naturally leads to positivity, as we schematically
depict in Figure~\ref{fig:reflection}.

\vskip 0.5cm

The price we paid for positivity is that the six-point comb channel again misses important dynamical
data, namely the STT-STT-STT coefficients. One may now wonder how far one has to push the study of
multi-point functions in order to combine positivity with full access to dynamical data. For 3d theories
without internal symmetries, the answer is simple: it suffices to consider eight-point functions of a
single scalar field.\footnote{While there are no known unitary 3d CFTs without global symmetry, this statement applies, for example, to the singlet subsector of any theory.} And even in the 3d Ising model in which fields transform in the two irreducible
representations of a global $\mathbb{Z}_2$ symmetry, there is actually a finite-reflection positive
system of eight-point correlators to consider that, barring unknown selection rules, contains all the
CFT data of the Ising model.\footnote{Instead, for a theory with continuous global symmetry, an
infinite number of irreducible representations exists and so we always need an infinitely sized set
of correlators to manifest all the CFT data. Nonetheless, higher-point functions can still play a
useful role by repackaging several irreps. in a single correlator.} These are the eight-point functions
containing the two lightest $\mathbb{Z}_2$ odd scalars $\sigma$ and $\sigma'$ along with the
$\mathbb{Z}_2$ even scalar $\epsilon$,
\begin{equation}
\langle \sigma \sigma \sigma \sigma \sigma \sigma \sigma \sigma \rangle\,, \quad
\langle \sigma \sigma \sigma \sigma \epsilon \epsilon\epsilon \epsilon \rangle \,,
\quad  \langle \sigma \sigma \sigma \sigma \sigma' \sigma' \sigma' \sigma' \rangle \,, \quad \dots\quad  .
\end{equation}
Indeed, taking OPEs pairwise in the eight-point functions, we see that all four-point functions with
arbitrary operators are contained and hence all possible OPE coefficients with three generic STTs appear.
\smallskip

We conclude from the previous paragraphs that going from the traditional mixed correlator four-point
bootstrap to consider multi-point extensions represents a qualitatively novel and promising leap ahead
since the multi-point bootstrap naturally combines the constraints from infinitely many four-point
crossing equations. While this is true for any number $N>4$ of external points, the positivity
features that are vital for a controlled SDP based numerical bootstrap require $N$ to be even, making
$N=6$ the simplest example. In this paper we take the first steps in developing the multi-point bootstrap
program by formulating and solving the six-point crossing equation as an SDP, with the goal of extracting
bounds on new CFT data, and perhaps strengthening the four-point bounds. To make progress, we take advantage
of the significant kinematic
simplification that come with the restriction to 1-dimensional CFTs. While this eliminates the presence
of spinning OPE coefficients (since there are no continuous rotations and hence no STTs), most of the
richness and structure of the higher-dimensional case is preserved: As we will show below, we are still
solving crossing for an infinite number of four-point functions! Furthermore, 1-dimensional CFTs are
interesting objects on their own. Most importantly they can describe e.g. boundaries in 2-dimensional CFTs and line defects of higher
dimensional theories. Moreover, they can also appear on the boundary of 2-dimensional QFTs in AdS
\cite{Paulos:2016fap}, or simply as a kinematical restriction of higher dimensional CFTs, see e.g.  
\cite{Hogervorst:2013kva}. In this context, the higher-point bootstrap gives us access to  operators in more complicated sectors: We can study several global symmetry representations simultaneously, different transverse spins for line defects and focus on triple-twist operators for models related to generalised free fields. It also has the chance to improve the usual four-point bounds which notoriously struggle to isolate theories in one dimension.

\paragraph{Summary.}

Before we end this introduction we want to summarise our main results while describing the structure of
the paper. In Section~\ref{sec:sixptcomb}, we review the kinematics and OPE structure of one-dimensional
six-point functions and show how the comb-channel crossing equation naturally leads to an SDP. We move
on to find a reformulation of this SDP, which we dub the \textit{descendant-space} SDP, that only
requires differential relations between conformal blocks instead of their explicit evaluations. This
description emerges from expressing the six-point function in terms of products of four-point
functions (as in Figure~\ref{fig:reflection}), and it turns out to be crucial in establishing a
robust numerical algorithm. We also develop some important necessary conditions for positivity,
related to the asymptotics at large scaling dimension.

In Section~\ref{sec:noidentity}, we consider the toy problem of maximising the gap in a six-point function
where no identity operator is exchanged in the OPE of two external scalars. After some simple but illustrative
analytical bounds, we develop a numerical truncation scheme which can be fed to a standard SDP solver, and then
show how to rigorously confirm the positivity of the functionals obtained numerically. Our numerical results for
the gap maximisation without identity are presented in Subsection~\ref{sec:noidresults}. These corroborate and
reinforce our analytical expectation: The obtained numerical bound appears to be saturated by a simple crossing
symmetric function for which the gap is given by $9/5$ of the external dimension, see Figure~\ref{fig:NO1}.

With the numerical machinery in place, we use Section~\ref{sec:GFFbounds} to study a simple but more
realistic class of examples in which the identity field appears in the OPE of two external scalars
$\phi$. The section starts by outlining how to constrain triple-twist data in a rather general setup,
assuming that we have solved or bootstrapped the four-point function of the scalar $\phi$. This program
is then carried out concretely in two examples. First, we take as an input the four-point function of a
generalised free fermion (GFF) of dimension $\Delta_\psi$ and bound the
gap of the six-point function. We find a bound which approaches $3\Delta_\psi+3$: the leading triple-twist
operator for the GFF theory, see Figure~\ref{fig:GFF+}. Then, we consider the same problem with the
four-point function of a generalised free boson (GFB) as input. Surprisingly, the resulting bound on the
conformal dimension of the leading triple-twist operator seems to approach a value strictly above
$3\Delta_\phi$, the leading triple-twist operator of GFB (Figure~\ref{fig:GFB+}). We explain this result
in terms of a perturbative deformation of GFB: A $\Phi^6$ contact interaction in AdS$_2$, which can
increase the six-point gap without altering the four-point function. A similar interpretation for the
GFF result suggests a sign constraint on the leading fermionic sextic coupling. The section ends with a discussion on OPE-coefficient maximisation as well as bounds on full four-point correlators. As a concrete example, we provide a numerical bound on the GFF four-point function of three $\psi$ insertions and the lowest lying triple-twist operator (Figure \ref{fig:4PBound}).

We conclude in Section~\ref{sec:conclusions} and present an outlook to some immediate and more long-term
goals. Certain technical aspects of the numerical implementation and auxiliary analytical results are
relegated to several appendices.

\section{Six-point Crossing Equation in the Comb-Channel}
\label{sec:sixptcomb}

The goal of this section is to set up an infinite dimensional semi-definite program (SDP) dual to a crossing
equation of a six-point function in the comb-channel expansion. We begin by reviewing the kinematics and basic
properties of six-point functions and their conformal block expansions in Subsection~\ref{sec:sixpt-kinematics}.
An associated crossing equation is then spelled out in Subsection~\ref{sec:sixpt-crossing} and it is analysed
by rewriting it in terms of an infinite matrix whose rows and columns are labelled by the primaries exchanged
in the OPE of two external operators. While this perspective is close to the established multi-correlator
four-point bootstrap, we shall argue that it is preferable to pass to a different, yet equivalent, formulation.
As explained in Subsection~\ref{sec:demonstrateequiv}, the rows and columns of the matrices occurring in our
second formulation are labelled by descendants of the operators exchanged in the OPE of three external
operators. In Subsection~\ref{sec:toeplitzarc}, we establish asymptotic properties of these new matrices
which will be of key importance in the construction of positive definite combinations in the subsequent
sections. In the final Subsection~\ref{sec:commentson}, we present some conceptual remarks that provide a
broader perspective on our new approach to six-point bootstrap. While Subsections~\ref{sec:sixpt-kinematics}
to \ref{sec:toeplitzarc} lay out the foundation for the rest of this work, Subsection~\ref{sec:commentson}
can safely be skipped on a first reading.

\subsection{Kinematics and conventions}
\label{sec:sixpt-kinematics}

The protagonist of this paper is a six-point function of identical scalar operators $\phi(x_i)$ with
scaling dimension $\Delta_\phi$. All fields are inserted at points $x_i \in \mathbb{R}$ on the line
with the ordering prescription $x_i<x_{i+1}$. Their six-point function is denoted by
\begin{equation}
\mathcal{G}(x_1,\dots,x_6) \equiv \langle \phi(x_1)\phi(x_2)\phi(x_3) \phi(x_4)\phi(x_5)\phi(x_6) \rangle \,.
\end{equation}
In one dimension, four-point functions are parameterised by a single cross-ratio, and each extra point
adds a single subsequent degree of freedom. Thus, the correlator $\mathcal{G}$ is effectively a function
of three variables, which we choose to be
\begin{equation} \label{eq:CR}
    \chi_1=\frac{x_{12}x_{34}}{x_{13}x_{24}}\,, \quad \chi_2=\frac{x_{23}x_{45}}{x_{24}x_{35}} \,, \quad \chi_3=\frac{x_{34}x_{56}}{x_{35}x_{46}} \,,
\end{equation}
and $x_{ij}=x_i-x_j$. With the specified ordering of operators on the line, the cross-ratios satisfy $0<\chi_i<1$. For covariance
under conformal transformations, the correlator must be the product of a conformally invariant function of
the cross-ratios with an appropriate dimensionful prefactor $\mathcal{L}$, referred to as the leg factor,
\begin{equation}
    \mathcal{G}(x_1,\dots,x_6)=\mathcal{L}(x_i) \mathcal{G}(\chi_1,\chi_2,\chi_3)\,.
\end{equation}
By abuse of notation, we use the same letter $\mathcal{G}$ to denote both the entire correlator and
the associated invariant function of cross-ratios. For the leg factor we adopt the same choice as in
\cite{Rosenhaus:2018zqn},
\begin{equation}\label{eq:legg}
    \mathcal{L}(x_i)=\chi_2^{-\Delta_\phi}x_{12}^{-2\Delta_\phi}x_{34}^{-2\Delta_\phi}x_{56}^{-2\Delta_\phi}
    \,.
\end{equation}

Six-point correlators admit different conformal block decompositions which depend on the order in which
the OPEs are performed. There are two channel topologies, snowflake and comb, which each give rise to
several possible channels. For example, taking the OPEs in disjoint pairs, as indicated by the brackets
on the left hand side of equation \eqref{eq:snowCBexpansion}, leads to the snowflake expansion
\begin{equation}
\label{eq:snowCBexpansion}
    \langle (\phi_1\phi_2)(\phi_3 \phi_4)(\phi_5\phi_6) \rangle = \mathcal{L}(x_i) \sum_{\mathcal{O}_a,\mathcal{O}_b,\mathcal{O}c} C_{\phi \phi
    \mathcal{O}_a}C_{\phi \phi \mathcal{O}_b}C_{\phi \phi
    \mathcal{O}_c}C_{\mathcal{O}_a \mathcal{O}_b \mathcal{O}_c}
    G^{12,34,56}_{\Delta_a,\Delta_b,\Delta_c}(\chi_i)\,.
\end{equation}
Here, we used the shorthand notation $\phi_i \equiv \phi(x_i)$. On the right hand side, we introduced the
OPE coefficients $C$, and the conformal blocks $G_{\Delta_a,\Delta_b,\Delta_c}(\chi_i)$, whose topology
and channel is identified by the superscript, in this case the $12,34,56$ snowflake-channel. We note that
the theory dependent data in eq.~\eqref{eq:snowCBexpansion} contains the three-point functions $\langle\mathcal{O}_a
\mathcal{O}_b \mathcal{O}_c\rangle$ for any triple of operators that can appear in the OPE of two
external scalars $\phi$ along with three OPE coefficients that appear in the block expansion of
the four-point function already. This makes the snowflake-channel an interesting one, especially for higher
dimensional theories where is gives access to OPE coefficients of three fields in STT representations,
as discussed in the introduction. On the other hand, the snowflake-channel does not have the
positivity features that are needed for the numerical conformal bootstrap.

In order to develop our positive semi-definite six-point bootstrap program, we shall therefore
focus on the comb-topology. One particular conformal block expansion with comb-topology is given by
\begin{equation}
\label{eq:combCBexpansion}
    \langle \left((\phi_1\phi_2)\phi_3\right) \left(\phi_4(\phi_5\phi_6)\right \rangle =
    \mathcal{L}(x_i) \sum_{\mathcal{O}_1,\mathcal{O}_2,\mathcal{O}_3}
    C_{\phi \phi \mathcal{O}_1}C_{\mathcal{O}_1 \phi \mathcal{O}_2}C_{\mathcal{O}_2 \phi \mathcal{O}_3}
    C_{\phi \phi \mathcal{O}_3} G^{123,456}_{\Delta_1,\Delta_2,\Delta_3}(\chi_i)\,,
\end{equation}
where we distinguish the comb from the snowflake through the superscript with a single comma separating
two ordered triplets $123,456$. To specify a channel, one has to fix the order in which the OPEs in each
triplet are performed, as we did on the left hand side of the previous equation. In our notation for
blocks this is achieved by paying attention to the order of the fields in each of the triplets. For
the expansion at hand, the superscript $123,456$ encodes the OPE channel $((12)3)(4(56))$ whereas
a superscript $132,456$ would label the channel $((13)2)(4(56))$, to give just one example.

Let us note that comb-channel expansion \eqref{eq:combCBexpansion} includes two pairs of OPE
coefficients. The OPE coefficients of the first pair arise from the OPE of two external scalars.
These coefficients can be studied within the four-point bootstrap. In addition, we now have a
pair of new OPE coefficients in which (only) one field is an external scalar. While this is not
as generic as the new OPE coefficient we saw in the snowflake expansion above, the pairwise
structure of OPE coefficients in the comb topology is the feature that enables us to set
up a positive semi-definite bootstrap.

There is one important aspect of the new OPE coefficients in the comb-channel that we must pay
some attention to, namely their dependence on ordering. The OPE of two identical scalars $\phi$
only contains parity even fields and hence the order of operators is irrelevant for the OPE
coefficients $C_{\phi \phi \mathcal{O}_a}$ that appear in the snowflake and the comb-channel.
The same is actually true for the OPE coefficients $C_{\mathcal{O}_a \mathcal{O}_b \mathcal{O}_c}$
that appear with the central vertex in the snowflake-channel for six identical external fields.
Things are different for the OPE coefficients $C_{\mathcal{O}_1 \phi \mathcal{O}_2}$ with a
single external field $\phi$. In this case both parity even and parity odd fields can appear
in the operator product. Parity is the one-dimensional avatar of spin and leads to subtle minus
signs in correlators, see for example \cite{Homrich:2019cbt}. On the level of OPE coefficients,
these signs arise upon reordering of the fields as
\begin{equation}
C_{123} = (-1)^{\sigma_{123}} C_{321}\,, \quad \sigma_{123}\equiv P_1 + P_2 + P_3 \mod 2\,.
\end{equation}
Here, $P_i$ denotes the parity of the $i^\textrm{th}$ operator with the convention that $P=0$
for even and $P=1$ for odd operators.

Having discussed at length the OPE coefficients in the expansions \eqref{eq:snowCBexpansion}
and \eqref{eq:combCBexpansion}, we also briefly want to comment on the conformal blocks. While
in general space-time dimensions the physically relevant case where multiple spinning operators
are exchanged is still not understood in full generality.\footnote{See however \cite{Goncalves:2019znr,Parikh:2019dvm,Fortin:2019zkm,Poland:2021xjs,Buric:2020dyz,Buric:2021ttm,
Buric:2021kgy,Buric:2021ywo,Fortin:2023xqq} for recent progress in this direction.} in one dimension there are
drastic simplifications due to the kinematics and the absence of spin. Indeed, \cite{Rosenhaus:2018zqn}
provided closed form expressions in the form of multi-variable hypergeometric series expansions for
all 1d conformal blocks of comb-topology. In the six-point case with identical external operators,
we have
\begin{equation}
\label{eq:combblock}
   G^{123,456}_{\Delta_i}(\chi_i) = \chi_1^{\Delta_1}\chi_2^{\Delta_2}\chi_3^{\Delta_3}
   \setlength\arraycolsep{0.5pt}
{} F_K\left[\begin{matrix}\Delta_1,\Delta_1+\Delta_2-\Delta_\phi,\Delta_2+\Delta_3-\Delta_\phi,\Delta_3\\2\Delta_1,
2\Delta_2,2\Delta_3\end{matrix}; \chi_1,\chi_2,\chi_3\right]\,,
\end{equation}
where the comb-function $F_K$ is defined through the triple series
\begin{equation}
\label{eq:combfunction}
    \setlength\arraycolsep{0.5pt}
{} F_K\left[\begin{matrix}a_1,b_1,b_2,a_2\\c_1,c_2,c_3\end{matrix}; \chi_1,\chi_2,\chi_3\right]
\equiv \sum_{n_i=0}^{\infty}
\frac{(a_1)_{n_1}(b_1)_{n_1+n_2}(b_2)_{n_2+n_3}(a_2)_{n_3}}{(c_1)_{n_1}(c_2)_{n_2}(c_3)_{n_3}}
\frac{\chi_1^{n_1}}{n_1!}\frac{\chi_2^{n_2}}{n_2!}\frac{\chi_3^{n_3}}{n_3!}\,.
\end{equation}
These functions have power law behaviour in each of the $\chi_i$ variables, making the identification
of the dimensions of exchanged operators $\mathcal{O}_1,\mathcal{O}_2,\mathcal{O}_3$ straightforward.

\subsection{The crossing equation and positive semi-definiteness}
\label{sec:sixpt-crossing}

This subsection describes the comb-topology crossing equations and derives an SDP dual to them.
More concretely, we shall argue that crossing symmetry allows us to construct families of symmetric
bilinear forms $B^k_{\Delta_2}$ (defined in eq.~\eqref{def:Bk}) with respect to which certain vectors
of OPE coefficients (defined in eq.~\eqref{eq:defCOO}) are null. This statement is concisely captured
by the main equation of this subsection, namely eq. \eqref{eq:crossingB}. It also has an obvious SDP
dual: The task of constructing positive definite linear combinations of said bilinear forms to rule
out the existence of solutions to the crossing equations.

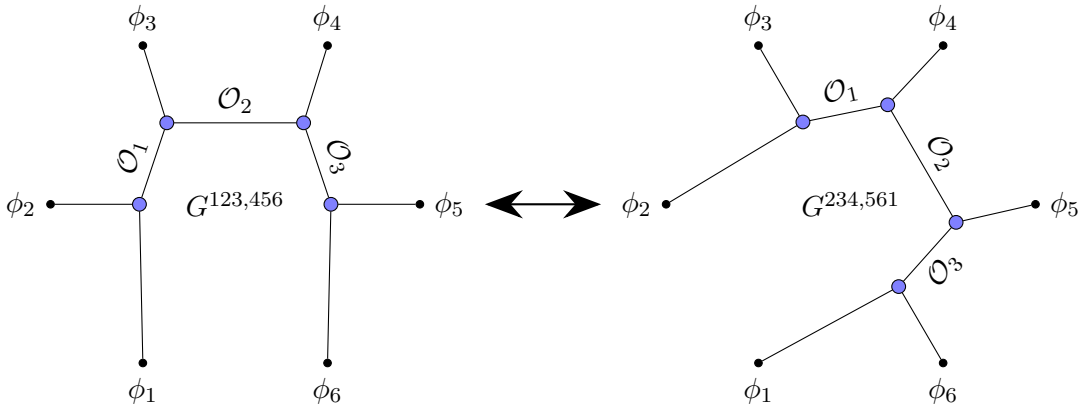
\begin{figure}[ht]
    \centering
    \begin{tikzpicture}[scale=0.9]
   \newdimen\R
   \R=2.7cm
   \foreach \x/\l/\p in
     { 60/{$\phi_4$}/above,
      120/{$\phi_3$}/above,
      180/{$\phi_2$}/left,
      240/{$\phi_1$}/below,
      300/{$\phi_6$}/below,
      360/{$\phi_5$}/right
     }
     \node[inner sep=1pt,draw,circle,fill,label={\p:\l}] (\x) at (\x:\R) {};
      \node[shape=circle,fill=blue!50,draw=black, scale = 0.5] (V1) at (-1.4,0) {};
        \node[shape=circle,fill=blue!50,draw=black, scale = 0.5] (V4) at (1.4,0) {};
        \node[shape=circle,fill=blue!50,draw=black, scale = 0.5] (V3) at (1,1.2) {};
        \node[shape=circle,fill=blue!50,draw=black, scale = 0.5] (V2) at (-1,1.2) {};

        \draw (240) -- (V1);
        \draw (180) -- (V1);
        \draw (120) -- (V2);
        \draw (60) -- (V3);
        \draw (360) -- (V4);
        \draw (300) -- (V4);
        \path[-,sloped, draw] (V1) edge node[above] {$\mathcal{O}_1$} (V2);
        \path[-, draw] (V2) edge node[above] {$\mathcal{O}_2$} (V3);
        \path[-, sloped, draw] (V3) edge node[above] {$\mathcal{O}_3$} (V4);

        \node[] (T) at (0,0) {$G^{123,456}$};

        \def\x{9};
        \def\l{1};
        \node (A1) at (3.5,0) {};
        \node (A2) at (5.5,0) {};
        \draw[{Stealth[length=5mm]}-{Stealth[length=5mm]},thick] (A1) -- (A2);

        \begin{scope}[xshift=9 cm]
            \newdimen\R
   \R=2.7cm
   \foreach \x/\l/\p in
     { 60/{$\phi_4$}/above,
      120/{$\phi_3$}/above,
      180/{$\phi_2$}/left,
      240/{$\phi_1$}/below,
      300/{$\phi_6$}/below,
      360/{$\phi_5$}/right
     }
     \node[inner sep=1pt,draw,circle,fill,label={\p:\l}] (\x) at (\x:\R) {};
      \node[shape=circle,fill=blue!50,draw=black, scale = 0.5] (V1) at ({-1.4*cos(60)},{1.4*sin(60)}) {};
        \node[shape=circle,fill=blue!50,draw=black, scale = 0.5] (V4) at ({1.4*cos(60)},{-1.4*sin(60)}) {};
        \node[shape=circle,fill=blue!50,draw=black, scale = 0.5] (V3) at ({cos(60)+1.2*sin(60)},{1.2 * cos(60)-sin(60)}) {};
        \node[shape=circle,fill=blue!50,draw=black, scale = 0.5] (V2) at ({-cos(60) + 1.2*sin(60)},{1.2*cos(60)+sin(60)}) {};

        \draw (180) -- (V1);
        \draw (120) -- (V1);
        \draw (60) -- (V2);
        \draw (360) -- (V3);
        \draw (300) -- (V4);
        \draw (240) -- (V4);
        \path[-, sloped,draw] (V1) edge node[above] {$\mathcal{O}_1$} (V2);
        \path[-,sloped, draw] (V2) edge node[above] {$\mathcal{O}_2$} (V3);
        \path[-,sloped, draw] (V3) edge node[below] {$\mathcal{O}_3$} (V4);

        \node[] (T) at (0,0) {$G^{234,561}$};
        \end{scope}
    \end{tikzpicture}
        \caption{The OPE diagram for the crossing of six-point comb-channel expansions.}  \label{fig:combcrossing}
\end{figure}

In one dimension, it is natural to consider OPE channels that share the same cyclic ordering of points
to derive crossing equations. All permutations that preserve cyclic ordering are generated single cycle
permutation that shifts $x_i \to x_{i+1}$ with the periodic identification $x_7 = x_1$. The associated
crossing equation that relates comb-channel expansions $123,456 = ((12)3)(4(56)$ and $234,561=((23)4)
(5(61))$ is depicted graphically in Figure~\ref{fig:combcrossing}. The corresponding conformal blocks
are related by
\begin{equation}\label{eq:CrossingBlock}
    \mathcal{L}(x_i)G^{234,561}(\chi_j(x_{i})) = \mathcal{L}(x_{i+1})G^{123,456}(\chi_j(x_{i+1}))\,.
\end{equation}
Here, $\mathcal{L}(x_i)$ is the leg factor we defined by eq.\ \eqref{eq:legg} and the superscripts in our notation of the comb-channel blocks follow the conventions we introduced after
eq.\ \eqref{eq:combCBexpansion}.

Upon the cyclic shift $x_i \to x_{i+1}$, the cross ratios we introduced in eq.
\eqref{eq:CR} behave as
\begin{align}
    \chi_j(x_{i+1}) = \chi_{j+1}(x_i) \, ,
\end{align}
for $j=1,2,3$, provided we introduce $\chi_4$ as
\begin{align}
    \chi_4(x_i) \equiv  \chi_3(x_{i+1}) =
    \frac{ \chi_1 + \chi_2 + \chi_3 - \chi_1 \chi_3-1}{\chi_1 + \chi_2-1}\,.
\end{align}
In order to write down the crossing equation we depicted in Figure \eqref{fig:combcrossing}, we shall introduce
a shorthand for the ratio of leg factors on the two sides of the crossing equation,
\begin{equation}
Q(\chi_j)\equiv \frac{\mathcal{L}(x_{i+1})}{\mathcal{L}(x_{i})} =
\left(\frac{\chi _1^2 \chi _2 \chi _3}
{\left(\chi _1+\chi _2+\chi _3-\chi _1 \chi _3-1\right){}^2}\right)^{\Delta_\phi}\,.
\end{equation}
Note that, by construction of the leg factors, the ratio $Q(\chi_i)$ depends on the insertion points $x_i$ only
through the cross ratios $\chi_j$, but we will henceforth omit this dependence. It will also be convenient to abbreviate products of OPE coefficients
as
\begin{equation}\label{eq:defCOO}
    C^{\mathcal{O}_j}_{\mathcal{O}_2} \equiv C_{\phi \phi \mathcal{O}_j}
    C_{\mathcal{O}_j \phi \mathcal{O}_2}\,.
\end{equation}
In applications to our crossing equation, the index $j$ assumes the values $j=1,3$. After this preparation,
the crossing equation under consideration can be written in the form
\begin{equation}\label{eq:crossing_eq}
    \sum_{\mathcal{O}_1,\mathcal{O}_2,\mathcal{O}_3} C^{\mathcal{O}_1}_{\mathcal{O}_2}
     G^{123,456}_{\Delta_1,\Delta_2,\Delta_3}(\chi_1,\chi_2,\chi_3) C^{\mathcal{O}_3}_{\mathcal{O}_2} =
    \sum_{\mathcal{O}_1,\mathcal{O}_2,\mathcal{O}_3}
    Q C^{\mathcal{O}_1}_{\mathcal{O}_2}
    G^{123,456}_{\Delta_1,\Delta_2,\Delta_3}(\chi_2,\chi_3,\chi_4) C^{\mathcal{O}_3}_{\mathcal{O}_2}
     \,.
\end{equation}
The weights $\Delta_j$ that label the conformal blocks in eq.~\eqref{eq:crossing_eq} are the conformal weights of the operators $\mathcal{O}_j$ that are summed over. Before we rewrite the crossing
equation in the standard form of the numerical conformal bootstrap, we shall introduce the so-called
\textit{crossing vector} as
\begin{align}
    \tilde{F}_{\Delta_1,\Delta_2,\Delta_3}(\chi_1,\chi_2,\chi_3)=G^{123,456}_{\Delta_1,\Delta_2,\Delta_3}
    (\chi_1,\chi_2,\chi_3) - Q G^{123,456}_{\Delta_1,\Delta_2,\Delta_3}(\chi_2,\chi_3,\chi_4)\ .
\end{align}
This allows us to bring eq.\ \eqref{eq:crossing_eq} into the following very concise form
\begin{equation} \label{eq:crossingtF}
    \sum_{\mathcal{O}_1,\mathcal{O}_2,\mathcal{O}_3} C^{\mathcal{O}_1}_{\mathcal{O}_2}
    \tilde{F}_{\Delta_1,\Delta_2,\Delta_3}(\chi_1,\chi_2,\chi_3)C^{\mathcal{O}_3}_{\mathcal{O}_2} =0\,.
\end{equation}
Writing the crossing equation in this way makes manifest that it can be seen as a sum of matrices
labelled by $\Delta_2$, which are contracted with two copies of the vector $C_{\mathcal{O}_2} \equiv
(C_{\mathcal{O}_2}^\mathcal{O})_\mathcal{O}$ whose components are labelled by primaries $\mathcal{O}$
that appear in the OPE of two external scalars $\phi$. Since we are contracting the crossing
vector $\tilde{F}_{\Delta_1,\Delta_2,\Delta_3}(\chi_1,\chi_2,\chi_3)$ with a symmetric tensor, we
can symmetrise with respect to the indices $\Delta_1$ and $\Delta_3$ to deduce
\begin{align}\label{eq:crossingF}
    \sum\limits_{\mathcal{O}_2} \left(\sum\limits_{\mathcal{O}_1 \mathcal{O}_3}
    C^{\mathcal{O}_1}_{\mathcal{O}_2} F_{\Delta_1 \Delta_2 \Delta_3}
    (\chi_1,\chi_2,\chi_3)C^{\mathcal{O}_3}_{\mathcal{O}_2}\right) = 0\ .
\end{align}
In passing from equation \eqref{eq:crossingtF} to the new crossing equation \eqref{eq:crossingF}, we
have introduced the symmetrised crossing vector $F$ as
\begin{align}\label{eq:defF}
     F_{\Delta_1,\Delta_2,\Delta_3}\equiv \tilde{F}_{\Delta_1,\Delta_2,\Delta_3}+
     \tilde{F}_{\Delta_3,\Delta_2,\Delta_1}.
\end{align}
The symmetrization proceedure we applied in order to get to the crossing equation \eqref{eq:crossingF}
can also be understood as the result of adding a second crossing equation to the original one that was
given in eq.~\eqref{eq:crossing_eq}. The second equation we need to add is obtained from the first
one by exchanging $\Delta_1$ and $\Delta_3$. Hence, our symmetrisation boils down to adding the
crossing equation for the comb-channels $654,123$ and $543,216$.

Following the usual lore of the numerical conformal bootstrap, we do not directly study
eq.~\eqref{eq:crossingF} as a functional equation for arbitrary $\chi_i$. Instead, we choose some
finite set of linear functionals $\{\alpha_k\}_{k=1}^{N_\alpha}$ acting on the crossing vector $F$.
These functionals then allow us to focus on the finite subset
\begin{align}\label{eq:crossingalpha}
    \sum\limits_{\mathcal{O}_2} \left(\sum\limits_{\mathcal{O}_1 \mathcal{O}_3}
    C^{\mathcal{O}_1}_{\mathcal{O}_2} \alpha_k(F_{\Delta_1 \Delta_2 \Delta_3})
    C^{\mathcal{O}_3}_{\mathcal{O}_2}\right) = 0\,, && k = 1, \dots, N_\alpha\,,
\end{align}
of the a priori infinitely many constraints imposed by the crossing equation \eqref{eq:crossingF}.
An obvious candidate for the choice of functionals $\alpha$ is to apply differential operators
$\mathcal{D}_k$ in the three cross-ratios and to evaluate the result for some fixed kinematics \footnote{See however \cite{Paulos:2019fkw,Ghosh:2023onl} for recent usage of other classes of functionals in the numerical bootstrap.}.
The canonical prescription for the evaluation is the crossing symmetric point which is $\chi_j
= 1/3$ in our case. We shall adopt these choices throughout the rest of the work.

Thanks to the symmetrisation procedure above, we can now define a family of \emph{symmetric} bilinear
forms $\{B_{\Delta_2}^k\}_{k=1}^{N_\alpha}$. The arguments of these forms are vectors $C =
(C^\mathcal{O})_\mathcal{O}$ whose components are labelled by the primary fields $\mathcal{O}$
that appear in the OPE of two external scalars $\phi$. Given any two such vectors $C_L$ and
$C_R$, the bilinear forms are introduced as
\begin{align}\label{def:Bk}
    B_{\Delta_2}^k(C_L,C_R):= \sum\limits_{\mathcal{O}_1 \mathcal{O}_3}
    C_L^{\mathcal{O}_1}\alpha_k(F_{\Delta_1 \Delta_2 \Delta_3})C_R^{\mathcal{O}_3}\,
\end{align}
for all $\Delta_2$ and $k=1, \dots, N_\alpha$. In other words, the bilinear forms $B^k_{\Delta_2}$ can
simply be thought of as matrices whose rows and columns are labelled by the operators exchanged in the
$\phi\times\phi$ OPE, namely
\begin{equation}(B^k_{\Delta_2})_{\Delta_1,\Delta_3}=\alpha_k(F_{\Delta_1 \Delta_2 \Delta_3})\,.
\end{equation}
With this definition, our crossing equation \eqref{eq:crossingalpha} finally takes the advocated form
\begin{align}\label{eq:crossingB}
    \sum\limits_{\mathcal{O}_2} B_{\Delta_2}^k(C_{\mathcal{O}_2},C_{\mathcal{O}_2}) = 0\,, && k = 1,
    \dots, N_\alpha.
\end{align}
Equation \eqref{eq:crossingB} is the starting point of the SDP based six-point bootstrap: As in the
four-point bootstrap, the existence of positive definite linear combinations of the bilinear forms
$B^k$ puts rigorous constraints on the spectra of CFTs. For example, if one can find numerical
coefficients $c_k, k = 1, \dots, N_\alpha,$ such that
\begin{equation}\label{eq:PrimarySpaceSDP}
\sum_{k}c_k B^k_{\Delta_2} \succ 0\quad \text{for all} \quad \Delta_2\geq \Delta_2^*\,,
\end{equation}
one can conclude that the leading operator in the $\mathcal{O}_2$ OPE exchange must have conformal
weight $\Delta_2 < \Delta_2^*$. This follows by contracting with the vector $C_{\mathcal{O}_2}$ of
eq.~\eqref{eq:defCOO} and using equation \eqref{eq:crossingB}. Put differently, the existence of
numerical coefficients $c_k$ that satisfy the condition \eqref{eq:PrimarySpaceSDP} provides an
upper bound of $\Delta_2^*$ on the leading operator that is exchanged in the middle leg.

The proposition for a numerical six-point bootstrap discussed so far is close to the usual four-point
bootstrap formulated directly in terms of OPE coefficients and conformal blocks. However, the numerical
search for solutions of eq.\ \eqref{eq:PrimarySpaceSDP} in the framework of this subsection is quite
challenging: If one naively discretises these equations by considering some finite list of possible
values for the weights $\Delta_1$, $\Delta_2$ and $\Delta_3$, one obtains large dense matrices that
lack a particularly useful asymptotic behaviour which would control the convergence of the
discretisation procedure. Finding an alternative formulation of eq.~\eqref{eq:crossingB} which
is more suitable for numerics, is the main goal of the next subsection.

\subsection{The descendant-space formulation of six-point crossing}
\label{sec:demonstrateequiv}

We now refine the ideas of the previous subsections and introduce a new set of bilinear forms
$b^k_{\Delta_2}$, see eq.~\eqref{eq:defM}, that will become central for the subsequent numerical
analysis. The transition from the bilinear forms $B^k_{\Delta_2}$ we introduced in the previous
Subsection~\ref{sec:sixpt-crossing} to the new forms $b^k_{\Delta_2}$ reduces the complexity of
the corresponding SDPs and it makes the constraints that we impose physically more transparent. Ultimately though, our new formulation turns out to be equivalent to that of the previous section.

The transition to our second formulation starts with the following representation of the comb
channel six-point blocks \eqref{eq:combblock},
\begin{align}\label{eq:Gscprod}
    G_{\Delta_1,\Delta_2,\Delta_3}(\chi_1,\chi_2,\chi_3) = \sum_n F_n(\Delta_1, \chi_1)
    \frac{\chi_2^{n+\Delta_2}}{(2 \Delta_2)_n n!} F_n(\Delta_3,\chi_3) \,,
\end{align}
where we have absorbed sums over $n_1$ and $n_3$ in eq.~\eqref{eq:combfunction} into the definition
of the functions
\begin{align}
\label{eq:phiphiphiO2n}
   F_n(\Delta_i,\chi_i) := (\Delta_i + \Delta_2 - \Delta_\phi)_{n}\chi_i^{\Delta_i
}\,_2F_1(\Delta_i,\Delta_i+\Delta_2-\Delta_\phi+n,2\Delta_i,\chi_i)\,.
\end{align}
Up to normalisation, this is the four-point conformal block for the exchange of $\mathcal{O}_i$ in a correlator of three operators with dimension $\Delta_\phi$ and one (primary) operator with dimension $\Delta_2 + n$. It is tempting to physically identify the operator of dimension $\Delta_2 + n$ with a descendant of $\mathcal{O}_2$. Section~\ref{sec:commentson} clarifies the exact relation to the correlator $\langle \phi \phi \phi \mathcal{O}_2^{(n)}\rangle$, where $\mathcal{O}_2^{(n)}$ is a level $n$ descendant of $\mathcal{O}_2$.

The functions $F_n$ possess a very helpful feature: Derivatives with respect to the cross ratio simply act by shifting the index $n$. Indeed, it is easy to see that
\begin{align}\label{eq:varthetaFn}
    \chi_1 \partial_{\chi_1} F_n(\Delta_1,\chi_1)  =  F_{n+1}(\Delta_1,\chi_1)
    - (\Delta_2+n-\Delta_\phi) F_n(\Delta_1,\chi_1) \,,
\end{align}
where crucially the coefficients in front of the $F_n$ functions on the r.h.s.~do not
depend on $\Delta_1$. Of course, there is a similar formula with $1 \to 3$. One can view eq.~\eqref{eq:varthetaFn} simply as a mathematical consequence of the hypergeometric nature of comb-channel blocks. However, the interpretation of $F_n$ in terms of correlators of descendants alluded to above allows us to assign some physical meaning to eq.~\eqref{eq:varthetaFn} in Section~\ref{sec:commentson}.

For our program, eq.\ \eqref{eq:varthetaFn} is very useful. Recall that the linear functionals
$\alpha$ we introduced in the previous subsection involve acting with some partial differential
operators in the three cross ratios. Now we see that the action of a such a differential operator
on the comb blocks preserves the bilinear structure of equation \eqref{eq:Gscprod}. Concretely,
acting with a differential operator $\mathcal{D}$ of order $\Lambda$ results in
\begin{align}
\label{eq:Mdef}
    \mathcal{D} G_{\Delta_1,\Delta_2,\Delta_3}(\chi_1,\chi_2,\chi_3) = \sum_{n,m}
    F_n(\Delta_1, \chi_1)(M[\mathcal{D}])_{n m} F_m(\Delta_3,\chi_3)\,,
\end{align}
where $M[\mathcal{D}]$ is a bandmatrix whose bandwith is bounded from above by the order
$\Lambda$ of $\mathcal{D}$ with respect to $\chi_1$ and $\chi_3$, and whose entries are functions
of \emph{only $\Delta_2$} and the cross-ratios. Consequently, if the $N_\alpha$ functionals $\alpha_k$
with which we act on the crossing equation are given by evaluating partial derivatives with respect
to the three cross-ratios at the crossing symmetric point
\begin{align}
    \chi_1 = \chi_2 = \chi_3 = 1/3\,,
\end{align}
then the bilinear form corresponding to a functional $\alpha_k$ of derivative order  $\Lambda_k$ has
the form
\begin{align}\label{eq:defM}
    B_{\Delta_2}^k(C_L,C_R) = \sum\limits_{n,m} F_n(C_L) M^k_{nm}(\Delta_2) F_m(C_R) =:
    b_{\Delta_2}^k(F(C_L),F(C_R))\,,
\end{align}
where the bandwidth of $M^k$ is bounded above by $\Lambda_k$,  The symbol $F(C)$ denotes the sequence
$(F_n(C))_{n \in \mathbb{N}}$ defined by
\begin{align}\label{eq:Fn(C)}
    F_n(C) := \sum\limits_{\mathcal{O}} F_n(\Delta_\mathcal{O},1/3)C^{\mathcal{O}}\,.
\end{align}
In other words, the bilinear form $B_{\Delta_2}^k$ is the pullback of a bilinear form $b_{\Delta_2}^k$
defined on some real sequence space with respect to the linear map $F$. Hence, $B_{\Delta_2}^k$ is
positive semi-definite if and only if $b_{\Delta_2}^k$ is positive semi-definite on the image of $F$.

Based on numerical tests, we consider it highly plausible that the map $F$ is surjective and that
therefore the positive semi-definiteness of $b_{\Delta_2}^k$ is equivalent to that of $B_{\Delta_2}^k$.
This strongly supports the claimed equivalence between the two pictures of crossing that we have presented.
Note, however, that independently of the surjectivity of the linear map $F$, a positive semi-definite linear
combination of the $b_{\Delta_2}^k$ always corresponds to a positive semi-definite combination of the
$B_{\Delta_2}^k$. Thus, establishing surjectivity more rigorously would only improve our a posteriori
understanding for why our methods are effective, but it is not necessary to justify their validity.

The main practical consequence of this discussion is that, instead of constructing positive
semi-definite linear combinations of the bilinear forms $B_{\Delta_2}^k$, we can equally well
work with the $b_{\Delta_2}^k$. Hence, our second formulation of the SDP problem for six-point
crossing amounts to finding numerical coefficients $c_k$ that satisfy
\begin{equation}
\label{eq:positivesemidefb}
\sum_{k}c_k b^k_{\Delta_2} \succ 0 \quad \text{for all} \quad  \quad \Delta_2\geq \Delta_2^*\.
\end{equation}
We shall refer to this formulation as \textit{descendant-space} SDP problem and to the objects $b^k$
it involves as \textit{descendant-space bilinear forms}. This is to be contrasted with the original
\textit{primary-space} SDP problem \eqref{eq:PrimarySpaceSDP} which is based on the \textit{primary-space
bilinear forms} $B^k$.

It turns out that the descendant-space formulation of the SDP problem has a number of important advantages
over the primary-space formulation. Let us briefly list them here.

\begin{enumerate}
    \item
    One immediate qualitative difference between descendant- and primary-space is that in the descendant-space we
    do not need to evaluate the six-point block themselves, only establish differential relations between them.
    This is conceptually quite different from what is done in the standard four-point bootstrap, and suggests that
    a higher-dimensional generalisation might be possible, even without full control over the blocks.
    \item
    A second difference is that the linear map \eqref{eq:Fn(C)} sends the continuous parameter
    $\Delta_\mathcal{O}$ to the discrete label $n$. Any attempt to numerically construct positive semi-definite
    linear combinations of the bilinear forms that we consider necessarily has to address their infinite dimensional
    nature, in practice by some truncation to finite submatrices. In primary-space, this truncation requires both
    cutting-off the matrices at a maximal scaling dimension $\Delta_{cut}$ and a choice of discretisation of the
    interval $[0,\Delta_{cut}]$. In contrast to this, the discrete nature of the descendant-space label $n$ allows
    for a truncation of the bilinear forms to finite matrices simply by picking a maximal value for their indices.
    \footnote{In both cases, we still truncate and discretise with respect to the $\Delta_2$ variable.}
    \item
    The third and most important advantage of the descendant-space approach is that the bilinear forms $b^k$ it
    works with are banded. This suggests that they are much easier to handle numerically than the dense matrices
    $B^k$ encountered in primary-space.
\end{enumerate}
These advantages, however, could not be profited of, without firm control over the convergence of the truncation
procedure that we necessarily have to implement. Thus, we devote the next subsection to the asymptotic structure
of the banded matrices occurring in descendant-space.

\subsection{Large dimension asymptotics and the Toeplitz arc}\label{sec:toeplitzarc}

This subsection is organised in three logical steps that lead from simple asymptotic properties
of the descendant-space bilinear forms to an algorithm that stabilises the convergence of truncation
schemes for SDPs involving these bilinear forms. We begin by making observations about the asymptotics
of descendant-space bilinear forms for large $\Delta_2$ as well as far down their diagonal. Concretely,
these are captured by eqs.~\eqref{eq:asymptotics1} and \eqref{eq:asymptotics2}. This sets the stage for
a concise characterisation of the asymptotics of general large $\Delta_2$ and $n$ regimes with fixed
ratio $\Delta_2/n$. It is in this characterisation, that a curve of Toeplitz matrices enters that is
referred to in the subsection title. Let us recall that a matrix $\mathcal{T}$ is called Toeplitz if
it is of the form
 \begin{align}\label{eq:Toeplitz}
    \mathcal{T}_{ij} = \sum\limits_{n = -\infty}^\infty \delta_{i-j}^n a_n,
\end{align}
where $(a_n)_{n \in \mathbb{Z}}$ is a bounded family of complex numbers. In the final step, we establish
that positive semi-definiteness of the matrices constituting the Toeplitz arc is equivalent to positivity
of certain polynomials. This is the central result of this subsection and is concisely formulated in
eq.~\eqref{eq:asymptpoly}.
\medskip

Let us start by using the elementary fact that positive semi-definiteness is a basis independent
property so that we are free to conjugate the bilinear forms under consideration by an arbitrary
invertible matrix. Concretely, positive semi-definiteness of some linear combination of bilinear
forms $\{b_{\Delta_2}^k\}_{k=1}^{N_\alpha}$ associated, in the sense of Section
\ref{sec:demonstrateequiv}, to partial derivatives of order $\Lambda_k$, is most conveniently
shown after conjugation with the matrix
\begin{align}
    N_{ij} \equiv \sum\limits_{n=0}^\infty \delta^n_i \delta^n_j
    \sqrt{\frac{3^n (2 \Delta_2)_n n!}{(\Delta_2+n)^\Lambda}}
    && \text{where} && \Lambda \equiv \max\limits_{k=1,\dots,N_\alpha} \Lambda_k.
\end{align}
It follows directly from eq.~\eqref{eq:varthetaFn} that the matrix $\hat{M}^k$ obtained by
conjugating the $M^k$ defined in eq.~\eqref{eq:defM} with $N$, has the following asymptotic
properties
\begin{align}\label{eq:asymptotics1}
    \hat{M}^k_{nm}(\Delta_2) & \sim \Delta_2^{\Lambda_k - \Lambda -\frac{1}{2}|n-m|}
    \quad & \text{for} \quad \Delta_2 \gg 1 \quad \text{and} \quad n \ \text{fixed,}  \\[2mm]
      \label{eq:asymptotics2}
    \hat{M}^k_{nm}(\Delta_2) & \sim n^{\Lambda_k - \Lambda} \hspace*{15mm} &  \text{for} \quad
    n \gg 1 \quad \text{and} \quad \Delta_2 \ \text{fixed.}
\end{align}
To see this statement in action, let us consider as an example the functional associated to the
first derivative $\partial_{\chi_1}$. Acting with this functional on the crossing vector of
eq.~\eqref{eq:defF} produces the combination
\begin{align}\label{eq:dchi1}
    \partial_{\chi_1} F_{\Delta_1,\Delta_2,\Delta_3}=
    -18 \Delta_\phi G_{\Delta_1,\Delta_2,\Delta_3} + \partial_{\chi_1}G_{\Delta_1,\Delta_2,\Delta_3} +
    \partial_{\chi_3}G_{\Delta_1,\Delta_2,\Delta_3}
\end{align} of derivatives of the comb-channel block evaluated at the crossing symmetric point.
Using eqs.~\eqref{eq:Gscprod} and \eqref{eq:varthetaFn}, this can be rewritten as
\begin{align}
    \partial_{\chi_1} F_{\Delta_1,\Delta_2,\Delta_3}= \hspace{-0.2 cm}\sum\limits_{n,m=0}^\infty
    \hspace{-0.2 cm} F_n\left(\Delta_1,\frac{1}{3}\right) \frac{ \delta_{n-1}^{m}+\delta_{n+1}^{m}- 2
    \delta^m_{n}(\Delta_2 + n + 2 \Delta_\phi) }{3^{n-1} ( 2\Delta_2)_n n!} F_m\left(\Delta_1,\frac{1}{3}\right)\,.
\end{align}
Hence, the matrix associated to this functional by eq.~\eqref{eq:defM} is
\begin{align}
    M_{nm}^{\partial_{\chi_1}} = \frac{ \delta_{n-1}^{m}+\delta_{n+1}^{m}- 2 \delta^m_{n}(\Delta_2 + n + 2 \Delta_\phi) }{3^{n-1} ( 2\Delta_2)_n n!} \,.
\end{align}
After conjugation by $N$, defined here with $\Lambda = 1$, we end up with
\begin{align}\label{eq:Mhatdelchi1}
   \hat{M}_{nm}^{\partial_{\chi_1}} = 3 \sqrt{3} \sqrt{\frac{n+1}{2 \Delta_2+n+1}}(\delta_{n-1}^{m}+\delta_{n+1}^{m}) -\frac{6(\Delta_2 + n + 2 \Delta_\phi)}{2 \Delta_2+n}\delta_n^m\,
\end{align}
which has the anticipated asymptotic properties \eqref{eq:asymptotics1} and \eqref{eq:asymptotics2} for
$\Lambda - \Lambda_k = 0$.
\bigskip

As announced in the beginning of this subsection, let us now move on to the more general case where both $\Delta_2$ and $n$ are taken to infinity with the ratio $\Delta_2/n$ fixed. To describe this regime, we associate to each derivative functional $\alpha_k$ the following one-parameter family of real symmetric Toeplitz matrices:
\begin{align}\label{eq:asymptoep}
    \mathcal{T}^k_{ij}(\theta) \equiv \lim\limits_{n \rightarrow \infty} \hat{M}^k_{n+i,n+j}(n \tan(\theta))
    \,, && \theta \in [0,\pi/2[\,.
\end{align}
Clearly, eqs.~\eqref{eq:asymptotics1} and \eqref{eq:asymptotics2} imply that $\mathcal{T}^k$ vanishes for
$\Lambda_k < \Lambda$ and that $\mathcal{T}^k$ approaches a multiple of the identity matrix as $\theta$
approaches $\pi/2$. For $\Lambda_k = \Lambda$ and generic values of $\theta$, however, one obtains a
banded Toeplitz matrix with bandwidth $\Lambda_k$. In the example of the single derivative with respect
to the cross ratio $\chi_1$, this Toeplitz matrix takes the form
\begin{align}\label{eq:exampleT}
   \mathcal{T}^{\partial_{\chi_1}}_{ij}(\theta) =
   \sqrt{\frac{3}{1+2 \tan (\theta )}} (\delta^i_{j-1}+\delta^i_{j+1})-
   \frac{6 (1+\tan (\theta ))}{1+2 \tan (\theta )}  \delta^i_j\,.
\end{align}

A necessary condition for the positive semi-definiteness of a linear combination of functionals for all values of
$\Delta_2$ larger than some arbitrary finite value $\Delta_*$ is that the corresponding linear combination of asymptotic
Toeplitz matrices is positive semi-definite for all values of $\theta$, i.e. our SDP condition \eqref{eq:positivesemidefb}
implies that
\begin{align}
\mathcal{T}(\theta) \equiv \sum\limits_{k=1}^{N_\alpha} c_k \mathcal{T}^k(\theta) \succeq 0\, \quad
\text{ for all } \quad  \theta \in [0,\pi/2[ \ .
\end{align}
What makes this necessary criterion for positive semi-definiteness especially useful is that it is easy
to check. To do so, we rely on the following basic result of the theory of Toeplitz operators (see for
instance section 1.5 of \cite{BoettcherGrudsky}).

\begin{theorem*}[Spectra of Toeplitz operators.]\label{theo:Toeplitz} Let $\mathcal{T}$ be the Toeplitz
operator, i.e. an infinite matrix of the form \eqref{eq:Toeplitz}. Define the symbol of $\mathcal{T}$ as
the curve
\begin{align} \label{eq:Tpos}
    \gamma_\mathcal{T}: [0,2 \pi] \rightarrow \mathbb{C}, \, t \mapsto \sum\limits_{n = -\infty}^\infty a_n e^{i n t}.
\end{align}
Then, $\mathcal{T}$ is a Fredholm operator on $\ell^p(\mathbb{C})$ (for $1\leq p \leq \infty$) iff
$0 \notin \gamma_\mathcal{T}([0,2 \pi]) $. In that case, the Fredholm Index of $\mathcal{T}$ is minus
the winding number of $\gamma_\mathcal{T}$ around $0$. Thus, the essential spectrum of $\mathcal{T}$
is $\gamma_\mathcal{T}([0,2 \pi])$. The spectrum of $\mathcal{T}$ is the set of points around which
$\gamma_{\mathcal{T}}$ has a non trivial winding.
\end{theorem*}

\noindent This theorem implies that the Toeplitz matrix $\mathcal{T}(\theta)$ from equation
\eqref{eq:asymptoep} is positive definite iff its symbol
\begin{align}\label{eq:gamma}
    \gamma_{\mathcal{T}(\theta)}(t) = \mathcal{T}_{00}(\theta) + 2 \sum\limits_{n=1}^\Lambda
    \mathcal{T}_{0n}(\theta) \cos(n t)
\end{align}
is positive for all $t \in [0,2\pi]$. It is actually possible to rephrase this as a positivity
condition on a polynomial in two variables. In a first step we shall employ the transformation
\begin{align}
    \cos(t) = \frac{x^2 - 1}{1 + x^2}\,,
\end{align}
to turn the trigonometric dependence in $t$ into a rational dependence on the new variable $x$.
Using the Chebyshev polynomials $T_n$ the positivity condition on the symbol \eqref{eq:gamma}
can be rewritten as
\begin{align}
    \gamma_{\mathcal{T}(\theta)}(x) \equiv \mathcal{T}_{00}(\theta) + 2
    \sum\limits_{n=1}^\Lambda \mathcal{T}_{0n}(\theta)
    T_n\left(\frac{x^2-1}{1+x^2}\right) > 0 \,, \quad \text{for all} \quad  x \in \mathbb{R}\,,
\end{align}
which in turn is equivalent to the positivity of the polynomial
\begin{align}
    p_{\mathcal{T}(\theta)}(x) \equiv (1+x^2)^\Lambda \gamma_{\mathcal{T(\theta)}}(x) \,.
\end{align}
Clearly, $p_{\mathcal{T}}(x)$ is linear in $\mathcal{T}$, so
\begin{align}
     p_{\mathcal{T}(\theta)}(x) = \sum\limits_{k=1}^{N_\alpha} c_k p_{\mathcal{T}^k(\theta)}(x).
\end{align}
Now we also want to turn the dependence on the variable $\theta$ into a rational function. To this
end, we can perform the reparametrisation
\begin{align}\label{eq:ytrafo}
    \tan(\theta) = \frac{(1 + y^2)^2 - 1}{2}.
\end{align}
This indeed guarantees that the dependence of the functions $p_{T^k(\theta)}(x)$ on the position
$\theta$ along the arc becomes rational too. By multiplying with an appropriate positive factor,
one ends up with a polynomial dependence in the variable $y$. We denote by $P_k(x,y)$ the
resulting two variable polynomial associated to a functional $\alpha_k$.

To illustrate the construction of the polynomial $P$, let us consider once again our example of
the functional that is given by the derivative with respect to $\chi_1$. From the explicit expression
for the arc of asymptotic Toeplitz matrices associated to $\partial_{\chi_1}$ that we spelled out in
eq.~\eqref{eq:exampleT}, we find
\begin{align}
    p_{\mathcal{T}^{\partial_{\chi_1}}(\theta)}(x) = 6 \sqrt{3} \left(x^2-1\right)
    \sqrt{\frac{1}{2 \tan (\theta )+1}}-\frac{6 \left(x^2+1\right) (\tan (\theta )+1)}{2 \tan (\theta )+1}.
\end{align}
Applying transformation \eqref{eq:ytrafo} and multiplying by $(1 + y^2)^2$, we obtain the
following polynomial
\begin{align*}
    P_{\partial_{\chi_1}}(x,y) = -3 x^2 y^4+\left(6 \sqrt{3}-6\right) x^2 y^2+\left(6 \sqrt{3}-6\right)
    x^2-3 y^4-\left(6+6 \sqrt{3}\right) y^2-6 \sqrt{3}-6.
\end{align*}
Thereby, we have now understood how to rephrase the positivity condition \eqref{eq:Tpos} as a
positivity condition for the associated polynomial $P(x,y)$. Since we argued before that the
positivity condition \eqref{eq:Tpos} was necessary for positive semi-definiteness of some linear
combination of functionals, we now conclude that our SDP condition \eqref{eq:positivesemidefb}
implies that
\begin{align}
\label{eq:asymptpoly}
P(x,y) \equiv  \sum\limits_{k=1}^{N_\alpha} c_k P^k(x,y) \ge 0 \quad \text{ for all } \quad
  (x,y) \in \mathbb{R}^2 \,
\end{align}
i.e.~in order to construct positive semi-definite linear combinations of the infinite dimensional
bilinear forms $\{b^k\}_{k=1}^{N_\alpha}$, one has to construct non-negative linear combinations
of the associated two-variable polynomials $P^k$. A sufficient (but famously not necessary condition)
for this is that the polynomial is a sum of squares (SOS). The task of producing a linear combination
of a set of polynomials which is an SOS, is equivalent to an SDP and can thus conveniently be
incorporated in an SDP based numerical bootstrap algorithm (see Appendix~\ref{sec:Polynomial SOS}
for details).

\subsection{Comments on the descendant-space formulation}\label{sec:commentson}
\def\bn{\underline{n}}
\def\bm{\underline{m}}

Our discussion above and in particular the passage from primary-space to descendant-space SDP
appeared rather technical since we based the reformulation on the particular property \eqref{eq:Gscprod}
of six-point comb channel blocks in $d=1$ dimension. Our main aim in this final subsection is to show
that the rewriting we proposed above does actually not depend on the concrete blocks but rather on
features of the crossing equation we consider. More concretely, we shall highlight that the
descendant-space SDP is based on a crossing between two distinct decompositions of a six-point
function into four-point correlators instead of the typical OPE decomposition into three-point
functions.

As usual, we begin our discussion of the crossing equation with two different insertions of the
identity operator inside the six-point correlator, i.e.
\begin{align}\label{eq:sixpointdecompositionintofour}
&\sum\limits_{\mathcal{O},\bn,\bm} \langle \phi_6 \phi_5 \phi_4
\mathcal{O}_{\bn} \rangle \mathcal{N}^{\bn\bm }
\langle \mathcal{O}_{\bm}^\dagger \phi_3 \phi_2 \phi_1\rangle =
\sum\limits_{\mathcal{O},\bn,\bm} \langle \phi_4 \phi_3 \phi_2
\mathcal{O}_{\bn} \rangle \mathcal{N}^{\bn, \bm} \langle
\mathcal{O}_{\bm}^\dagger \phi_6 \phi_5 \phi_1\rangle\ .
\end{align}
Here, the sum extends over all primaries $\mathcal{O}$ of the theory and over a basis of descendants
that are labelled by some index $\bn$. In a one-dimensional theory, $\bn = n = 0,1,2,\dots$ is just a
non-negative integer that counts the number of derivatives that are needed in order to construct
$\mathcal{O}_n$ from the corresponding primary. With this choice of basis in the space of descendants,
it is easy to compute the normalisation matrix $\mathcal{N}$. Concretely,
\begin{align}
    \mathcal{O}_n = P^n \mathcal{O} &&
    \text{gives} &&\mathcal{N}^{nm} = \frac{\delta_{nm}}{(2 \Delta_\mathcal{O})_n n!}\,,
\end{align}
with $P=\partial_x$, the momentum generator. Here, we assumed that the primary fields are normalised canonically. The decomposition formula
\eqref{eq:sixpointdecompositionintofour} focuses on a symmetric split of the fields into two
groups of three fields. This is similar to the symmetric split of four-point functions. In
the case of four fields, such a symmetric split is actually the only non-trivial one. For higher point
correlations, however, there exists other non-trivial decompositions (i.e.~the decompositions into a three- and a five-point function). One advantage of the decomposition into four-point correlators is that it contains new data beyond that of the $\phi \times \phi$ OPE. More importantly, the symmetric decomposition leads to a crossing equation that has a clear SDP dual, namely the descendant-space SDP introduced in Section \ref{sec:demonstrateequiv}.

To establish this, let us further analyse the quantities that appear in the crossing equation
\eqref{eq:sixpointdecompositionintofour}. Through Ward identities, four-point functions that
involve the descendants $\mathcal{O}_n$ of a primary field $\mathcal{O}$ can be reduced to derivatives of the
underlying correlators of $\mathcal{O}$. If we were analysing the decomposition of four-point
correlators into three-point functions, the application of Ward identities would allow us to sum the contributions of the descendants into conformal blocks. The corresponding SDPs hence do not have to deal with matrices labelled by descendants as in eq.~\eqref{eq:sixpointdecompositionintofour}, but rather with kinematically fixed functions multiplied by an unknown number, which is the square of an OPE coefficient. In the
case of six-point functions, the theory dependent four-point functions of primaries are certainly not
just numbers but rather functions of some cross ratios. These functions are the unknowns
in eq.~\eqref{eq:sixpointdecompositionintofour} and contain all dynamical data. We show in Appendix~\ref{app:decomp6to4} how to directly derive by application of Ward identities and
conformal transformations that
\begin{align}\label{eq:PnKnidmain}
    \sum\limits_{n=0}^\infty\frac{\langle \phi_6 \phi_5 \phi_4 \mathcal{O}_n \rangle \langle \mathcal{O}^\dagger_n \phi_3 \phi_2 \phi_1\rangle}{(2 \Delta_\mathcal{O})_n n!}
    = \sum\limits_{n=0}^\infty\frac{( \bar \chi_1  \chi_2 \bar \chi_3 )^{ \Delta_\mathcal{O} + n - \Delta_\phi}\partial_{\bar\chi_1}^n f(\bar\chi_1) \partial_{\bar\chi_3}^n f(\bar\chi_3)}{(x_{12}^ 2x_{34}^2x_{56}^2)^{\Delta_\phi}(2 \Delta_\mathcal{O})_n n!} ,
\end{align}
where
\begin{align}
    \bar \chi \equiv \frac{1}{\chi} && f(\bar\chi) \equiv \langle \phi(\infty)\mathcal{O}(\bar\chi) \phi(1)\phi(0)\rangle.
\end{align}
Eq.~\eqref{eq:PnKnidmain} implies that in terms of
\begin{align}
    F_n(\chi) \equiv (-1)^n(\bar \chi)^{\Delta_\mathcal{O} + n - \Delta_\phi} \partial_{\bar \chi}^n f(\bar \chi),
\end{align}
which satisfies the recurrence relation
\begin{align}\label{eq:recursionrel}
   \chi \partial_\chi F_n(\chi) = F_{n+1}(\chi) - (\Delta_\mathcal{O} + n - \Delta_\phi) F_n(\chi)
\end{align}
we can write
\begin{align}\label{eq:mainPnKnid}
     \sum\limits_{n=0}^\infty\frac{\langle \phi_6 \phi_5 \phi_4 P^n \mathcal{O} \rangle \langle \mathcal{O}^\dagger K^n \phi_3 \phi_2 \phi_1\rangle}{(2 \Delta_\mathcal{O})_n n!}
    =\mathcal{L}(x_i)\sum\limits_{n=0}^\infty F_n(\chi_1) \frac{\chi_2 ^{ \Delta_\mathcal{O} + n}}{(2\Delta_\mathcal{O})_n n!}F_n(\chi_3) ,
\end{align}
with
\begin{align}
    \mathcal{L}(x_i)=(x_{12}^ 2x_{34}^2x_{56}^2\chi_2 )^{-\Delta_\phi}.
\end{align}
Comparing this to Section~\ref{sec:demonstrateequiv}, it is clear that eq.~\eqref{eq:mainPnKnid} is
exactly a decomposition of the six-point correlator in the form of eq.~\eqref{eq:Gscprod}, while the
relation eq.~\eqref{eq:recursionrel} coincides with the relation eq.~\eqref{eq:varthetaFn} on the
level of individual blocks. However, the derivation of the descendant-space  SDP dual to six-point
crossing in Section~\ref{sec:demonstrateequiv} only used this decomposition and the recurrence
relation. Hence, applying eq.~\eqref{eq:mainPnKnid} on both sides of
\eqref{eq:sixpointdecompositionintofour} and repeating the steps of
Section~\ref{sec:demonstrateequiv} reproduces the descendant-space SDP.

Even though it just reproduces the SDP formulated in Section~\ref{sec:demonstrateequiv}, we
consider this alternative perspective as an important complement to the derivation given in
Section~\ref{sec:demonstrateequiv}.
The reason for this is threefold.

Firstly, the derivation establishes that our ability to formulate the descendant-space
SDP is not just a consequence of coincidental special properties of the comb-channel blocks.
On the contrary, it makes manifest that exact knowledge of the comb-channel blocks is not
crucial and highlights the actually necessary physical input.

Secondly, it thus shows which conceptual aspects of our approach could potentially be useful in
other settings beyond the 1d conformal six-point correlators we consider here. For instance, one
could in principle follow the same steps as in Appendix~\ref{app:decomp6to4} to find a higher
dimensional formulation of numerical six-point bootstrap, that does not require the evaluation
of six-point blocks. In attempting to pursue this, however, one has to overcome two technical hurdles.
Clearly, the structure of the projectors onto a conformal multiplet becomes
more complicated in higher dimensions i.e.~it is less obvious what the optimal choice for a basis of descendants would be in $d>2$. Relatedly, it is not clear what the correct choice of functionals should be if one would like to maintain the band-structure that makes the descendant-space
matrices so amenable to numerical implementation in 1d.
Recall that the band structure arises in 1d because acting with a differential operator of degree $\Lambda$ on a level $n$ descendant, independently of $n$ can only produce a linear combination of $\Lambda+1$ descendants (namely those at levels $n, n+1 ,\dots n+\Lambda$).
In higher $d$ however, the action of a generic degree $\Lambda$ differential operator on a level $n$ descendant can produce a linear combination of up to $(n+\Lambda)^d - (\text{max}(0,n-1))^d$ states (since at level $k$ there are $k^d$ states). Hence, one should expect that a careful choice of basis both for the descendants and for the functionals will be important for the generalisation to higher space-time dimension.

Finally, we would like to point out that the perspective taken in this subsection also makes
it clearer how to extract the four-point crossing content of the six-point crossing equation.
We provide some comments on this in Appendix~\ref{app:from4to3to8to5}.

\section{First Bounds: Gap Maximisation without Identity}
\label{sec:noidentity}
 Having established an SDP formulation for six-point crossing \eqref{eq:positivesemidefb}, as well as the necessary
 asymptotic conditions \eqref{eq:asymptpoly}, we now study a simple bootstrap problem where we can compare numerical
 results to concrete analytical expectations. We dub this first problem ``gap maximisation without identity''.

Gap maximisation without identity amounts to finding the largest possible value for the conformal dimensions
$\Delta_2$ of the leading operator $\mathcal{O}_2$ in the triple operator product $(\phi \times \phi)\times \phi
\sim \mathcal{O}_2$ of the external field $\phi$, subject to crossing symmetry and unitarity. Of course, the
typical scenario is that the $\phi\times\phi$ OPE contains the identity field. In this case we would expect
that the leading operator in the triple OPE is simply given by $\mathcal{O}_2=\phi$. Hence, it might seem most
natural to assume that $\phi$ is exchanged in the middle channel and to study the gap above it, and we will
indeed do so in Section~\ref{sec:GFFbounds}. On the other hand, nothing stops us from considering the
technically somewhat simpler situation where the leading operator $\mathcal{O}_1$ in the double OPE
$\phi \times \phi$ is nontrivial and where the OPE of this operator $\mathcal{O}_1$ with $\phi$ produces
some leading contribution $\mathcal{O}_2 \neq \phi$. In this setup, it makes sense to directly study bounds
on the leading exchange $\mathcal{O}_2$.\footnote{Some intuition can be gained by considering the four-point
analogue of this problem. What is the upper bound on the dimension $\Delta_{\mathcal{O}}$ in a four-point
function $\langle \phi \phi \phi \phi \rangle$, where $\phi\times\phi\sim \mathcal{O}+ \dots$? Running the
standard numerical four-point
bootstrap indicates that $\Delta_{\mathcal{O},\text{max}}= 4\Delta_\phi/3$ \cite{Esterlis:2016psv}. This can
be identified as a solution to crossing without an identity exchange which obstructs the existence of bounds
on certain OPE coefficients in 1d CFT \cite{Antunes:2021abs,Cordova:2022pbl}. It can be written as the correlator
of a $:\varphi^3:$ type correlator in GFF where each constituent field is Wick contracted with exactly one field
in the other 3 points, in complete analogy to the six-point solution we will describe below.}

 As is standard in optimisation problems, it is useful to formulate both a primal and a dual approach to gap maximisation
 without identity. Concretely, these take the following form\bigbreak

\noindent \underline{Gap Maximisation without identity (primal formulation):} Find the maximal value $\Delta^P_{\text{opt.}}$
for which there exists a family $K^{\mathcal{O}_1 \mathcal{O}_3}(\Delta_2) = C^{\mathcal{O}_1}_{\mathcal{O}_2}
C^{\mathcal{O}_3}_{\mathcal{O}_2}$ of positive semi-definite kernels pa\-ra\-me\-trised by $\Delta_2 \in
\mathbb{R}_{>0}$ such that
\begin{align}
  G(\chi_i) =  \sum\limits_{\mathcal{O}_1, \mathcal{O}_2, \mathcal{O}_3} C^{\mathcal{O}_1}_{\mathcal{O}_2} C^{\mathcal{O}_3}_{\mathcal{O}_2} G^{123,456}_{\Delta_1 \Delta_2 \Delta_3}(\chi_i)
\end{align}
is a crossing symmetric function and $K(\Delta_2) = 0$ for all $\Delta_2 < \Delta^P_{\text{opt.}}$. \bigbreak

\noindent \underline{Gap Maximisation without identity (dual formulation):} Find the minimal value $\Delta^D_{\text{opt.}}$
for which there exists a functional $\alpha$ such that the corresponding family $b_{\Delta_2}$ of bilinear forms
parameterised by $\Delta_2 \in \mathbb{R}_{>0} $ is positive semi-definite for all $\Delta_2 > \Delta^D_{\text{opt.}}$. \bigbreak

In practice, we address the primal problem by constructing an exact solution to crossing by hand and reading off the value $\Delta^P_*$ of the leading term $\mathcal{O}_2$ in the middle channel. This value
may not be $\Delta^P_{\text{opt.}}$, but it clearly sets a lower bound $\Delta^P_* \leq \Delta^P_{\text{opt.}}$. Similarly
the solutions to the dual problem that we construct via SDP do not necessarily give $\Delta^D_{\text{opt.}}$, but rather an upper
bound $\Delta^D_* \geq \Delta^D_{\text{opt.}}$, i.e.
\begin{equation}
\Delta^P_* \leq \Delta^P_{\text{opt.}} \leq \Delta^D_{\text{opt.}} \leq \Delta^D_*\ \ .
\end{equation}
If the determined values $\Delta^P_*$ and $\Delta^D_*$ are close together.\footnote{If there was a large duality gap $\Delta^D_{\text{opt.}}-\Delta^P_{\text{opt.}}$ it would of course be impossible to find $\Delta^D_*$ and $\Delta^P_*$ that are close to each other. It is however a standard result of finite dimensional SDP that $\Delta^D_{\text{opt.}} = \Delta^P_{\text{opt.}}$. The results presented in this work a posteriori suggest that the duality gap is also small or even vanishing for the infinite dimensional SDP that we study.}, they are good approximations
to the solution of the primary and the dual problem. Since we mostly discuss $\Delta_*^D$ and only refer to $\Delta_*^P$ in relatively small portions of the text, we usually use the short notation
\begin{align}
    \Delta_* \equiv \Delta_*^D.
\end{align}

In the first subsection, we shall construct an exact solution to crossing and read off the corresponding
value $\Delta^P_*$. Then, we turn to the dual problem in Subsection 3.2. After discussing a simple analytical
bound that arises from first order derivative functionals, we shall describe how to write the dual problem
in a numerically treatable way that can be fed to a standard SDP solver, in our case \texttt{SDPA-GMP} \cite{SDPA,SDPA2,SDPA3}, see Subsection
\ref{sec:SDPAinput}. In Subsection~\ref{sec:checkingSDPA}, we elaborate on how to thoroughly check
positivity of the numerically constructed functionals. The final Subsection~\ref{sec:noidresults}
discusses the concrete numerical results and their comparison to the analytical expectation. In particular
we shall see that the values $\Delta^P_*$ and $\Delta^D_*$ we determine from the exact solution and the
numerical study of four-derivative functionals are quite close already.

 \subsection{Exact solution to crossing - primal approach}
 \label{sec:Primal}

Before discussing the numerical dual approach to gap maximisation without identity, we present the best
candidate for an optimal solution to crossing that we managed to explicitly construct. This primal
candidate is meant to serve as a reference point for the dual results discussed in the subsequent
subsections.

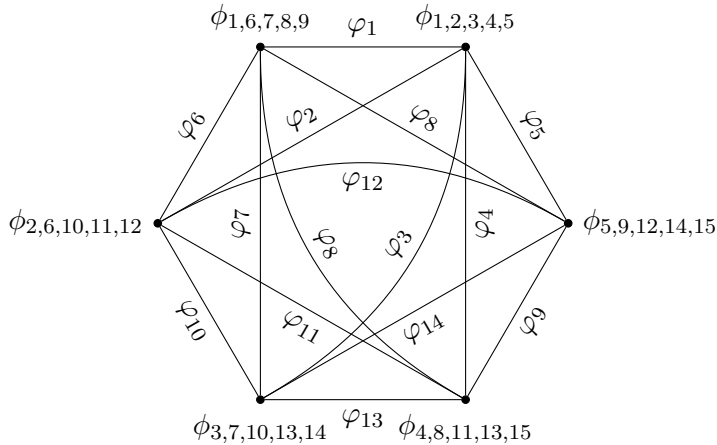
\begin{figure}[ht]
    \centering
    \begin{tikzpicture}
   \newdimen\R
   \R=2.7cm
   \foreach \x/\l/\p in
     { 60/{$\phi_{1,2,3,4,5}$}/above,
      120/{$\phi_{1,6,7,8,9}$}/above,
      180/{$\phi_{2,6,10,11,12}$}/left,
      240/{$\phi_{3,7,10,13,14}$}/below,
      300/{$\phi_{4,8,11,13,15}$}/below,
      360/{$\phi_{5,9,12,14,15}$}/right
     }
     \node[inner sep=1pt,circle,draw,fill,label={\p:\l}] (\x) at (\x:\R) {};
     \path[-, draw,sloped] (60) edge node[above] {$\varphi_1$} (120); {test};
     \path[-, draw,sloped] (60) edge node[above] {$\varphi_5$} (360); {test};
     \path[-, draw,sloped] (300) edge node[below] {$\varphi_9$} (360); {test};
    \path[-, draw,sloped] (300) edge node[below] {$\varphi_{13}$} (240); {test};
    \path[-, draw,sloped] (240) edge node[below] {$\varphi_{10}$} (180); {test};
    \path[-, draw,sloped] (120) edge node[above] {$\varphi_6$} (180); {test};

    \path[bend right = 30, draw,sloped] (360) edge node[below] {$\varphi_{12}$} (180); {test};
     \path[bend left = 30, draw,sloped] (60) edge node[above] {$\varphi_3$} (240); {test};
     \path[bend right = 30, draw,sloped] (120) edge node[above] {$\varphi_8$} (300); {test};

     \path[-, draw,sloped] (60) edge node[above] {$\varphi_2$} (180); {test};
     \path[-, draw,sloped] (240) edge node[above] {$\varphi_7$} (120); {test};
     \path[-, draw,sloped] (180) edge node[below] {$\varphi_{11}$} (300); {test};
     \path[-, draw,sloped] (240) edge node[below] {$\varphi_{14}$} (360); {test};
     \path[-, draw,sloped] (300) edge node[below] {$\varphi_{4}$} (60); {test};
     \path[-, draw,sloped] (360) edge node[above] {$\varphi_{8}$} (120); {test};
    \end{tikzpicture}
    \caption{Wick contractions leading to the correlator \eqref{eq:corr}.}
    \label{fig:Wick}
\end{figure}

The correlator in question appears in the direct sum of 15 decoupled copies of GFF, with corresponding
scalar fields $\{\varphi_i\}_{i=1}^{15}$ of equal scaling dimension $\Delta_{\varphi}$. In this simple
CFT, consider the composite operators
\begin{align}
    \phi_{a,b,c,d,e} = \varphi_a \varphi_b \varphi_c \varphi_d \varphi_e,
\end{align}
with $\{a,b,c,d,e\}$ pairwise distinct. These operators all have the same dimension $\Delta_\phi = 5
\Delta_\varphi$. The correlator
\begin{align}\label{eq:corr}
    \langle \phi_{1,2,3,4,5} \phi_{1,6,7,8,9} \phi_{2,6,10,11,12}\phi_{3,7,10,13,14}\phi_{4,8,11,13,15} \phi_{5,9,12,14,15}  \rangle = \mathcal{L}(x_i) \mathcal{G}(\chi_j)
\end{align}
can be computed easily by applying Wick contractions between all pairs of external operators. The pattern
of these contractions shown in Figure~\ref{fig:Wick}.
After splitting off the leg factor $\mathcal{L}$ defined in eq.~\eqref{eq:legg} and expressing
the remaining function $\mathcal{G}$ in terms of the cross ratios \eqref{eq:CR}, we obtain
\begin{equation}
\label{eq:noidentcorr}
\small
    \mathcal{G}(\chi_j) = \frac{\chi_1^{8 \Delta_\varphi}\chi_2^{9 \Delta_\varphi}
    \chi_3^{8 \Delta_\varphi}}{\big((1-\chi_1)(1-\chi_2)(1-\chi_3)(1-\chi_1-\chi_2)
    (1-\chi_2-\chi_3)(1-\chi_1-\chi_2-\chi_3+\chi_1 \chi_3)\big)^{2\Delta_\varphi}}\,.
\end{equation}
This expression can now be expanded in a power series. From the exponent of the leading term in
each of the cross ratios $\chi_j$, we can read off the conformal dimension $\Delta_{\mathcal{O}_j^*}$ of the
leading primary $\mathcal{O}_j^*$ exchanged in the $j$th internal edge of the comb. Concretely we find
\begin{equation}
 \Delta_{\mathcal{O}_1^*} = 8\Delta_{\phi}/5 \quad \text{and} \quad \Delta_{\mathcal{O}_2^*}=9\Delta_{\phi}/5\ .
\end{equation}
A more careful expansion in terms of the conformal blocks \eqref{eq:combblock} allows us to read
off the OPE coefficients, which we then check to satisfy unitarity, as described in Appendix
\ref{app:blockexpansion}. We hence conclude that
\begin{equation} \label{eq:primalbound}
\Delta^P_\text{opt.} \geq  \Delta^P_* = 9\Delta_{\phi}/5 \, .
\end{equation}
There is another way to arrive at this lower bound for the primal approach, in terms of
$\mathbb{Z}_n$ chiral twist fields $\sigma_n$. Let us recall that for a parent theory of
central charge $c$ the conformal dimension of such a chiral twist field is given by
\begin{equation} \Delta_{\sigma_n} = \frac{c}{4}\left( \frac{1}{n} - \frac{1}{n^2}\right)\ .
\end{equation}
In order to obtain a non-vanishing six-point function without identity exchange in the two
    outer channels, we consider the case of a $\mathbb{Z}_6$ orbifold and set $\phi = \sigma_6$.
The operator product of two such twist fields contains a twist field $\mathcal{O}_1^* = \sigma_3$
of order three while the leading contribution to the intermediate exchange in the middle channel
is given by a twist field $\mathcal{O}_2^* = \sigma_2$. Application of the general formula for the
conformal dimension of twist fields gives
\begin{equation}
\Delta_\phi =  \frac{5}{36} \frac{c}{4}\ ,
\quad \Delta_{\mathcal{O}_1^*} = \frac{2}{9} \frac{c}{4}= \frac{8}{5}{\Delta_\phi}
\ , \quad \Delta_{\mathcal{O}_2^*} =  \frac{1}{4} \frac{c}{4} = \frac{9}{5} \Delta_\phi \ .
\end{equation}
The results from this short analysis confirm the conclusion \eqref{eq:primalbound} of the first
elementary analysis in terms of generalised free fields, but they do not improve on the lower bound we obtain
for the value of $\Delta^P_\text{opt.}$. This concludes our brief discussion of exact solutions to the
six-point crossing equation and the primal approach to gap maximisation without identity.

\subsection{Analytical and numerical bounds - dual approach}

In this section, we construct functionals which provide a rigorous upper bound on the value of
$\Delta^D_{\text{opt.}}$. Taking advantage of the ``no identity'' assumption, we construct a simple one-derivative
functional which provides an upper bound that can be easily checked analytically. Then, we consider the
semi-definite problem \eqref{eq:positivesemidefb}, which we discretise by introducing a grid and a
cutoff for the variable $\Delta_2$. In addition, the asymptotic constraints \eqref{eq:asymptpoly} are
included. This puts the problem in a format amenable to the semi-definite solver
\texttt{SDPA-GMP}.\footnote{For our truncated scheme, where we have many large matrices with numerical
entries, we found \texttt{SDPA-GMP} to be faster than \texttt{SDPB} \cite{Simmons-Duffin:2015qma}. We note that for most computations performed in this paper \texttt{SDPA-QD} would be sufficient. However, we found it convenient to use the extra-flexibility of the arbitrary precision solver \texttt{SDPA-GMP} throughout.
We also note that it is possible to formulate the problem in a way where the $\Delta_2$ dependence
is treated as a polynomial, the type of problem \texttt{SDPB} was designed to tackle. We will
return to this point in future work.}

\subsubsection{An analytical toy bound}\label{sec:toybound}
Let us consider a simple functional composed of first order derivatives $\partial_{\chi_1}$ and
$\partial_{\chi_2}$. Since we have already computed the derivative of the crossing vector
$F$ with respect to the cross ratio $\chi_1$ in eq.~\eqref{eq:dchi1}, it only remains to state a
formula for the derivative with respect to $\chi_2$. Acting on the crossing vector \eqref{eq:defF}
with $\partial_{\chi_2}$ and evaluating at the crossing symmetric point gives
\begin{align}
  \partial_{\chi_2}F_{\Delta_1,\Delta_2,\Delta_3}=  -21 \Delta_\phi G_{\Delta_1,\Delta_2,\Delta_3}
  + \frac{1}{2}\left(\partial_{\chi_1}+\partial_{\chi_3}\right)G_{\Delta_1,\Delta_2,\Delta_3} +\partial_{\chi_2}G_{\Delta_1,\Delta_2,\Delta_3}\,.
\end{align}
We can combine this formula with our expression \eqref{eq:dchi1} for the $\chi_1$-derivative of the
crossing vector to obtain
\begin{align}\label{eq:x2halfx1G}
 \left(\partial_{\chi_2} - \frac{1}{2} \partial_{\chi_1} \right)F_{\Delta_1,\Delta_2,\Delta_3}
 = -12 \Delta_\phi G_{\Delta_1,\Delta_2,\Delta_3} + \partial_{\chi_2}G_{\Delta_1,\Delta_2,\Delta_3}\,.
\end{align}
Inserting our representation ~\eqref{eq:Gscprod} for the comb channel six-point blocks $G$, we can
further evaluate this expression,
\begin{align}\label{eq:x2halfx1}
 \left(\partial_{\chi_2} - \frac{1}{2} \partial_{\chi_1} \right)F_{\Delta_1,\Delta_2,\Delta_3}
 =\sum\limits_{n=0}^\infty  \frac{F_n(\Delta_1, 1/3)(n+\Delta_2 -
 4 \Delta_\phi)F_n(\Delta_3,1/3)}{3^{n+h_2-1} (2 h_2)_n n!}\,.
\end{align}
From this equation, we may read off the matrix $M$ that characterises the bilinear form, see
eq.~\eqref{eq:Mdef}. In descendant-space the matrix that is associated to $\partial_{\chi_2} -
1/2 \partial_{\chi_1}$ is diagonal and, for $\Delta_2 > 4 \Delta_\phi$, the diagonal has only
positive entries. Hence, we can directly deduce an upper bound of $\Delta^D_\text{opt.} \leq 4 \Delta_\phi$.
If we allow ourselves to  also consider non-diagonal matrices in descendant-space, we can actually
push this simple analytical bound even down to $\Delta^D_{\text{opt.}} \leq 2 \Delta_\phi$, still using only
the two linearly independent first derivatives $\partial_{\chi_1}$ and $\partial_{\chi_2}$, though
now the linear combination $\partial_{\chi_2} - \partial_{\chi_1}$. To see this, let us restrict
to the special case $\Delta_\phi = 1$. In eq.~\eqref{eq:Mhatdelchi1}, we gave an expression for
the normalised descendant-space matrix of the functional corresponding to the derivative
$\partial_{\chi_1}$. Analogously, one obtains
\begin{align}\label{eq:Mhat1minus2}
  \hat{M}^{\partial_{\chi_2} - \partial_{\chi_1}}_{nm} =  \frac{6 (\Delta_2+n+2)}{2 \Delta_2+n}
  \delta_n^m-\frac{18}{2 \Delta_2+n} -\frac{3}{2} \sqrt{3} \sqrt{\frac{n+1}{2 \Delta_2+n+1}}
  (\delta_{n-1}^m + \delta_{n+1}^m)
\end{align}
for the normalised descendant-space functional corresponding to $\partial_{\chi_2} - \partial_{\chi_1}$.
We can show that this matrix is positive semi-definite by constructing auxiliary matrices with only
finitely many entries. This will also serve as a warm-up for a similar method to check the positivity of functionals with arbitrary derivative order, to be discussed in Section~\ref{sec:checkingSDPA}. For the simple functional at hand, it is enough to define the matrix

\begin{align}
\label{eq:Msimple}
     M^{(n)}_{ij} \equiv
    \begin{cases}
         \frac{1}{2}(1+\delta_{n,0}) \hat{M}^{\partial_{\chi_2} - \partial_{\chi_1}}_{ij}    &  i,j = n+1\\
        \hat{M}^{\partial_{\chi_2} - \partial_{\chi_1}}_{ij}   & n + 1< \max(i,j) \land \min(i,j)  \leq n + 99 \\
       \frac{1}{2}  \hat{M}^{\partial_{\chi_2} - \partial_{\chi_1}}_{ij}   & i = j = n +100 \\
        0  & \text{else}
    \end{cases}
    \,,
\end{align}
or more visually:

\begin{align} \small  M_{ij}^{(n)}= \left(\begin{array}{cc|ccccc|cc}
    \phantom{1}&\phantom{1} & & & & & & \phantom{1}&\phantom{1} \\\hline
    & & \frac{1}{2}(1+\delta_{n,0}) \hat{M}_{n+1,n+1} &\hat{M}_{n+1,n+2} &0 &\dots &0 & &\\
    & & \hat{M}_{n+2,n+1} & \hat{M}_{n+2,n+2} &\ddots &\ddots &\vdots & & \\
    & & 0 & \ddots &\ddots & \ddots &0 & & \\
    & & \vdots  &\ddots &\ddots & \hat{M}_{n+99,n+99} &\hat{M}_{n+99,n+100} & &\\
    & & 0 &\dots & 0 &\hat{M}_{n+100,n+99} &\frac{1}{2} \hat{M}_{n+100,n+100} & &  \\\hline
    & & & & & & & &
  \end{array}\right) \,,
\end{align}
where we suppressed the superscript ${\partial_{\chi_2} - \partial_{\chi_1}}$ for simplicity of
notations, and the empty blocks correspond to entries which take the value $0$. By construction,
we can recover the infinite dimensional matrix $\hat M$ from $\hat
M^{(n)}$ as
\begin{align}\label{eq:Msum}
    \hat{M}^{\partial_{\chi_2} - \partial_{\chi_1}}_{ij} =
    \sum\limits_{n = 0}^\infty M_{ij}^{(99 n)}\,.
\end{align}
To show that $\hat{M}^{\partial_{\chi_2} - \partial_{\chi_1}}$ is positive semi-definite, it is
sufficient to establish positive semi-definiteness for each summand on the r.h.s.~of eq.~\eqref{eq:Msum}.
This is accomplished in two steps. First we shall analyse the matrix $M^{(0)}$ and then in a second step
we show that $M^{(n)}$ for any $\Delta_2$ is positive semi-definite for $n \geq 99$.

\begin{figure}[ht]
	\centering \includegraphics[width=350pt]{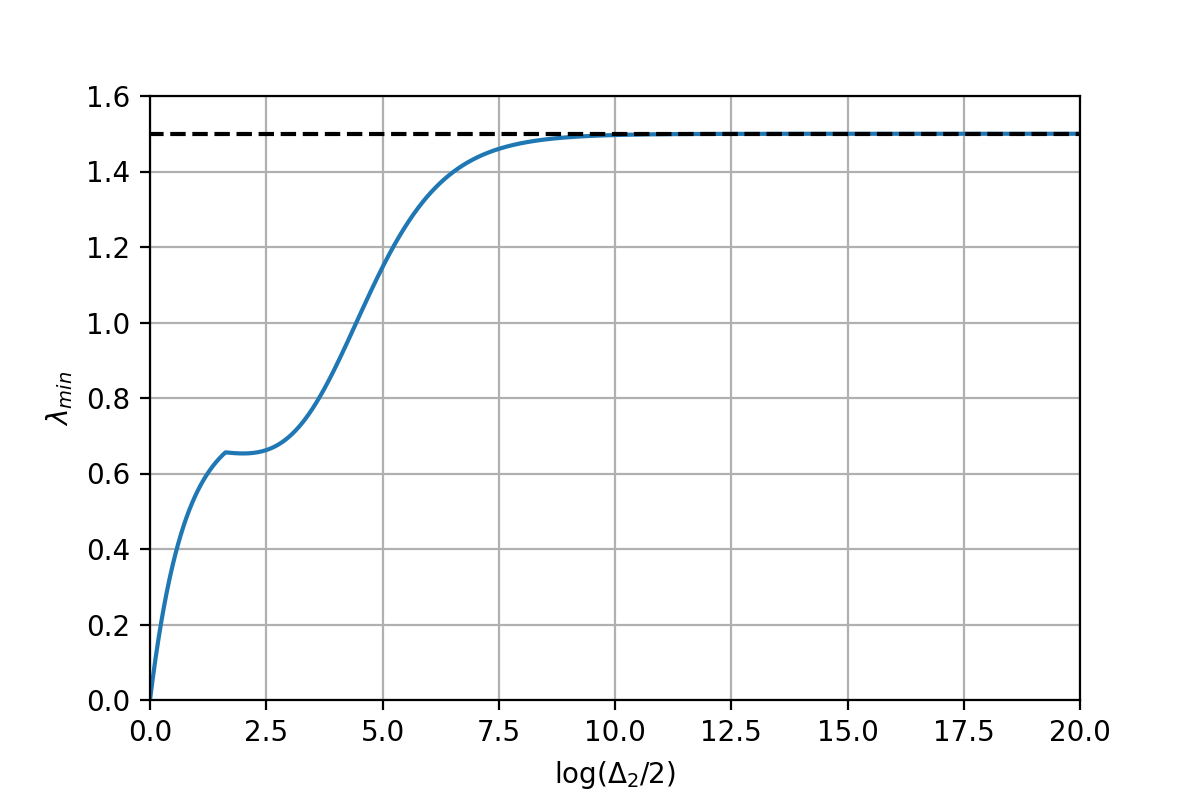}
	\caption{In blue, we plot the minimal non zero eigenvalue $\lambda_{\textrm{min}}$ as a
 function of $\Delta_2$ on a logarithmic scale. The black dashed line corresponds to the
 asymptotic large $\Delta_2$ value.
		\label{fig:M0lown}
	}
\end{figure}

Addressing $M^{(0)}$, let us start by noting that the limit $\Delta_2 \rightarrow \infty$ of the
matrix $M^{(0)}$ is a diagonal matrix whose smallest non-zero eigenvalue $\lambda_{\textrm{min}}
= \lambda_{\textrm{min}}(\infty)$ is given by $\lambda_{\textrm{min}}=1.5$, as can be read off
directly from eq.~\eqref{eq:Mhat1minus2}. With this knowledge of the asymptotics, we may
demonstrate the positive semi-definiteness of $M_{ij}^{(0)}$ for all $\Delta_2 > 2$ by considering
the smallest non-zero eigenvalue as a function of $\Delta_2$. One can compute and plot this minimal
eigenvalue $\lambda_{\textrm{min}}(\Delta_2)$ in the regime $\Delta_2 \geq 2$ starting from
$\Delta_2=2$ up until one reaches values of $\Delta_2$ where the minimal eigenvalue of $M^{(0)}$
reaches its large $\Delta_2$ asymptotics value $\lambda_{\textrm{min}}(\Delta_2) \sim 1,5$. This
plot can be seen in Figure~\ref{fig:M0lown}. It shows very clearly that the smallest eigenvalue
of $M^{(0)}$ is indeed positive and hence that $M_{ij}^{(0)}$ is a positive semi-definite matrix
for all values $\Delta_2 > 2$.

Before we consider general $M_{ij}^{(n)}$ with $n \ge 99$, let us briefly comment on the asymptotic
Toeplitz arc of $\hat{M}^{\partial_{\chi_2}-\partial_{\chi_1}}$ in terms of the matrices $M^{(n)}$.
In other words, let us consider the spectrum of the matrices
\begin{align}
    M^{(\infty)}(r) := \lim\limits_{n \rightarrow \infty} M^{(n)}(\Delta_2 = n r)\,.
\end{align}
Clearly, it is a necessary condition for the positivity property of  $\hat{M}^{\partial_{\chi_2}-
\partial_{\chi_1}}$ in question that $ M^{(\infty)}(r)$ is positive semi-definite for all $r > 0$.
Note that $r \rightarrow \infty$ is just the large $\Delta_2$ limit, where we expect the smallest
non vanishing eigenvalue to take the value $\lambda_{\textrm{min}}(\infty) = 1.5$. On the other
hand, $r = 0$ corresponds to sending $n$ to infinity at finite $\Delta_2$. The smallest non
vanishing eigenvalue of the $r = 0$ matrix is $\lambda_{\textrm{min}}(r=0) = 0.75$. having
analysed these two extremal values of $r$, it now remains to plot the minimal eigenvalue of
$M^{(\infty)}$ as a function of $r$. This plot can be seen in Figure~\ref{fig:combcrossing}.
It shows clearly that the minimal eigenvalue of the matrix $M^{(\infty)}(r)$ is positive for
all values of $r$

\begin{figure}[ht]
	\centering \includegraphics[width=350pt]{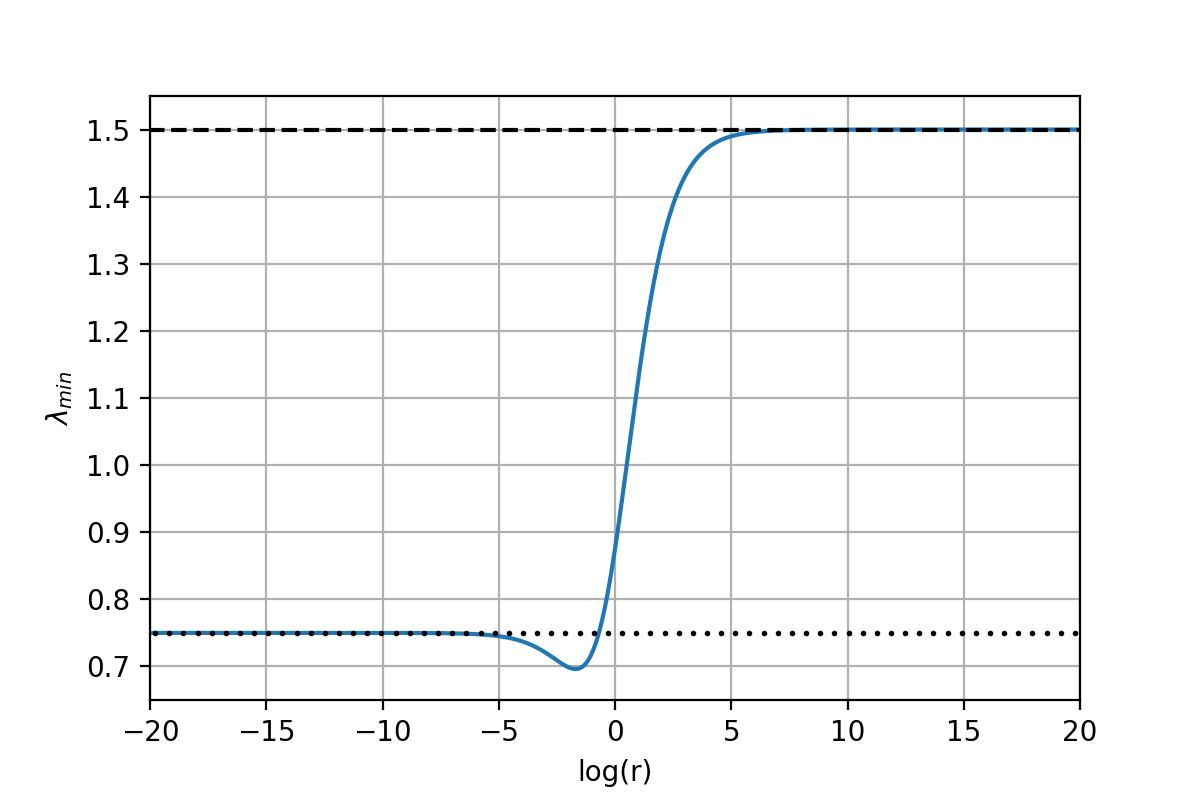}
	\caption{ In blue, we plot the minimal non zero eigenvalue of $M^{(\infty)}$ as a function of the ratio $r$ between $n$ and $\Delta_2$ on  a logarithmic scale. The black dotted line corresponds to the asymptotic $n\to \infty$ limit, while the black dashed line represents the $\Delta_2 \to \infty$ regime.
		\label{fig:Minfr}
	}
\end{figure}

\begin{figure}[ht]
	\centering \includegraphics[width=300pt]{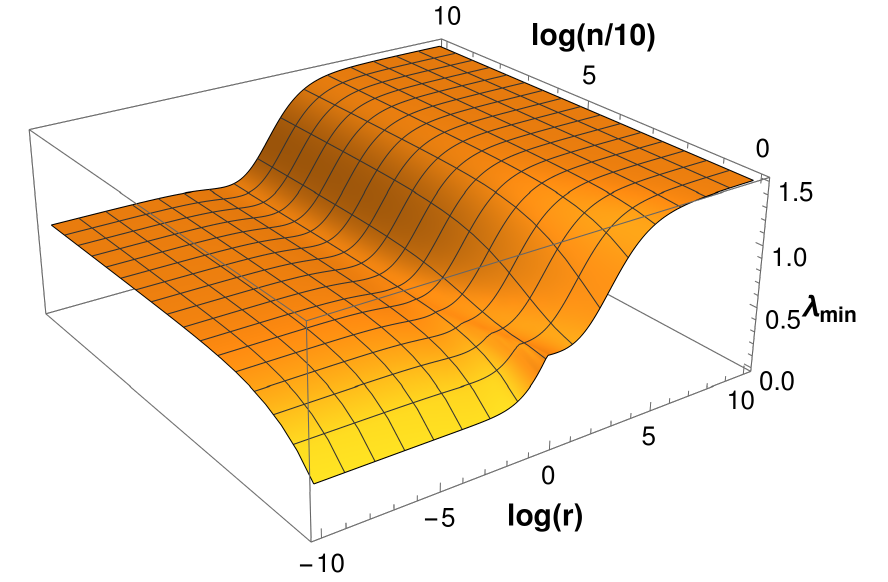}
	\caption{ Here, we illustrate the surface of minimal non-vanishing eigenvalues of the matrices $M^{(n)}(\Delta_2 = n r)$ as functions of $r$ and $n$.
		\label{fig:eigenvaluesurface}
	}
\end{figure}

Let us finally comment on the spectrum of $M^{(n)}$ for finite $n$. Though we only need to show
$M^{(n)}(\Delta_2) \succeq 0$ for $n \ge 100$ and $\Delta_2 > 2$, we choose to consider all
$\Delta_2 > 0$ and $n \ge 10$, as this is slightly more instructive. The minimal eigenvalue of
$M^{(n)}(n r)$ for small $n$ is a function of $r$ about which we only know the asymptotic
value $\lambda_{\textrm{min}}(\infty)=1.5$ for sufficiently large $\Delta_2$. Thus, in order
to demonstrate positivity of the minimal eigenvalue, we need to compute $\lambda_{\textrm{min}}(r)$
for $r \geq 0$ up until values of $r$ at which the minimal eigenvalue reaches its asymptotic
value. As $n$ increases, however, the $r$-profile for the minimal eigenvalue of $M^{(n)}(n r)$
approaches the function plotted in Figure~\ref{fig:Minfr}. Thus, only a finite range of $r$ and
$n$ has to be plotted explicitly to conclude that $M^{(n)}(\Delta_2)$ is positive semi-definite
for all $\Delta_2 > 0$ and $n \ge 10$. This plot can be found in Figure~\ref{fig:eigenvaluesurface},
which finishes our argument for the positivity of the functional that is associated with
$\partial_{\chi_2} - \partial_{\chi_1}$, for all values $\Delta_2 \geq 2$ and with the
conformal weight of the external field set to $\Delta_\phi =1$.

\subsubsection{Producing the input for \texttt{SDPA}}\label{sec:SDPAinput}

We now want to automatise the task of producing positive functionals by implementing the
discretised version of the SDP \eqref{eq:positivesemidefb} alluded to above.
In particular, we will describe the explicit input that we feed into \texttt{SDPA} in order
to solve the SDP dual to gap maximisation without identity. The \texttt{Mathematica13.2}
notebook that we wrote for this purpose can be found as an ancillary file in this arXiv
submission\footnote{We use \texttt{SPDA-GMP} 7.1.3 with the following parameters (with minor modifications depending on the specific SDP at hand): Large \texttt{maxIteration} (e.g.~5000) to ensure that the solver does not stop before convergence, \texttt{epsilonStar} and \texttt{epsilonDash} varying between $10^{-10}$ to $10^{-20}$ depending on the specific input SDP,  Large \texttt{lambdaStar} ($10^5$) and order one \texttt{omegaStar} ($2$), \texttt{lowerBound}$=0$ and \texttt{upperBound}$=0$ so that SDPA terminates once it knows that a certain assumption on the gap can be excluded or cannot be excluded, \texttt{betaStar}=$0.01$, \texttt{betaBar}=$0.02$, \texttt{gammaStar}=$0.9$, \texttt{precision}=128 (some runs also with 256).} It implements the following six steps:
\smallskip

\noindent
\textbf{1. Read in the external parameters.} In the first section of the notebook, we
input key parameters, like the highest derivative order to be considered (\texttt{$\Lambda$}),
the total number of functionals (\texttt{Nfunc}), the gap that we would like to exclude
(\texttt{gap}), the truncation size (\texttt{DTrunc}), the discrete range of $\Delta_2$
values on which we explicitly impose positivity (\texttt{h2ran}) and a choice of an
overall sign (\texttt{sign}).
\smallskip

\noindent
\textbf{2. Generate a basis of independent derivative functionals.} Next, we generate a
basis of linearly independent derivative functionals with a total number of derivatives
less or equal to \texttt{$\Lambda$}. Concretely, we apply partial derivatives to the
function defined in eq.~\eqref{eq:defF} and evaluate them at the crossing symmetric
point, without explicitly evaluating the conformal blocks $G$. We then treat different
derivatives of $G$ as linearly independent. Thus each derivative functional corresponds
to a specific linear combination of these linearly independent functions and we can
determine a basis of this space directly by Gaussian elimination.
\smallskip

\noindent
\textbf{3. Transform to the descendant-space basis.} Now, we compute the normalised
descendant-space matrices associated to the selected derivative functionals. As explained
in Section~\ref{sec:demonstrateequiv}, to achieve this, one simply has to act with the
derivatives on the blocks written in the form of eq.~\eqref{eq:Gscprod} and then apply
eq.~\eqref{eq:varthetaFn}. Once we have computed formulas for arbitrary entries of the
descendant-space  matrices this way, we explicitly evaluate the upper left $\texttt{DTrunc}
\times \texttt{DTrunc}$ block of each functional.
\smallskip

\noindent
\textbf{4. Compute the asymptotic arc polynomials.} Then, we  use the expressions generated
in step 3 in order to compute the asymptotic arc polynomials as described in Section
\ref{sec:toeplitzarc}.
\smallskip

\noindent
\textbf{5. Generate an SDP to construct SOS combinations of the asymptotic arc polynomials.}
The final piece of input data that we need to compute for \texttt{SDPA} is the set of matrices
that realise the SDP formulation of the task of constructing SOS linear combinations of the
asymptotic arc polynomials. We achieve this through the minimalistic approach described in
Appendix~\ref{sec:Polynomial SOS}, and use the result in the final step.
\smallskip

\noindent
\textbf{6. Formulate the gap maximisation as an SDP and export \texttt{SDPA} input file.}
\texttt{SDPA} solves the following semi-definite optimisation problem: Given $N+1$
(block-diagonal) matrices $\{\textbf{F}_i\}_{i=0}^N$ and an objective $\{c_i\}_{i=1}^N$,
 find coefficients $\{x_i\}_{i=1}^N$ such that
    \begin{align}
        \sum\limits_{i=1}^N c_i x_i \text{ is minimised and } \sum\limits_{i=1}^N x_i
        \textbf{F}_i - \textbf{F}_0 \succeq 0\,.
    \end{align}
In its formulation (\ref{eq:positivesemidefb}), however, the dual to gap maximisation without
identity is not directly in this form. Instead it is rather a family of Linear Matrix
Inequalities (LMI)\footnote{A LMI is the problem of finding $x_i$ for given matrices
$\{\textbf{F}_i\}_{i=0}^N$ s.t.~$\sum\limits_{i=1}^N x_i \textbf{F}_i- \textbf{F}_0
\succeq 0$ } which express the constraint that the linear combination
of functionals under consideration should be positive semi-definite above the chosen gap.
The existence (or not) of solutions to these LMI allows us to exclude the gap (or not). Of course, we could simply transform the LMI to an optimisation
problem that can be solved by \texttt{SDPA} by choosing an arbitrary objective function
    (say for instance a vanishing objective). But as suggested in the \texttt{SDPA} manual
    \cite{SDPAman}, it is preferable to instead add an identity matrix $(\textbf{F}_{N+1})_{ij}
    = \delta_{ij}$ and consider the optimisation problem of finding $\{x_i\}_{i=1}^{N+1}$ such
    that
    \begin{align}
         x_{N+1} \text{ is minimised and } \sum\limits_{i=1}^{N+1} x_i \textbf{F}_i - \textbf{F}_0 \succeq 0 \,.
    \end{align}
    If this problem has a solution with negative objective, we are guaranteed that
    \begin{align}
        \textbf{F}\equiv\sum\limits_{i=1}^{N} x_i \textbf{F}_i - \textbf{F}_0 \succeq 0
    \end{align}
    and moreover, $0<- x_{N+1}$ is a lower bound on the smallest eigenvalue of \textbf{F}.

Apart from the identity that is added to realise the LMI, there are three qualitatively different
matrices $\textbf{F}_i$ that we feed into \texttt{SDPA}.

First of all we have the matrices $\textbf{F}_1, \dots, \textbf{F}_{\texttt{Nfunc}-1}$ corresponding to
$\texttt{Nfunc}-1$ of the functionals under consideration. Each of these is a large block diagonal matrix
whose blocks are: For each value $h_2>\texttt{gap}$ in \texttt{h2ran}, a $\texttt{DTrunc}\times\texttt{DTrunc}$
matrix corresponding to the upper left corner of the full descendant-space  matrix, evaluated at $\Delta_2 = h_2$;
and the coefficient matrix of the associated asymptotic arc polynomial obtained through the SOS SDP described in step 5.

The role of $\textbf{F}_0$ is played by the matrix corresponding to the one derivative functional described
in eqs.~\eqref{eq:x2halfx1G} and \eqref{eq:x2halfx1}, which is diagonal in the descendant space basis. As the
coefficient of $\textbf{F}_0$ is not variable, we in principle need to consider also the case where $\textbf{F}_0$
is multiplied by a sign in order to have access to the most general linear combination (at high enough derivative
order the choice of the sign does not seem to matter anymore). This sign is determined through the parameter
\texttt{sign} defined in the first section of the notebook. Up to the sign, $\textbf{F}_0$ is on the same footing
as the other $\textbf{F}_i$.

In addition to $\textbf{F}_0, \dots, \textbf{F}_{\texttt{Nfunc}-1}$, we also have to include $\textbf{F}_i$
corresponding to redundancies of the SOS SDP (see Appendix~\ref{sec:Polynomial SOS}). These have vanishing
$\texttt{DTrunc}\times\texttt{DTrunc}$ blocks for all $h_2$ in \texttt{h2ran} above $\texttt{gap}$ and
a block that is one of the matrices $\texttt{K}[\![j]\!]$ satisfying eq.~\eqref{eq:vker}.

\subsubsection{Checking the output of \texttt{SDPA}}
\label{sec:checkingSDPA}
A priori, the functionals that we determine numerically are not guaranteed to fulfil the full set of positivity
constraints that are needed to establish bounds on CFT data. After all, we only ask \texttt{SDPA} to construct
functionals that are positive semi-definite asymptotically in the sense of Section~\ref{sec:toeplitzarc} and for
a specific truncation in the matrix size and discretisation of the $\Delta_2$-range, \texttt{h2ran}.

There are several criteria which we use to rigorously establish the full positivity of the functionals. Every
bound claimed here has been checked using some combination of these.

The simplest criterion which can sometimes be used to rigorously establish the positivity of a given functional
is the Gershgorin circle theorem:
\begin{theorem*}[Gershgorin]
Let $M$ be a $n \times n$ matrix with complex entries and $r_i := \sum\limits_{j \neq i} |M_{ij}|$. Then the spectrum of $M$ is a subset of the union
\begin{align}
    \bigcup\limits_{i=1}^n B_{r_i}(M_{ii})
\end{align}
of all Gershgorin circles $B_{r_i}(M_{ii}):=\{z \in  \mathbb{C} : |z-M_{ii}| \leq r_i\}$.
\end{theorem*}
Since we have concrete analytic expressions for the entries of the infinite-dimensional matrix corresponding to
a given functional, we can simply compute the distance of all of its Gershgorin circles from the negative real
axis and then take a minimum. If this minimum is strictly bigger than zero, one has established positive
semi-definiteness of the bilinear form in question. This simple criterion works well for functionals involving
only a small number of derivatives, as one would expect: In the simplest case of just first order derivatives,
for example, the asymptotic Toeplitz matrix corresponding to the functional is just tridiagonal. But, as a
consequence of Theorem~\ref{theo:Toeplitz}, the Gerschgorin criterion is not only sufficient for positive
semi-definiteness but even necessary in the case of tridiagonal Toeplitz matrices.

For functionals involving a larger number of derivatives, the Gerschgorin criterion is however too restrictive.
The remainder of this section aims at describing the more generally applicable higher derivative extension of
the ideas described in Section~\ref{sec:toybound}, where we demonstrated the positivity of a particular
first derivative functional.

In order to show positive semi-definiteness of an infinite band matrix $M$ of bandwidth $\Lambda$, we introduce
an auxiliary $n_c \times n_c$ matrix $C$, which we refer to as the \emph{cutting matrix}. We demand that $n_c
\ge \Lambda +1$. To this matrix, we associate the matrix $\overline{C}$ defined by
\begin{align}
    \overline{C}_{ij} \equiv 1 - C_{ij}\,.
\end{align}
Next, we choose two integers $n_l > n_c$ and $n_h > 2 n_c $. These choices let us define matrices $M^{(0,C)}$
and $\{M^{(n,C)}\}_{n>0}$ as
\begin{align}
    M^{(0,C)}_{ij} \equiv
    \begin{cases}
        M_{ij}  & \min(i,j)  \leq  n_l - n_c \\
        M_{ij} \overline{C}_{i-(n_l-n_c),j-(n_l-n_c)} & n_l-n_c < i,j \leq n_l  \\
        0  &  \text{else}
    \end{cases}
\end{align}
and
\begin{align}
    M^{(n,C)}_{ij} \equiv
    \begin{cases}
         M_{ij} C_{i-n,j-n}  & n < i,j \leq n+ n_c\\
        M_{ij}  & n+ n_c < \max(i,j) \land \min(i,j)  \leq n + n_h - n_c \\
        M_{ij} \overline{C}_{i-(n+n_h- n_c),j-(n+n_h- n_c)} & n+n_h- n_c < i,j \leq n+n_h \\
        0  & \text{else}
    \end{cases}\,.
\end{align}
Note that we already discussed a special case with $n_c=1$ in eq.~\eqref{eq:Msimple}. With these more general
definitions, we conclude that
\begin{align}
    M_{ij} = M^{(0,C)}_{ij} + \sum\limits_{\ell=0}^\infty M^{(\ell (n_h - n_c) + n_l- n_c,n_c)}_{ij}\,.
\end{align}
Thus, if we are able to show for some specific choice of the cutting matrix $C$ and of integers $ n_h,n_l$
that $M^{(0,C)}$ and $M^{(n,C)}$ are positive semi-definite for all $n \ge n_l - n_c$ then we can
conclude that $M$ is positive semi-definite. In principle, a suitable cutting matrix $C$ tailored specifically
to the matrix $M$ at hand, could be constructed using SDP. However, we found it sufficient to simply always
take an Ansatz
\begin{align}
    C_{ij} = a + (1-2 a)\frac{i+j-2}{2 n_c-2}\,,
\end{align}
for some constant $a$ together with sufficiently large integers $n_c$, $n_l$ and $n_h$.

In practice we perform these checks in an automated iterative procedure consisting of the following four
steps:
\begin{enumerate}
    \item Use \texttt{SDPA} to construct a functional that is positive on a certain grid \texttt{h2ran}
    of $\Delta_2$ values up to a certain cutoff \texttt{DTrunc}. Then initiate testing by proceeding to step 2.
    \item Refine the grid by complementing it with the midpoints of all neighbouring grid-elements. If the
    functional is negative somewhere on the refined grid, replace \texttt{h2ran} by the refined grid and
    return to step 1. Otherwise proceed with step 3.
    \item Increase \texttt{DTrunc} by some fixed value and check whether the functional stays positive on
    the refined grid for the new value of \texttt{DTrunc}. If so, continue with the final step. Otherwise,
    return to step 1 with the new value of \texttt{DTrunc}.
    \item Generate plots of the minimal eigenvalues of the cut matrices, similar to those in Section
   ~\ref{sec:toybound}, that allow the user to visually verify full positivity.
\end{enumerate}

Ideally, one would avoid this verification process altogether by treating both $\Delta_2$ and $n$ as polynomial variables, and subsequently imposing polynomial inequalities. While this seems natural and straightforward for the $\Delta_2$ dependence, some extra thought is required for the $n$ dependence, which controls the size of the matrices we work with. However, the cut decomposition described here might be of use not just to check positivity but to impose it systematically. We hope to return to this point in the near future.

\subsubsection{Discussion of numerical results}
\label{sec:noidresults}
\begin{figure}[ht]
	\centering \includegraphics[width=350pt]{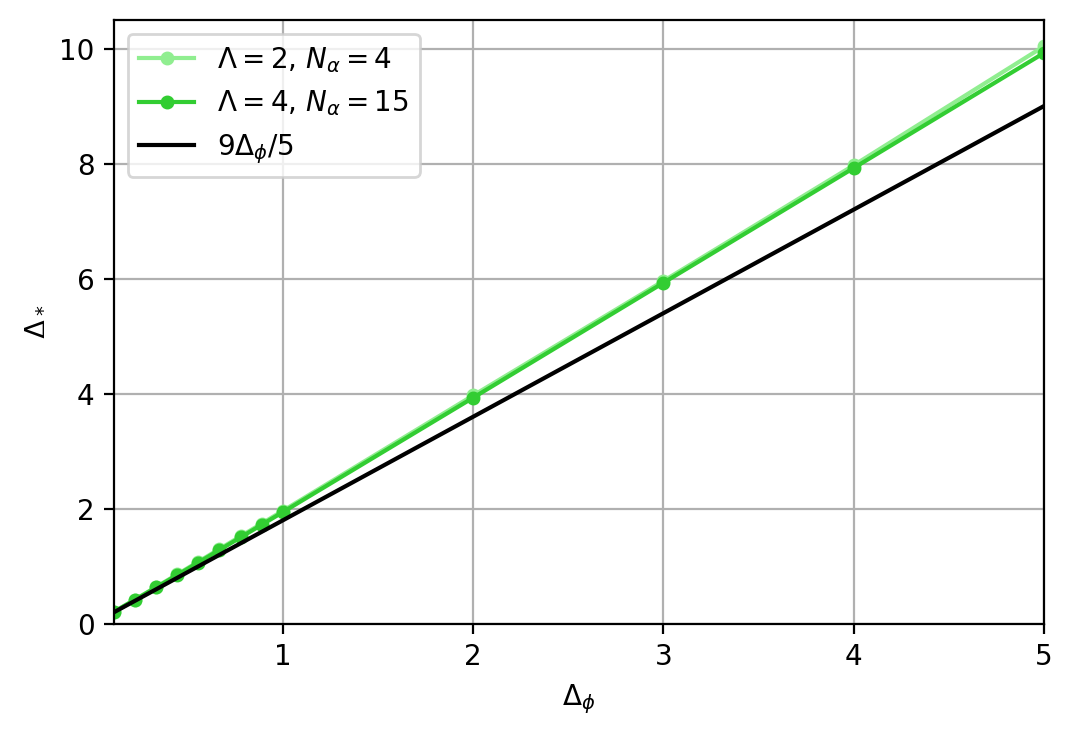}
	\caption{Bound on the dimension of the leading exchange on the $\mathcal{O}_2$ channel: $\Delta_*$ as a function of the external dimension $\Delta_\phi$. In green, we plot the upper bounds derived from the numerical SDP approach, with darker colours corresponding to higher derivative order $\Lambda$. In black, we plot the primal solution constructed in Section~\ref{sec:Primal}, which agrees remarkably well with the numerical bound for small values of $\Delta_\phi$.}
 \label{fig:NO1}
\end{figure}

Having established the methods to obtain linear combinations of functionals whose positivity we can
thoroughly check, we now present the resulting numerical bounds for the problem of gap maximisation
without identity. In Figure~\ref{fig:NO1}, we plot the maximum allowed value for $\Delta_2=\Delta_*$,
the dimension of the leading operator in the $\mathcal{O}_2$ channel, as a function of the external
dimension $\Delta_\phi$. We present results for several values of $0<\Delta_\phi\leq5$, with a more
refined grid in the range $0<\Delta_\phi\leq1$. The bound is roughly linear, and completely smooth.

\begin{figure}[ht]
	\centering \includegraphics[width=350pt]{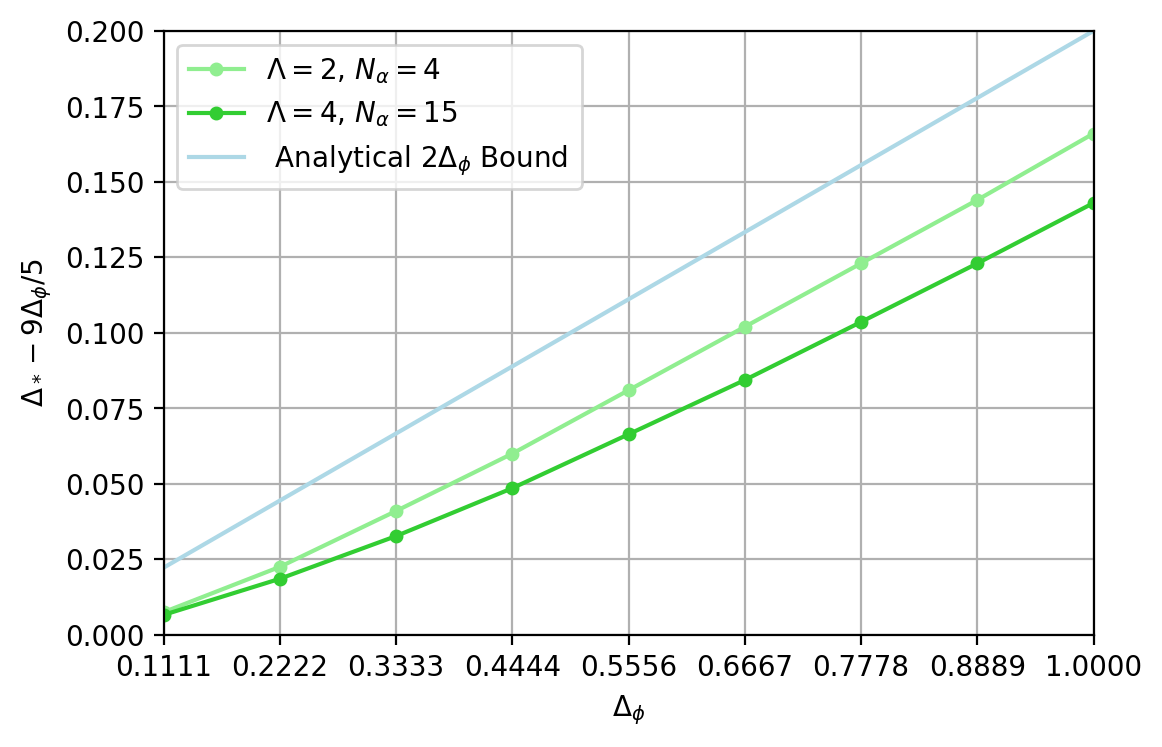}
	\caption{Difference between the small $\Delta_\phi$ data-points plotted in Figure~\ref{fig:NO1}
 and the best primal solution to gap maximisation without identity that we were able to construct. The
 thin blue line represents the analytical toy bound of Section~\ref{sec:toybound}.}
 \label{fig:NO1lowh}
\end{figure}

Remarkably, even at the lowest derivative order $\Lambda=2$ (the lightest shade of green in the plot) we
find percent-level agreement with the primal solution $\Delta_2=9\Delta_\phi/5$ discussed in section
\ref{sec:Primal}. In fact, even at this derivative order all the points with $0<\Delta_\phi\leq1$ give
a bound strictly stronger than $\Delta_2<2\Delta_\phi$, which leads us to conjecture that $\Delta_2=
9\Delta_\phi/5$ actually saturates the bound on the gap maximisation without identity problem. For
small $\Delta_\phi$ the bound is actually so close to the conjectured optimum $\Delta^P_\ast =
9 \Delta_\phi/5$ that the difference between the green and the black lines cannot be resolved in
Figure~\ref{fig:NO1}. Therefore, we plot the difference between the obtained bounds and $\Delta^P_\ast$
in the range $0 < \Delta_\phi < 1$ in the separate Figure~\ref{fig:NO1lowh}.

Subsequent increases of the derivative order $\Lambda$ lead only to a slight (but finite) improvement
to the bound, which is unsurprising, given how close the $\Lambda=2$ answer already is to the primal
solution we constructed. This good agreement at low derivative order and slow improvement with
$\Lambda$ seems to simply be an artefact of the ``no-identity'' problem, as we will find a more
familiar behaviour in the GFF-type problems considered in the next section.

\section{Bounds on Theories with GF Four-Point Functions}
\label{sec:GFFbounds}
With all the numerical machinery in place, we now tackle more realistic bootstrap problems in
which the operator product of the external field $\phi$ with itself contains the identity operator. The appearance of the identity
implies that in the OPE limit the $\mathcal{O}_2$ channel is dominated by $\phi$, i.e.\ the external
operator. Including also the first subleading contributions we have
$$ (\phi\times\phi)\times\phi\sim(\mathbf{1}+\mathcal{O}_1^*+\dots)\times \phi
\sim \phi+ \mathcal{O}_2^* + \dots$$
where we denoted the first nontrivial operator in the intermediate channels by $\mathcal{O}_j^*$.
Given this natural setup, an immediate quantity to consider is the gap above $\phi$ in the middle
channel, i.e.\ the conformal dimension $\Delta_*$ of the leading non-trivial field $\mathcal{O}_2^*$.
To obtain a bound on this quantity, we should solve the following positivity conditions\footnote{We prefer to formulate
equation \eqref{eq:positivityconditions} in terms of the primary space bilinear-forms $B$, as this seems to be more natural for the point that is made here. We could have also chosen to use the descendant-space formulation and replace $B$ by $b$.}
\begin{align}\label{eq:positivityconditions}
\sum_{k}c_k B^k_{\Delta_\phi} \succ 0\,, \quad \quad \sum_{k}c_k B^k_{\Delta_2} \succ 0\, \quad
\textrm{for all} \quad \Delta_2\geq \Delta_*\,.
\end{align}
In is important to note, however, that the six-point function $\mathcal{G}$ now contains contributions
from the intermediate exchange of $\mathcal{O}_2 = \phi$ that are qualitatively different form the
remaining terms in the block expansion. Let us collect all the terms in the conformal blocks expansion
\eqref{eq:combCBexpansion} that involve six-point blocks with $\Delta_2 = \Delta_\phi$ in a new function
\begin{equation}
\label{eq:phiblock}
\mathcal{G}_{\textrm{4-pt}}(\chi_1,\chi_2,\chi_3)\equiv \sum_{\mathcal{O}_1,\mathcal{O}_3}
C_\phi^{\mathcal{O}_1} C_\phi^{\mathcal{O}_3} G^{123,456}_{\Delta_1,\Delta_\phi,\Delta_3} =
\sum_{\mathcal{O}_1,\mathcal{O}_3} C_{\phi \phi \mathcal{O}_1}^2
C_{\phi \phi \mathcal{O}_3}^2 G^{123,456}_{\Delta_1,\Delta_\phi,\Delta_3}\, \ .
\end{equation}
We stress that this function depends on all three six-point cross ratios. The reason we denoted it
with a subscript `4-pt' is that it only involves CFT data contained in the four-point function of
$\phi$. One should also observe that the product of OPE coefficients appearing in this special contribution to the full six-point function is
manifestly  positive.

These observations have important consequences for the choice of strategy that should be implemented
to ensure
\begin{equation}
\label{eq:G4ptpos}
\alpha(\mathcal{G}_{\textrm{4-pt}}) =    \sum_k c_k \alpha_k(\mathcal{G}_{\textrm{4-pt}}) \geq 0 \,.
\end{equation}
First of all, note that the transformation to descendant-space discussed in Section~\ref{sec:demonstrateequiv}
preserves the positivity property of the primary-space OPE vector $C_{\phi \phi \mathcal{O}}^2$. Thus, both in
primary- and descendant-space, ensuring eq.~\eqref{eq:G4ptpos} does not require positive semi-definiteness
of the bilinear form $B_{\Delta_\phi}$ but only co-positivity i.e.~the property $B_{\Delta_\phi}(C,C) \ge
0$ for all OPE vectors $C$ that are componentwise positive (see e.g.~\cite{CoPositiveReview} for a review on
co-positive programming). For instance, every
matrix with only positive entries is manifestly co-positive, and of course every positive semi-definite matrix 
is co-positive. However, solving a co-positive program is generally NP-hard.
Nonetheless, it might be sufficient to relax co-positivity to componentwise positivity or positive 
semi-definiteness of the bilinear forms to construct valid bootstrap bounds, which of course cannot be stronger 
than in the co-positive case.

 We leave
further exploration of this question to future work and in the following take the perspective
that one has solved or
bootstrapped the four-point function of $\phi$, and hence knows all non-vanishing OPE coefficients
$C_{\phi \phi \mathcal{O}}^2$. In this scenario, we can just compute $\mathcal{G}_{\textrm{4-pt}}$
and directly impose eq.\ \eqref{eq:G4ptpos} as a constraint on the choice of the functionals. A
simple case where we are in this situation is for generalised free theories, which we will proceed
to analyse in the subsections below.

\subsection{Bounds on the triple-twist gap for GFF}
Let us first consider the generalised free fermion theory with a real fermionic field $\psi$. In
this case, the spectrum in the $\psi\times\psi$ OPE contains operators of conformal dimension
$\Delta_n^{\textrm{F}}=2\Delta_\psi+2n+1$ with $n\in \mathbb{Z}_{\geq0}$, and the corresponding
OPE coefficients are given in eq.~\eqref{eq:OPEpsi}.

\begin{figure}[ht]
	\centering \includegraphics[width=350pt]{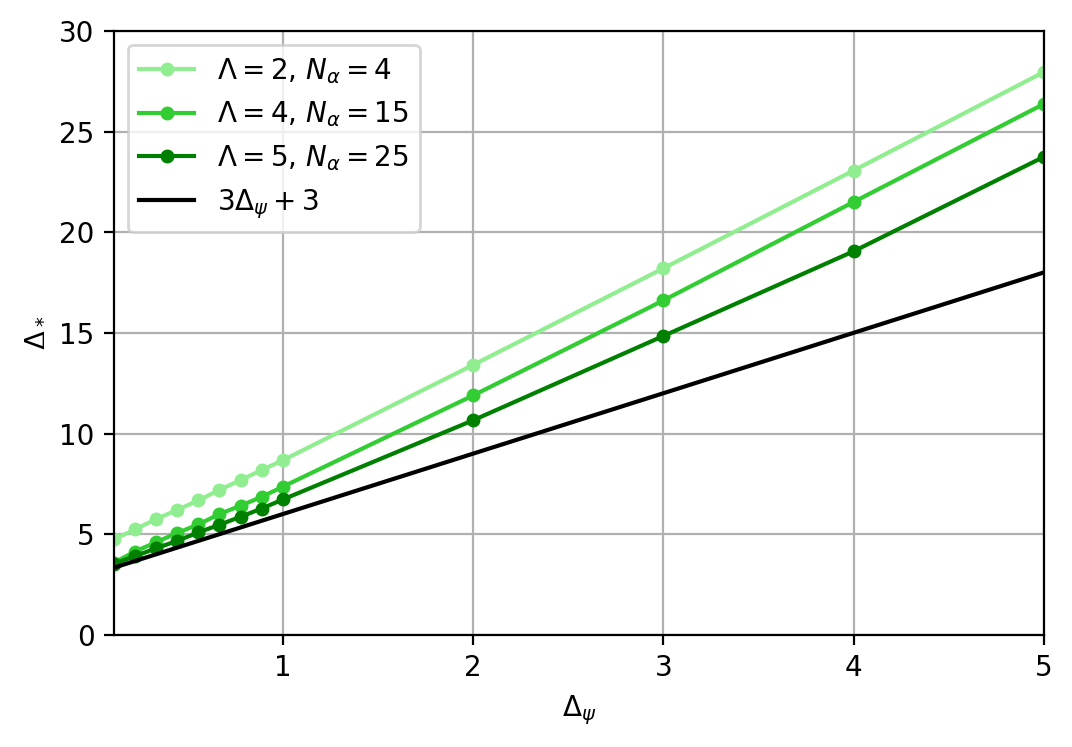}
	\caption{Bound on the dimension of the subleading exchange on the $\mathcal{O}_2$ channel:
        $\Delta_*$ as a function of the external dimension $\Delta_\psi$. In green, we plot the upper
        bounds derived from the numerical SDP approach, with darker colours corresponding to higher
        derivative order $\Lambda$. In black, we plot the expected gap from the leading triple-twist
        operator in GFF theory, which agrees remarkably well with the numerical bound for low
        values of $\Delta_\psi$.\label{fig:GFF+}
	}
\end{figure}

We can then easily reconstruct the function $\mathcal{G}_{\textrm{4-pt}}$ we defined in
eq.~\eqref{eq:phiblock} and which now takes the following explicit form
\begin{align}
\label{eq:G4ptGFF}
   &\mathcal{G}_{\textrm{4-pt}}^{\textrm{GFF}}(\chi_i)/\chi_{2}^{\Delta_\psi}= 1
   -\chi_1^{2\Delta_\psi} + \left(\frac{\chi_1}{1-\chi_1}\right)^{2\Delta_\psi}
   -\chi_3^{2\Delta_\psi} + \left(\frac{\chi_3}{1-\chi_3}\right)^{2\Delta_\psi}
   + \left(\frac{\chi_1 \chi_3}{1-\chi_2}\right)^{2\Delta_\psi}\nonumber \\[2mm]
  &  - \left(\frac{\chi_1 \chi_3}{1-\chi_1-\chi_2} \right)^{2\Delta_\psi}
  - \left(\frac{\chi_1 \chi_3}{1-\chi_3-\chi_2} \right)^{2\Delta_\psi}
  + \left(\frac{\chi_1 \chi_3}{1-\chi_1-\chi_2-\chi_3 + \chi_1\chi_3} \right)^{2\Delta_\psi} \,.
\end{align}
This piece of the full correlator can also be identified directly with a Wick contraction computation
of the six-point function, keeping track of the usual fermionic minus signs. Each of the nine terms
above can be identified with a Wick contraction where exactly one of the propagators crosses from the
set of points $\{x_1,x_2,x_3\}$ to the set $\{x_4,x_5,x_6\}$. We can finally use eq.~\eqref{eq:G4ptGFF}
to add a constraint to our SDP according to eq.~\eqref{eq:G4ptpos}. Our results for the bound on the
dimension $\Delta_*$ of the first operator $\mathcal{O}_2^*$ above the operator $\psi$ in the
$\mathcal{O}_2$ channel are shown in Figure~\ref{fig:GFF+}.

 As in Section~\ref{sec:noidresults}, we present results for several values of $0<\Delta_\psi\leq5$,
 with a more refined grid in the range $0<\Delta_\psi\leq1$. At fixed derivative order $\Lambda$, the
 bound is roughly linear in the interval $0<\Delta_\psi\leq1$ but starts bending upward for larger
 values of $\Delta_\psi$. Nonetheless, the bound is completely smooth, and the non-linearity seems
 to simply be caused by a slower convergence at larger values of $\Delta_\psi$. In this case, as
 $\Lambda$ increases, and the tone of green gets darker, there is a more substantial improvement
 of the bounds when compared to the ``no-identity'' case. For reference, we note that at the highest derivative order $\Lambda=5$ each value of $\Delta_\psi$ took a few days on a single core of a 2.8 GHz \texttt{AMD EPYC 7402} 24-core processor, and is roughly 50 to 100 times slower than the smaller value $\Lambda=4$. The majority of this time is taken by solving the actual SDP. We expect that making full use of the sparsity structure of the matrix, this can be sped up by several orders of magnitude.

Returning to the value of the bound itself, we see that it is always above
 the black line. The latter corresponds to the value $\Delta^P_*= 3\Delta_\psi+3$ which is the
 gap for the $\mathcal{O}_2$ exchange in the GFF six-point function. The spectrum of non-trivial fields that are exchanged in the $\mathcal{O}_2$ channel of GFF can
be read off easily from the following function in which we subtract all contributions from the
exchange or $\psi$ and its descendants,
\begin{align}
\label{eq:GFF6pt}
   &(\mathcal{G}^{\textrm{GFF}}-\mathcal{G}_{\textrm{4-pt}}^{\textrm{GFF}})/\chi_{2}^{\Delta_\psi}
   =\left(\frac{\chi_1 \chi_2 \chi_3}{(1-\chi_1-\chi_2)(1-\chi_3)} \right)^{2\Delta_\psi}
   +\left(\frac{\chi_1 \chi_2 \chi_3}{(1-\chi_3-\chi_2)(1-\chi_1)} \right)^{2\Delta_\psi} \nonumber \\[2mm]
  & - \left(\frac{\chi_1 \chi_2 \chi_3}{(1-\chi_1)(1-\chi_2)(1-\chi_3)} \right)^{2\Delta_\psi}
  -\left(\frac{\chi_1 \chi_2 \chi_3}{(1-\chi_1-\chi_2)(1-\chi_2-\chi_3)} \right)^{2\Delta_\psi}\\[2mm]
  & -\left(\frac{\chi_1 \chi_2 \chi_3}{1-\chi_1-\chi_2-\chi_3+ \chi_1 \chi_3} \right)^{2\Delta_\psi} + \left(\frac{\chi_1 \chi_2 \chi_3}{(1-\chi_1-\chi_2-\chi_3+ \chi_1 \chi_3)(1-\chi_2)} \right)^{2\Delta_\psi} \,. \nonumber
\end{align}
Expanding this function in conformal blocks \eqref{eq:combblock}, and matching the Taylor series
at small $\chi_i$ allows us to read off the spectrum of primaries in all channels, see
Appendix~\ref{app:blockexpansion} for more details. In particular the first non-trivial operator
$\mathcal{O}_2^*$ indeed has dimension $\Delta^P_\ast = 3\Delta_\psi+3$. This was to be expected
also form the Fock space realisation of GFF. Note that in GFF the OPE of the single field $\psi$
and the two-field/double-twist operators $\mathcal{O}_1$, only contains $\psi$ itself and
three-field/triple-twist operators. Because of fermionic statistics, simple powers of $\psi$ and its derivatives
vanish and one has to insert derivatives of different orders to obtain a non-zero operator. In the
case of three-field/triple-twist, the lowest non-trivial operator is obviously given by
\begin{equation}
    \mathcal{O}_2^* = \psi (\partial \psi) (\partial^2 \psi)\, .
\end{equation}
Here normal ordering is understood. The conformal dimension of this operator is indeed given by
$\Delta(\mathcal{O}_2^*) = 3\Delta_\psi +3$, in agreement with the result we obtained by expanding
the function~\eqref{eq:GFF6pt} in the cross ratio $\chi_2$. More details on how to count multi-twist
operators in GFF can be found in Appendix~\ref{app:countmulti}.

While the numerics have yet to fully converge, we conjecture that the bound on the dimension of
$\mathcal{O}_2^*$ is saturated by $\Delta^P_\ast = 3\Delta_\psi+3$, the GFF value. At first glance
our conjecture that the optimal value $\Delta^D_{\textrm{opt}}$ of the dual SDP problem coincides
with the free field result, i.e.\ $\Delta^D_{\textrm{opt}} = \Delta^P_*$ might seem tautological,
since we assume the CFT data of GFF in the $\psi\times\psi$ OPE. However, the bound on the dimension
$\Delta^D_{\textrm{opt}}$ in the dual SDP problem is answering a non-trivial question: Assuming
that the four-point function of $\psi$ matches GFF, is it true that the six-point function of
$\psi$ must agree with GFF as well?

In order to build some more intuition into how this could indeed fail to be the case, let us make
use of the description of GFF in terms of a free massive Majorana fermion $\Psi$ in AdS$_2$. If we turn on
a sextic coupling $g_6$ in AdS, as shown in the schematic EFT action\footnote{While the distribution
of derivatives in the sextic coupling is schematic, it is indeed true that the leading interaction
for a single Majorana fermion has six derivatives. Furthermore, note that a single coupling $g_6$ is irrelevant and presumably incompatible with Regge boundedness and therefore the action has to be understood in the EFT sense.}
\begin{equation} \label{eq:FAdS}
    S[\Psi] = \int_{AdS_2} \sqrt{g} \,d^2x \left(\bar{\Psi}(\slashed{\nabla}-m)\Psi
    + g_6 (\nabla\Psi)^6 + \dots \right)\,,
\end{equation}
while requiring quartic couplings to be zero, we change the triple-twist dimensions at leading order
in $g_6$ without touching the four-point function of $\psi$, due to the contribution of the six-point
contact diagram. In this language, the conjecture we formulated in the previous paragraph can be
reformulated as follows: Within the family of theories described by the action \eqref{eq:FAdS}, the
$g_6=0$ point (GFF) is extremal, saturating the bound.\footnote{It would be interesting to prove
extremality of the GFF six-point correlator using higher-point analytic functionals extending the
work of \cite{Mazac:2016qev,Mazac:2018mdx,Ghosh:2021ruh,Ghosh:2023lwe, Paulos:2020zxx, Ferrero:2019luz,
Caron-Huot:2020adz}}

The reformulation in terms of AdS also implies that our analysis could lead to bounds on the sign
of the irrelevant sextic coupling, since linearity of the anomalous dimensions in $g_6$ could either
push us above or below the extremal value. This would be the six-point generalisation of the bounds
on the sign of the $(\partial \phi)^4$ couplings obtained with the four-point bootstrap in
\cite{Antunes:2021abs}. While such sign constraints on irrelevant quartic couplings are well
understood from many other points of view \cite{Adams:2006sv,Hartman:2015lfa,Caron-Huot:2021enk,
Knop:2022viy}, to the best of our knowledge, sign constraint for a higher-point coupling would
be new.\footnote{We thank Sasha Zhiboedov for a discussion on this point.} See however \cite{Serra:2023nrn} for recent progress.

\subsection{Bounds on the triple-twist gap for GFB}
\label{sec:GFBbounds}

\begin{figure}[ht]
	\centering \includegraphics[width=350pt]{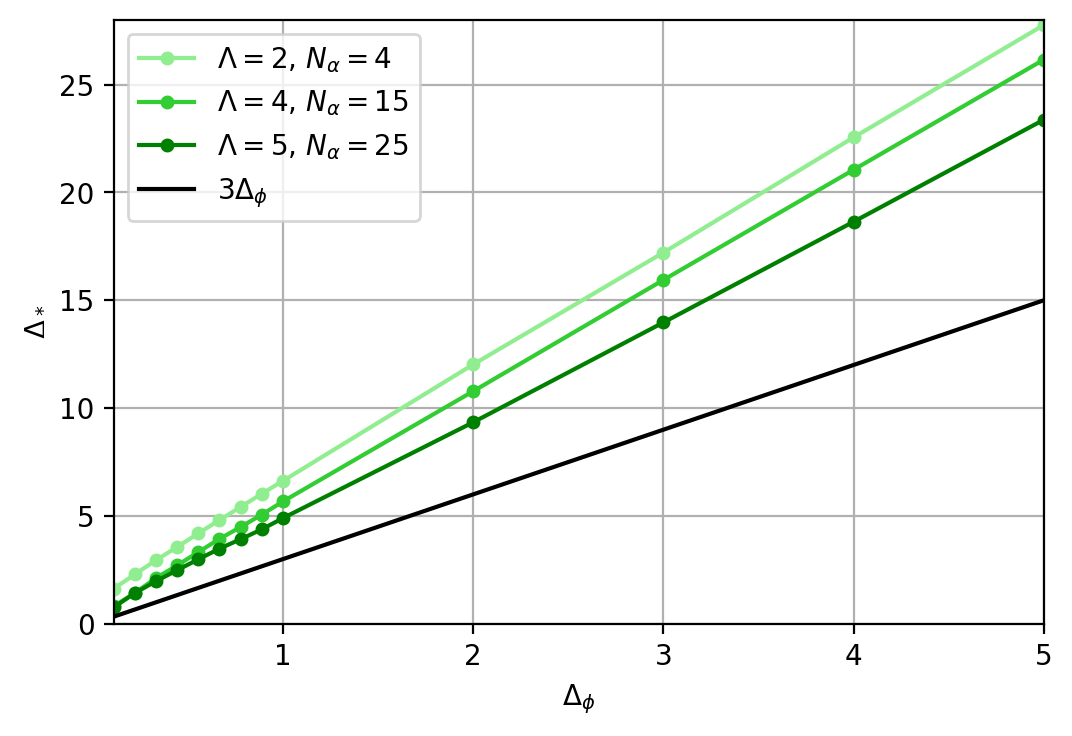}
	\caption{Bound on the dimension of the subleading exchange on the $\mathcal{O}_2$ channel:
 $\Delta_*$ as a function of the external dimension $\Delta_\phi$. In green, we plot the upper bounds
 derived from the numerical SDP approach, with darker colours corresponding to higher derivative
 order $\Lambda$. In black, we plot the expected gap from the leading triple-twist operator in GFB
 theory, which appears to be strictly below the numerical bound, even as $\Lambda$ is increased.
		\label{fig:GFB+}
	}
\end{figure}

It is straightforward to modify the above analysis to the case of generalised free bosons. For
this theory, the OPE of the external field $\phi$ with itself contains operators of dimensions
$\Delta_n^{\textrm{B}}=2\Delta_\psi+2n$ with $n\in \mathbb{Z}_{\geq0}$, and the corresponding
OPE coefficients are given in eq.~\eqref{eq:OPEboson}. Once again, we impose the positivity
condition on the $\Delta_\phi$ block by requiring eq.\ \eqref{eq:G4ptpos} on
\begin{align}
\label{eq:G4ptGFB}
   &\mathcal{G}_{\textrm{4-pt}}^{\textrm{GFB}}(\chi_i)/\chi_{2}^{\Delta_\phi}= 1
   +\chi_1^{2\Delta_\phi} + \left(\frac{\chi_1}{1-\chi_1}\right)^{2\Delta_\phi}
   +\chi_3^{2\Delta_\phi} + \left(\frac{\chi_3}{1-\chi_3}\right)^{2\Delta_\phi}
   + \left(\frac{\chi_1 \chi_3}{1-\chi_2}\right)^{2\Delta_\phi}\nonumber \\[2mm]
   & +\left(\frac{\chi_1 \chi_3}{1-\chi_1-\chi_2} \right)^{2\Delta_\phi}
   + \left(\frac{\chi_1 \chi_3}{1-\chi_3-\chi_2} \right)^{2\Delta_\phi}
   + \left(\frac{\chi_1 \chi_3}{1-\chi_1-\chi_2-\chi_3 + \chi_1\chi_3} \right)^{2\Delta_\phi} \,,
\end{align}
which is reconstructed from the bosonic four-point data, and can of course be obtained by the
same Wick contractions as the GFF correlator above.  With this, we present the results of the
bound $\Delta_*$ on the dimension $\Delta_2^*$  of the first operator $\mathcal{O}_2^*$ above
$\phi$ in the $\mathcal{O}_2$ channel in Figure~\ref{fig:GFB+}.
Once again, we choose several values of $0<\Delta_\phi\leq5$, with a more refined
grid in the range $0<\Delta_\phi\leq1$. At fixed derivative order $\Lambda$, the bound is roughly
linear in the interval $0<\Delta_\phi\leq1$ and seems to remain so for larger values of
$\Delta_\phi$. The values of the bound are always well below their GFF counterparts. However,
as $\Lambda$ increases, and the tone of green gets darker, there is a slower improvement of
the bounds when compared to the GFF case. We see that the bound is always above the black
line, corresponding to the value $3\Delta_\phi$ which is the gap in the $\mathcal{O}_2$
sector of the GFB six-point function
\begin{align}
\label{eq:GFB6pt}
   &(\mathcal{G}^{\textrm{GFB}}-\mathcal{G}_{\textrm{4-pt}}^{\textrm{GFB}})/\chi_{2}^{\Delta_\phi}
   = \left(\frac{\chi_1 \chi_2 \chi_3}{(1-\chi_1-\chi_2)(1-\chi_3)} \right)^{2\Delta_\phi}
   +\left(\frac{\chi_1 \chi_2 \chi_3}{(1-\chi_3-\chi_2)(1-\chi_1)} \right)^{2\Delta_\phi} \nonumber \\[2mm]
  & +\left(\frac{\chi_1 \chi_2 \chi_3}{(1-\chi_1)(1-\chi_2)(1-\chi_3)} \right)^{2\Delta_\phi}
  +\left(\frac{\chi_1 \chi_2 \chi_3}{(1-\chi_1-\chi_2)(1-\chi_2-\chi_3)} \right)^{2\Delta_\phi}\\[2mm]
  & +\left(\frac{\chi_1 \chi_2 \chi_3}{1-\chi_1-\chi_2-\chi_3+ \chi_1 \chi_3} \right)^{2\Delta_\phi}
  + \left(\frac{\chi_1 \chi_2 \chi_3}{(1-\chi_1-\chi_2-\chi_3+ \chi_1 \chi_3)(1-\chi_2)} \right)^{2\Delta_\phi} \,,
\nonumber
\end{align}
which we expand in conformal blocks in Appendix~\ref{app:blockexpansion}. Somewhat surprisingly,
however, in this case it seems that the bound does not converge all the way to the $\phi^3$ operator
of GFB, meaning the extremal theory whose four-point function agrees with GFB might not be GFB! Once
again the AdS$_2$ perspective sheds some light on why this might be the case. Consider the following
action in AdS
\begin{equation}
    S= \int_{AdS_2} \sqrt{g} \,d^2x \left(\frac{1}{2} \nabla^{\mu}\Phi\nabla_{\mu}\Phi
    + \frac{1}{2} m^2 \Phi^2 + g_6 \Phi^6 + \dots\right)\,,
\end{equation}
where $\Phi$ is the bulk scalar dual to $\phi$ on the boundary and we initially fine tuned quartic
couplings to zero. As before, to leading order in $g_6$, the four-point function of $\phi$ is
unaltered, but there is a six-point contact diagram leading to anomalous dimensions of
triple-twists.\footnote{It is not hard to see, based on the spectral or Mellin representation of
the six-point contact diagram that the comb-channel block expansion of this object contains only
blocks of the form $G_{4\Delta_\phi+n_1,3\Delta_\phi+n_2,2\Delta_\phi+n_3}$ as well as $G_{2\Delta_\phi+n_1,3\Delta_\phi+n_2,4\Delta_\phi+n_3}$ and $\partial_{n_2} G_{2\Delta_\phi+n_1,3\Delta_\phi+n_2,2\Delta_\phi+n_3}$ which indeed corresponds to having anomalous
dimension for triple-twists but no anomalous dimensions for double-twists, at leading order.}
Unlike for fermions, though, the $\Phi^6$ coupling is relevant, and hence we don't expect bounds on
the sign of its coupling. In particular, we should be able to increase the dimension of $\phi^3$
above the GFB value consistently, at least perturbatively. In practice, the bound should be a
finite distance away from the GFB value, and a detailed explanation could involve finite coupling
dynamics (perhaps related to instability of the potential), or an altogether unphysical solution,
so it is hard to estimate the actual upper bound from this analysis. We emphasise, however, that
very similar phenomena were observed in \cite{Antunes:2021abs,Ghosh:2023lwe}, for the mixed
four-point system of $\phi$ and $\phi^2$, where certain bounds are not saturated by GFB, even
though one imposes by hand part of the GFB CFT data.

\subsection{Comments on bounding three- and four-point functions}

One of the main appeals of the six-point bootstrap in higher dimensions is the access to new
OPE coefficients that are not probed by the scalar four point bootstrap such as OPE coefficients
involving two spinning operators. In the 3d Ising model, for example, the six-point correlators
can access the OPE coefficient for $TT\epsilon$ in the six-point function of $\epsilon$ or the
OPE coefficient for $T\sigma S$, where $S$ is a $\mathbb{Z}_2$ odd spinning operator, in the
six-point function of $\sigma$. Hence, it would certainly be of interest to bound (products of)
OPE coefficients $C^{\mathcal{O}_2}_{\mathcal{O}_1}$ within the numerical six-point
bootstrap.

A very basic (but non-generic) example is the OPE coefficient $C^{\phi^3}_{\phi^2}$ in GFB.
In this case, the only operator $\mathcal{O}_1$ with a non-vanishing OPE coefficient
$C_{\phi \mathcal{O}_1 \phi^3}$ is $\mathcal{O}_1 = \mathcal{O}_1^* = \phi^2$, i.e. the
OPE coefficients for the GFB obey
\begin{equation} \label{eq:propGFB}
 C_{\phi \mathcal{O}_1 \mathcal{O}_2^*} = 0 \quad \text{ for all }
\quad \mathcal{O}_1 \neq \mathcal{O}^*_1\ .
\end{equation}
Indeed, for the GFB this OPE coefficient vanishes for all other double-twist operators
with more derivatives, as one may infer from eq.~\eqref{eq:GFB6pt} by expanding in conformal
blocks (see Appendix~\ref{app:blockexpansion}). We can now separate out the contribution from
$\mathcal{O}_2^* =\phi^3$ exchange in the middle channel and rewrite the six-point
crossing equation as
\begin{equation}
    (C^{\mathcal{O}^*_2}_{\mathcal{O}_1^*})^2 \alpha(F_{2\Delta_\phi,3\Delta_\phi,
    2\Delta_\phi})
    =-\alpha(\mathcal{G}_{\textrm{4-pt}}) - \sum_{\Delta_2>\Delta_2^*}
    \left( \sum_{\mathcal{O}_1,\mathcal{O}_3}C^{\mathcal{O}_2}_{\mathcal{O}_1}
    \alpha (F_{\Delta_1,\Delta_2,\Delta_3}) C^{\mathcal{O}_2}_{\mathcal{O}_3}\right) \,.
\end{equation}
Here, we have also applied a functional $\alpha$ to both sides of the equation. Let us agree
to normalise the functional $\alpha$ such that it $\alpha(F_{2\Delta_\phi,3\Delta_\phi,
2\Delta_\phi}) = 1$. If we now require positive semi-definiteness above the gap, as usual,
we can maximise $\alpha(\mathcal{G}_{\textrm{4-pt}})$ to obtain the bound
\begin{equation} \label{eq:Cbound}
(C^{\mathcal{O}^*_2}_{\mathcal{O}_1^*})^2  \leq - \max{\alpha(\mathcal{G}_{\textrm{4-pt}})} \,.
\end{equation}
Recall that the vector $C^{\mathcal{O}^*_2}_{\mathcal{O}_1^*}$ is itself a product of two
OPE coefficients so that we have now bounded a product of four OPE coefficients. But it is
easy to obtain a bound on the interesting factor $C_{\mathcal{O}_1^*\phi\mathcal{O}_2^*}^2$
by taking the resulting bound \eqref{eq:Cbound} and dividing it by a bound on $C_{\phi\phi
\mathcal{O}_1^*}^2$ which can be obtained from the usual four-point bootstrap. Note that
to normalise on $F_{2\Delta_\phi,3\Delta_\phi,2\Delta_\phi}$ we are working in the primary-space
basis, which makes it easy to isolate the contribution of the specific OPE coefficient we are interested in. However, the positivity conditions above the gap, are most conveniently implemented in the descendant space basis, as we have argued.
\medskip

In a more generic situation, however, property \eqref{eq:propGFB} is not satisfied and instead
infinitely many operators $\mathcal{O}_1$ will have a non-vanishing OPE coefficient $C_{\phi
\mathcal{O}_1 \mathcal{O}_2^*}$. Hence, accessing individual OPE coefficients will be more
challenging. Furthermore, directly extracting bounds on OPE coefficients from six-point crossing
might also be considered somewhat artificial. As we have argued in Section~\ref{sec:commentson},
from the six-point bootstrap perspective it is natural to view the comb-channel OPE as  a sum of
scalar products of four-point and not three-point correlators whose positivity constraints CFT
data. Consequently, the most straightforward quantity to bound in this framework is the full
four-point correlator and not an individual OPE coefficient.\footnote{See also \cite{Paulos:2020zxx,Paulos:2021jxx} for recent work on bounding the four-point correlators directly from the four-point bootstrap.}
\smallskip

For example, let us discuss how to produce an upper bound on the four-point function of three
external scalars $\phi$ with the order $p$ descendant of the primary $\mathcal{O}_2^*$, i.e. the correlator $\langle \phi \phi \phi P^n \mathcal{O}_2^* \rangle$.
More precisely, we will consider the closely related (see Section \ref{sec:commentson}) quantity
\begin{align}
F^{(p)}_{\mathcal{O}_2^*}\equiv
\sum\limits_{\mathcal{O}}C_{\mathcal{O}_2^*}^{\mathcal{O}} F_p(\Delta).
\end{align}
Here we have used the functions $F_n$ that were introduced in eq.\ \eqref{eq:phiphiphiO2n} and
we evaluated both sides at the crossing symmetric point $\chi_j = 1/3$ (we actually suppressed
the $\chi_j$-dependence entirely). Instead of thinking of $F^{(p)}_{\mathcal{O}_2^*}$ in terms of a descendant four-point correlator
evaluated at $\chi = 1/3$, we can equivalently think of it as a derivative of the four-point
function with the primary $\mathcal{O}_2^*$. Hence we have access to all derivatives of the
four-point correlator $\langle \phi \phi \phi \mathcal{O}_2^*\rangle$ at $\chi_j = 1/3$ and
can reconstruct its Taylor expansion around that point. To obtain the bounds, we simply
construct functionals $\alpha = \sum\limits_{k=1}^{N_\alpha} c_k \alpha_k$ such that, in
the notation of Section~\ref{sec:demonstrateequiv},
\begin{align}
    \left(\sum\limits_{k=1}^{N_\alpha} c_k M^k_{nm}(\Delta_2) - \delta_{\Delta_2}^{\Delta_2^*}\delta_{n}^{p}\delta_{m}^{p}\right)_{n,m} \succ 0\,, &&
    \text{ for all } \quad  \Delta_2 \geq \Delta_2^* \,,\label{eq:subtractedpositivity}
\end{align}
and maximise $\alpha(\mathcal{G}_{\text{4-pt}})$, i.e. the action of the functional $\alpha$
on the function $\mathcal{G}_{\text{4-pt}}$ we defined in eq.~\eqref{eq:phiblock}. The
crossing equation then implies
\begin{equation}
   0 = \alpha(\mathcal{G}_{\textrm{4-pt}}) +\left(F^{(p)}_{\mathcal{O}_2^*}\right)^2
   + \sum_{\Delta_2\ge\Delta_2^*} \left( \sum_{n,m} F^{(n)}_{\mathcal{O}_2}
   (\sum\limits_{k=1}^{N_\alpha} c_k M^k_{nm}(\Delta_2) -
   \delta_{\Delta_2}^{\Delta_2^*}\delta_{n}^{p}\delta_{m}^{p})F^{(m)}_{\mathcal{O}_2}\right)\,,
\end{equation}
and hence
\begin{align}
\left(F^{(p)}_{\mathcal{O}_2^*}\right)^2\leq - \alpha(\mathcal{G}_{\textrm{4-pt}})\,.
\end{align}
Just as for the bounds on OPE coefficients in the usual four-point bootstrap, changing the
minus sign in eq.~\eqref{eq:subtractedpositivity} to a plus allows one to analogously also
construct lower bounds on $\left(F^{(p)}_{\mathcal{O}_2^*}\right)^2$. Note that to get
non-trivial lower bounds, the assumption of a gap below $\Delta_2^*$ is not sufficient
and we need to also impose an additional gap above $\Delta_2^*$.

\begin{figure}[ht]
	\centering \includegraphics[width=350pt]{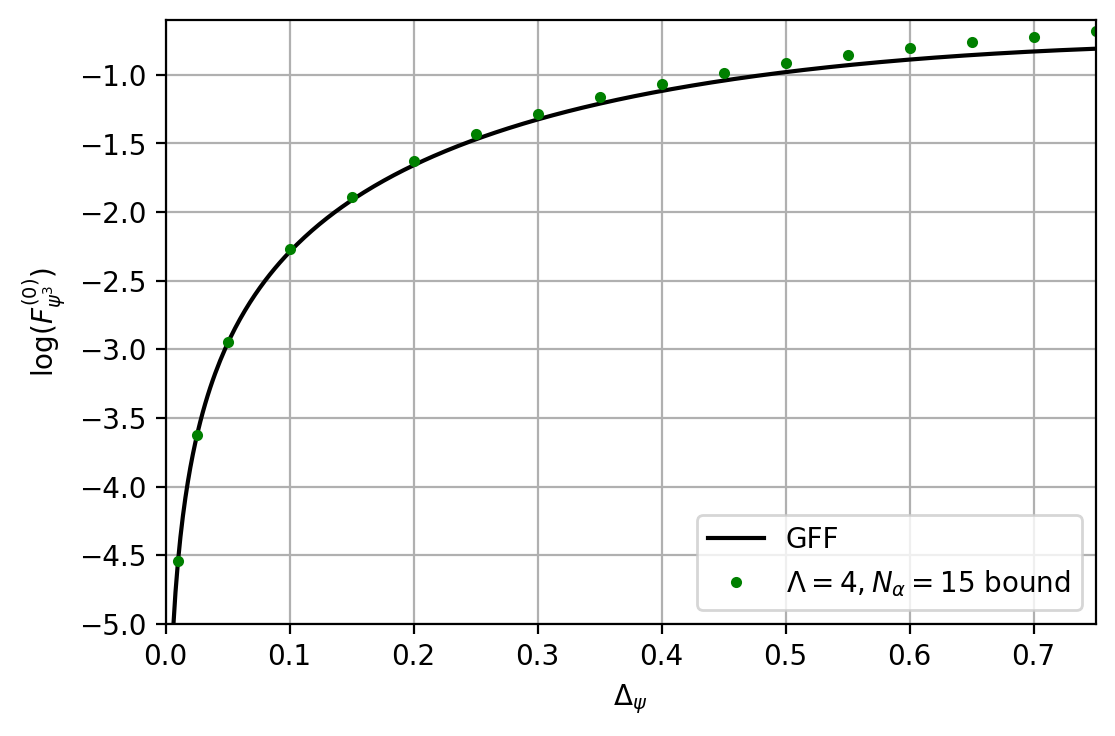}
	\caption{Bound on the $\langle \psi \psi \psi \mathcal{O}_2^*\rangle$ GFF four-point function evaluated at $\chi=1/3$. The green dots represent a numerical upper bound obtained at $\Lambda=4$. The black curve corresponds to the exact GFF result.
		\label{fig:4PBound}
	}
\end{figure}

As an application, let us consider an upper bound on $F^{(0)}_{\mathcal{O}_2^*}$ for the case of GFF, where $\Delta_2^* = 3 \Delta_\phi +3$ corresponds to the leading triple-twist and $\mathcal{G}_{\text{4-pt}}$ was given in eq.~\eqref{eq:G4ptGFF}.
Figure~\ref{fig:4PBound} shows this upper bound for $\Lambda=4$ and compares it to the exact result.
 We would like to stress that the scale of the plot is logarithmic -- the bootstrap bound is thus nearly saturated by GFF over several orders of magnitude. As expected, the bound worsens significantly as $\Delta_\psi$ increases, which can already be noticed in Figure~\ref{fig:4PBound}. We expect that a higher number of functionals would be necessary in order to extend the range of $\Delta_\psi$ where a reasonable agreement with the exact result is possible.

We leave a more systematic exploration of the six-point bounds on four-point functions for future work.
\section{Conclusions and Outlook}
\label{sec:conclusions}
In this paper, we took the first steps to establish a rigorous numerical bootstrap program for the six-point correlator. We wrote down the crossing equation in a way amenable to using SDP methods, albeit for infinite-dimensional matrices. Through the descendant-space approach, we made the infinite dimensional problem numerically manageable, due to good control over the large dimension asymptotics. Furthermore, in the descendant-space formulation we do not need to actually evaluate the six-point conformal blocks, we only use that they factorise into a sum of products of four-point blocks  in such a way that their derivatives can be written in terms of banded matrices. With this technology, we proved bounds on the leading dimensions in the $\mathcal{O}_2$ channel, for the problem without identity as well as for generalised free fermions and bosons. While the two first bounds where saturated by well-understood solutions to crossing, the third one already produced surprising results, which we tentatively interpreted in terms of non-trivial dynamics in AdS$_2$.
There are many clear and exciting future directions.

Firstly, there are some technical improvements to be made still in the realm of one-dimensional kinematics. It should be relatively straightforward to promote the treatment of $\Delta_2$ from a discretised grid of values to a polynomial approximation. In principle, in this formulation it should be possible to take better advantage of \texttt{SDPB}'s capabilities. Furthermore, a more systematic understanding of how to bound specific OPE coefficients would greatly increase the space of available observables. Following this logic, perhaps the extremal functionals associated to certain maximisation problems could contain further information about the full spectrum of the six-point function, in line with the extremal functional method of \cite{El-Showk:2012vjm,El-Showk:2016mxr}. In our case, the extremal functional is actually an infinite matrix: it seems rather suggestive to look at its spectrum of eigenvalues and eigenvectors, but further investigation is needed.

Secondly, regardless of technical improvements, there are already physically interesting problems to tackle in the one-dimensional setup. As we saw in Section~\ref{sec:GFBbounds}, non-trivial AdS$_2$ dynamics seems to emerge from the 1d
bootstrap. A more systematic exploration of deformations of GFB, in the spirit of \cite{Antunes:2021abs,Paulos:2019fkw,Ghosh:2023lwe} could also be performed for the six-point function. Another rich and important one-dimensional conformal system is the $1/2$-BPS Wilson line in $\mathcal{N}=4$ super-Yang--Mills theory \cite{Drukker:1999zq,Erickson:2000af,Drukker:2000rr}, which was recently studied in light of the bootstrap approach \cite{Liendo:2016ymz,Liendo:2018ukf,Cavaglia:2021bnz,Giombi:2018qox,Barrat:2021yvp,Giombi:2023zte,Cavaglia:2022qpg,Ferrero:2021bsb,Barrat:2021tpn,Barrat:2022eim}.
In this case, there is a structure of global charges from $R-$symmetry and the six-point function has access to CFT data in various sectors, capturing four-point functions with two external operators in all the representations in the tensor product of two fundamentals. After gaining some grip on the relevant super-conformal blocks, it would be interesting to study the six-points bounds and compare them to the mixed four-point correlator analysis. As a last example, to have an intrinsically one dimensional system, we can consider the long-range Ising model. This is a continuous family of CFTs labelled by the dimension $\Delta_\phi$ of the order parameter, which can take values in a certain range. The non-local nature of the model's kinetic term leads to non-renormalisation theorems on $\Delta_\phi$ and an associated shadow operator of dimension $1-\Delta_\phi$ which is interpreted as $\phi^3$ in the weakly coupled description. This gives some pieces of information which can be fed to the six-point bootstrap. While less numerical data is known about the one-dimensional version of the model as opposed to its two- and three-dimensional cousins, it would nonetheless be interesting to compare with results of the four-point single- and multi-correlator bootstrap \cite{Paulos:2015jfa,Behan:2017dwr,Behan:2017emf,Behan:2018hfx,Behan:2023ile}.

Perhaps the most important direction is to find a path to a higher-dimensional generalisation of our approach. As described in the introduction, there are many interesting observables which become more directly accessible in the six-point functions in higher dimensions. The two-dimensional case seems relatively straightforward, since the conformal blocks factorise into a left- and right- moving part, so the properties that lead to the descendant space basis are likely to remain unchanged. On the other hand, there will now be two quantum numbers associated to the operators in the $\mathcal{O}_2$ channel, so this must be taken into account, and will certainly increase the computational complexity of the problem. Instead, in 3 and 4 dimensions we have far less control over the conformal blocks. Nonetheless, factorisation of the six-point block into four-point blocks is understood in certain kinematics \cite{Buric:2021kgy}. If differential operators which relate the four-point constituents are used, it might be possible to run the conformal bootstrap without having to fully evaluate the six-point blocks in the first place, as we did in our descendant-basis formulation of the one-dimensional crossing equation.

Within the context of higher dimensional CFT there is one very interesting
extension that does actually not require much additional technical development, namely the defect bootstrap. The
traditional approach to bootstrapping $p$-dimensional defects in a $d$-dimensional CFT is based on bulk two-point
functions \cite{Liendo:2012hy,Billo:2016cpy}. The crossing equation for this setup arises from the comparison of
bulk- and defect channel conformal block expansions. While the scalar blocks for both expansions are well understood
by now, see e.g. \cite{Isachenkov:2018pef} for a complete theory of scalar bulk channel blocks and further references,
the bulk channel block expansion lacks the positivity features that are necessary for a SDP based numerical bootstrap.
Hence, so far, rigorous results could only be obtained through the analytical bootstrap, based on appropriate extensions
of the inversion formula, see in particular \cite{Lemos:2017vnx,Liendo:2019jpu}. Our present work opens a new
avenue towards and SDP-based numerical bootstrap for defect CFT. As was pointed out in \cite{Buric:2020zea},
defect channel crossing equations for four-point functions involving two bulk and two defect fields do possess
the positivity features SDP is based upon. While the relevant blocks are not known in closed form, the
corresponding blocks for three-point functions of one bulk and two defect fields are
\cite{Lauria:2020emq,Buric:2020zea,Buric:2022ucg}. According to the treatment we developed here, see in particular
the discussion of the descendant-space formulation in Subsection~\ref{sec:commentson}, all ingredients are in place
to code and run a positive semi-definite bootstrap for four-point functions of two bulk and two defect fields, at
least for all line and surface defects.

\section*{Acknowledgements}
We thank Till Bargheer, Julien Barrat, Carlos Bercini, Ilija Buric, Pedro Liendo, Jeremy Mann, Junchen Rong, Francesco
Russo, Slava Rychkov, Alessandro Vichi and Sasha Zhiboedov for useful comments and discussions. This
project received funding from the German Research Foundation DFG under Germany's Excellence Strategy
- EXC 2121 Quantum Universe - 390833306. SH is further supported by
the Studienstiftung des Deutschen Volkes.
\appendix
\section{Decomposition of six-point into four-point correlators}\label{app:decomp6to4}
The purpose of this appendix is to establish the identity
\begin{align}\label{eq:PnKnid}
    \sum\limits_{n=0}^\infty\frac{\langle \phi_6 \phi_5 \phi_4 P^n \mathcal{O} \rangle \langle \mathcal{O}^\dagger K^n \phi_3 \phi_2 \phi_1\rangle}{(2 \Delta_\mathcal{O})_n n!}
    = \sum\limits_{n=0}^\infty\frac{( \bar \chi_1  \chi_2 \bar \chi_3 )^{ \Delta_\mathcal{O} + n - \Delta_\phi}\partial_{\bar\chi_1}^n f(\bar\chi_1) \partial_{\bar\chi_3}^n f(\bar\chi_3)}{(x_{12}^ 2x_{34}^2x_{56}^2)^{\Delta_\phi}(2 \Delta_\mathcal{O})_n n!} ,
\end{align}
where
\begin{align}
    \bar \chi \equiv \frac{1}{\chi} && f(\bar\chi) \equiv \langle \phi(\infty)\mathcal{O}(\bar\chi) \phi(1)\phi(0)\rangle.
\end{align}
Three conformal transformations play a central role in the derivation. These are
\begin{align}
    \mu(z) &\equiv \frac{z-x_3}{z-x_4} x_4 && \partial_z \mu(z) =  \frac{x_{34}}{(z-x_4)^2} x_4 < 0\\
    \bar\chi_1(z) &\equiv\frac{(z-x_2)x_{13}}{(z-x_3)x_{12}}  && \partial_z\bar\chi_1(z) =\frac{x_{13}x_{23}}{(z-x_3)^2x_{12}} <0
    \\
    \bar\chi_3(z) &\equiv\frac{(z-x_5)x_{46}}{(z-x_4)x_{56}}  && \partial_z\bar\chi_3(z) = \frac{x_{45} x_{46}}{(z-x_5)^3x_{65}} > 0.
\end{align}
As indicated above, $\mu(z)	$ and $\bar \chi_1(z)$ are orientation reversing maps whereas $\bar \chi_3(z)$ preserves the orientation $x_i<x_{i+1}$. Note that the coincidence of our notation for the conformal maps $\bar \chi_1(z)$ and $\bar \chi_3(z)$ and the cross-ratios $\bar \chi_1$ and $\bar \chi_3$ is justified by
\begin{align}
    \bar\chi_1(x_4) = \bar\chi_1 && \bar\chi_3(x_3) = \bar\chi_3.
\end{align}
Before beginning the actual computation, let us qualitatively outline the strategy. Roughly speaking, we would like to insert $\mathcal{O}$ and $\mathcal{O}^\dagger$ into the six-point correlator at $x_3$ and $x_4$ respectively, so that we have a product of two four-point functions that depend on $\{x_6,x_5,x_4,x_3\}$ and $\{x_4,x_3,x_2,x_1\}$ only. Applying a conformal transformation on these two four-point functions, we could then further simplify to a product of two functions that depend only on $\chi_3$ and $\chi_1$ respectively multiplied by some simple prefactor.
There are two obstructions to this plan.
Firstly, in order to use the simple projector
\begin{align}
    \mathcal{P}_\mathcal{O}=\sum\limits_{n=0}^\infty \frac{|P^n \mathcal{O} \rangle \langle \mathcal{O}^\dagger K^n|}{(2 \Delta_\mathcal{O})_n n!},
\end{align}
we should insert $\mathcal{O}$ and $\mathcal{O}^\dagger$ at $0$ and $\infty$ not at $x_3$ and $x_4$. To tackle this obstruction, we apply the conformal involution $\mu(z)$ that exchanges $x_3 \leftrightarrow 0$, $x_4 \leftrightarrow \infty$ before inserting the projector $\mathcal{P}_\mathcal{O}$ and would then like to apply it a second time after inserting the projector to undo the first transformation. However, when we try to apply $\mu(z)$ after inserting the projector, we encounter the second obstruction: After inserting the projector, we no longer have a correlator of primaries, but a sum of products of correlators of descendants, which do not transform as nicely under conformal mappings. To overcome this second problem, we should first apply conformal Ward identities to express the descendant correlators in terms of primary correlators and only afterwards apply $\mu(z)$ a second time.

Let us now perform the steps sketched above. First, we apply $\mu(z)$ to the six-point correlator, obtaining
\begin{align}
    \langle \phi_6\phi_5\phi_4\phi_3\phi_2\phi_1\rangle = \Omega\langle \phi(\mu(x_1))\phi(\mu(x_2))\phi(0)\phi(\infty)\phi(\mu(x_5))\phi(\mu(x_6))\rangle,
\end{align}
where
\begin{align}
    \Omega = \left|\partial_z\mu(x_1)\partial_z\mu(x_2)\partial_z\mu(x_3)\lim_{x \rightarrow x_4}\frac{\partial_z\mu(x)}{\mu(x)^{2}} \partial_z\mu(x_5)\partial_z\mu(x_6) \right|^{\Delta_\phi}.
\end{align}
Thus, with the projectors inserted, we have
\begin{align}\label{eq:projectorsinsertec}
 &\sum\limits_{n=0}^\infty\frac{\langle \phi_6 \phi_5 \phi_4 P^n \mathcal{O} \rangle \langle \mathcal{O}^\dagger K^n \phi_3 \phi_2 \phi_1\rangle}{(2 \Delta_\mathcal{O})_n n!}  \\
    =&
    \Omega\sum\limits_{n=0}^\infty\frac{\langle\phi(\infty)\phi(\mu(x_5))\phi(\mu(x_6)) P^n\mathcal{O}\rangle \langle \mathcal{O}^\dagger K^n \phi(\mu(x_1))\phi(\mu(x_2))\phi(0)\rangle}{(2 \Delta_\mathcal{O})_n n!}.
\end{align}
Now it is time to use Ward identities and trade $P^n\mathcal{O}$ and $\mathcal{O}^\dagger K^n$ for
\begin{align}
   P^n |\mathcal{O}\rangle &= \partial_{z_3}^n\mathcal{O}(z_3) |0\rangle|_{z_3=0}, \\
   \langle \mathcal{O} | K^n  &= \langle 0 | (P^n \mathcal{O})^\dagger = \langle 0 | \partial_{z_1}^n\mathcal{O}(z_1)^\dagger|_{z_1=0} =  \partial_{z_1}^n z_1^{-2  \Delta_\mathcal{O}}\langle 0 | \mathcal{O}(1/z_1)|_{z_1=0}  ,
\end{align}
where for ease of notation, we will drop the $z_1 = 0$ and $z_3 = 0$ subscripts for the rest of the computation. Inserting these identities, we can apply $\mu(z)$ to deduce that
\begin{align}
\langle\phi(\infty)\phi(\mu(x_5))\phi(\mu(x_6)) P^n\mathcal{O}\rangle &= \partial_{z_3}^n \langle\phi(\infty)\phi(\mu(x_5))\phi(\mu(x_6)) \mathcal{O}(z_3)\rangle \\
&= \partial_{z_3}^n \Omega_{456}\langle \phi_6\phi_5\phi_4  \mathcal{O}(\mu(z_3))\rangle
\end{align}
and
\begin{align}
    \langle \mathcal{O}^\dagger K^n \phi(\mu(x_1))\phi(\mu(x_2))\phi(0)\rangle = \partial_{z_1}^n z_1^{-2  \Delta_\mathcal{O}} \Omega_{123}\langle  \mathcal{O}(1/z_1) \phi_3 \phi_2 \phi_1\rangle,
\end{align}
where
\begin{align}
    \Omega_{456} = \left|\lim_{x \rightarrow x_4}\frac{\partial_z\mu(x)}{\mu(x)^2}\partial_z\mu(x_5)\partial_z\mu(x_6)\right|^{-\Delta_\phi} |\partial_z \mu(z_3)|^{\Delta_\mathcal{O}}
\end{align}
and
\begin{align}
    \Omega_{123} = \left|\partial_z\mu(x_1)\partial_z\mu(x_2)\partial_z\mu(x_3)\right|^{-\Delta_\phi} |\partial_z \mu(1/z_1)|^{\Delta_\mathcal{O}}.
\end{align}
Next, we apply $\bar \chi_1(z)$ and $\bar \chi_3(z)$ to obtain
\begin{align}
    \partial_{z_3}^n \Omega_{456}\langle \phi_6\phi_5\phi_4  \mathcal{O}(\mu(z_3))\rangle &=  \partial_{z_3}^n \Omega_{456}\Omega_{\bar \chi_3}\langle \phi(\infty)\mathcal{O}(\bar\chi_3(\mu(z_3))) \phi(1)\phi(0)\rangle \\
    &= \partial_{z_3}^n \Omega_{456}\Omega_{\bar \chi_3} f(\bar\chi_3(\mu(z_3)))
\end{align}
and likewise
\begin{align}
   \partial_{z_1}^n z_1^{-2  \Delta_\mathcal{O}} \Omega_{123}\langle  \mathcal{O}(1/z_1) \phi_3 \phi_2 \phi_1\rangle = \partial_{z_1}^n z_1^{-2  \Delta_\mathcal{O}} \Omega_{123} \Omega_{\bar \chi_1} f(\bar\chi_1(\mu(1/z_1))),
\end{align}
where
\begin{align}
    \Omega_{\bar{\chi}_3} = \left|\partial_z \bar\chi_3(x_6)\partial_z \bar\chi_3(x_5)\lim\limits_{x \rightarrow x_4}\frac{\partial_z \bar\chi_3(x)}{\bar\chi_3(x)}\right|^{\Delta_\phi} |(\partial_z\bar\chi_3)(\mu(z_3))|^{\Delta_\mathcal{O}} =  \frac{|(\partial_z\bar\chi_3)(\mu(z_3))|^{\Delta_\mathcal{O}}}{|x_{45}x_{56}x_{64}|^{\Delta_\phi}}
\end{align}
and likewise
\begin{align}
    \Omega_{\bar{\chi}_1} = \frac{|(\partial_z\bar\chi_1)(\mu(1/z_1))|^{\Delta_\mathcal{O}}}{|x_{12}x_{23}x_{31}|^{\Delta_\phi}}.
\end{align}
Note that the products $\Omega_{123} \Omega_{\bar\chi_1}z_1^{-2  \Delta_\mathcal{O}} $ and $\Omega_{456} \Omega_{\bar\chi_3}$ are independent of $z_1$ and $z_3$ since
\begin{align}
    \Omega_{123} \Omega_{\bar\chi_1}z_1^{-2  \Delta_\mathcal{O}} =  \frac{|\partial_{z_1} \bar\chi_1(\mu(1/z_1))|^{\Delta_\mathcal{O}}}{|\partial_z\mu(x_1)\partial_z\mu(x_2)\partial_z\mu(x_3)x_{12}x_{23}x_{31}|^{\Delta_\phi}}
\end{align}
and
\begin{align}
    \Omega_{456} \Omega_{\bar\chi_3} =  \frac{|\partial_{z_3} \bar\chi_3(\mu(z_3))|^{\Delta_\mathcal{O}}}{\left|\lim\limits_{x \rightarrow x_4}\frac{\partial_z\mu(x)}{\mu(x)^2}\partial_z\mu(x_5)\partial_z\mu(x_6)x_{45}x_{56}x_{64}\right|^{\Delta_\phi}}
\end{align}
but $\bar\chi_1 (\mu(1/z_1))$ and $\bar\chi_3 (\mu(z_3))$ just depend linearly on $z_1$ and $z_3$
\begin{align}\label{eq:linear}
    \bar\chi_1 (\mu(1/z_1)) = \bar\chi_1 + x_4\frac{x_{13} x_{32}}{x_{12} x_{34}} z_1 && \bar\chi_3 (\mu(z_3)) = \bar\chi_3 + \frac{1}{x_4}\frac{x_{45} x_{46}}{x_{43} x_{56}} z_3.
\end{align}
Apart from ensuring that $\Omega_{123} \Omega_{\bar\chi_1}$ and $\Omega_{456} \Omega_{\bar\chi_3}$ are independent of $z_1$ and $z_3$, eq.~\eqref{eq:linear} also implies that
\begin{align}
    \partial_{z_1}^n f(\bar \chi_1(\mu(1/z_1)))|_{z_1 = 0}=
    \left(x_4\frac{x_{13} x_{32}}{x_{12} x_{34}}\right)^n\partial_{\bar\chi_1}^n f(\bar \chi_1) =
    \left|x_4\frac{x_{13} x_{32}}{x_{12} x_{34}}\right|^n(-1)^n\partial_{\bar\chi_1}^n f(\bar \chi_1)
\end{align}
and
\begin{align}
    \partial_{z_3}^n f(\bar \chi_3(\mu(z_3)))|_{z_3 = 0}=
    \left(\frac{1}{x_4}\frac{x_{45} x_{46}}{x_{43} x_{56}} \right)^n\partial_{\bar\chi_3}^n f(\bar \chi_3) = \left|\frac{1}{x_4}\frac{x_{45} x_{46}}{x_{43} x_{56}} \right|^n(-1)^n\partial_{\bar\chi_3}^n f(\bar \chi_3).
\end{align}
Thus, we conclude that
\begin{align}
\langle\phi(\infty)\phi(\mu(x_5))\phi(\mu(x_6)) P^n\mathcal{O}\rangle
=  \frac{\left|x_4\frac{x_{13} x_{32}}{x_{12} x_{34}}\right|^{\Delta_\mathcal{O}+n}(-1)^n\partial_{\bar\chi_1}^n f(\bar \chi_1)}{|\partial_z\mu(x_1)\partial_z\mu(x_2)\partial_z\mu(x_3)x_{12}x_{23}x_{31}|^{\Delta_\phi}}
\end{align}
and
\begin{align}
    \langle \mathcal{O}^\dagger K^n \phi(\mu(x_1))\phi(\mu(x_2))\phi(0)\rangle
    = \frac{\left|\frac{1}{x_4}\frac{x_{45} x_{46}}{x_{43} x_{56}} \right|^{\Delta_\mathcal{O}+n}(-1)^n\partial_{\bar\chi_3}^n f(\bar \chi_3)}{\left|\lim\limits_{x \rightarrow x_4}\frac{\partial_z\mu(x)}{\mu(x)^2}\partial_z\mu(x_5)\partial_z\mu(x_6)x_{45}x_{56}x_{64}\right|^{\Delta_\phi}}.,
\end{align}
Inserting these two expressions into eq.~\eqref{eq:projectorsinsertec} finally leads to
\begin{align}
    \sum\limits_{n=0}^\infty\frac{\langle \phi_6 \phi_5 \phi_4 P^n \mathcal{O} \rangle \langle \mathcal{O}^\dagger K^n \phi_3 \phi_2 \phi_1\rangle}{(2 \Delta_\mathcal{O})_n n!}
    &=\sum\limits_{n=0}^\infty\frac{\left|\frac{x_{13} x_{32}}{x_{12} x_{34}}\frac{x_{45} x_{46}}{x_{43} x_{56}} \right|^{\Delta_\mathcal{O}+n} \partial_{\bar\chi_1}^n f(\bar \chi_1) \partial_{\bar\chi_3}^n f(\bar \chi_3)}{|x_{12}x_{23}x_{31}x_{45}x_{56}x_{64}|^{\Delta_\phi}(2 \Delta_\mathcal{O})_n n!}\\
    &=\sum\limits_{n=0}^\infty\frac{( \bar \chi_1  \chi_2 \bar \chi_3 )^{ \Delta + n - \Delta_\phi}\partial_{\bar\chi_1}^n f(\bar\chi_1) \partial_{\bar\chi_3}^n f(\bar\chi_3)}{(x_{12}^ 2x_{34}^2x_{56}^2)^{\Delta_\phi}(2 \Delta_\mathcal{O})_n n!} ,
\end{align}
which is what we wanted to show.
\section{Comments on \texorpdfstring{$n$}{\textit{n}}-point crossing equations}\label{app:from4to3to8to5}
The aim of this appendix is to elaborate further on the relation between the six-point bootstrap and mixed correlator systems of four-point crossing equations. En passant, the discussion will naturally lead to some motivation for extending our programme from six- to eight-point correlators.

Our starting point is the search for a mixed four-point correlator interpretation of the crossing equation
\begin{align}\label{eq:sixpointdecompositionintofourAP}
\sum\limits_{\mathcal{O}} \sum\limits_{b} \langle \phi_1 \phi_2 \phi_3 \mathcal{O}^{b} \rangle  \langle \mathcal{O}^{b} \phi_4 \phi_5 \phi_6\rangle =\sum\limits_{\mathcal{O}} \sum\limits_{b} \langle \phi_2 \phi_3 \phi_4 \mathcal{O}^{b} \rangle \langle \mathcal{O}^{b} \phi_5 \phi_6 \phi_1\rangle \,,
\end{align}
for decompositions of a six-point function into products of four-point functions. Note that, as opposed to the main text (eq.~\eqref{eq:sixpointdecompositionintofour}), we find it more convenient to work with an orthonormal basis here, so that $\mathcal{N}_{ab} = \delta_{ab}$.

Inserting the identity in a carefully chosen way, we can directly see that this crossing equation is exactly equivalent to the full system of all s- to t-channel four-point crossing equations for correlators of two copies of $\phi$ and two arbitrary operators that appear in the $\phi\times\phi$ OPE.

Indeed, eq.~\eqref{eq:sixpointdecompositionintofourAP}
is equivalent to\footnote{Let us stress that this seemingly obvious claim is not trivial: There is a surjectivity statement of a map from a vector-space of real numbers labelled by operators occurring in the $\phi\times\phi$-OPE to a vector-space of real numbers labelled by descendants underlying it. Concretely, this map is induced by conformal blocks for $\langle\phi \phi \phi \mathcal{O}^a\rangle$ four-point correlators. In the 1d case, we have discussed it as the map from primary-space to descendant-space.}
\begin{align}\label{eq:sixpointdecompositionintothreeAP}
& \sum\limits_{\mathcal{O}_1,\mathcal{O},\mathcal{O}_3} \sum\limits_{a,b,c} \langle  \phi_2 \phi_3 \mathcal{O}_1^a\rangle \langle \mathcal{O}_1^a \phi_1\mathcal{O}^{b} \rangle  \langle \mathcal{O}^{b} \phi_4 \mathcal{O}^{c}_3\rangle \langle \mathcal{O}^{c}_3 \phi_5 \phi_6 \rangle \nonumber \\
=& \sum\limits_{\mathcal{O}_1,\mathcal{O},\mathcal{O}_3} \sum\limits_{a,b,c} \langle  \phi_2 \phi_3 \mathcal{O}_1^a\rangle \langle \mathcal{O}_1^a \phi_4\mathcal{O}^{b} \rangle  \langle \mathcal{O}^{b} \phi_1 \mathcal{O}^{c}_3\rangle \langle \mathcal{O}^{c}_3 \phi_5 \phi_6 \rangle\,,
\end{align}
which due to linear independence of the occurring three-point structures implies that
\begin{align}\label{eq:FourPointMixedAP}
\sum\limits_{\mathcal{O}} \sum\limits_{b}  \langle \mathcal{O}_1^a \phi_1\mathcal{O}^{b} \rangle  \langle \mathcal{O}^{b} \phi_4 \mathcal{O}^{c}_3\rangle
= \sum\limits_{\mathcal{O}} \sum\limits_{b} \langle \mathcal{O}_1^a \phi_4\mathcal{O}^{b} \rangle  \langle \mathcal{O}^{b} \phi_1 \mathcal{O}^{c}_3\rangle && \forall \mathcal{O}_1, \mathcal{O}_3 \subseteq \phi \times \phi \,.
\end{align}
Pictorially, we have said nothing more than
\begin{figure}[ht]
    \centering

    \begin{tikzpicture}
        \node (E2) at (-2,-1) {$\phi_1$};
        \node (E3) at (-1,1) {$\phi_3$};
        \node (E4) at (-2,1) {$\phi_2$};
        \node (E6) at (3,1) {$\phi_4$};
        \node (E7) at (2,-1) {$\phi_6$};
        \node (E8) at (3,-1) {$\phi_5$};
        \node[shape=circle,fill=blue!50,draw=black, scale = 0.5] (V1) at (0,0) {};
        \node[shape=circle,fill=blue!50,draw=black, scale = 0.5] (V3) at (1,0) {};

        \draw (E2) -- (V1);
        \draw (E3) -- (V1);
        \draw (E4) -- (V1);
        \draw (E6) -- (V3);
        \draw (E7) -- (V3);
        \draw (E8) -- (V3);
        \path[-, draw] (V1) edge node[above] {$\mathcal{O}$} (V3);

        \def\x{9};
        \def\l{1};
        \node (A1) at (0.5,-1) {};
        \node (A2) at (0.5,-3) {};
        \draw[{Stealth[length=5mm]}-{Stealth[length=5mm]},thick] (A1) -- (A2);

        \node (E2) at (-2,-5) {$\phi_1$};
        \node (E3) at (-1,-3) {$\phi_3$};
        \node (E4) at (-2,-3) {$\phi_2$};
        \node (E6) at (3,-3) {$\phi_4$};
        \node (E7) at (2,-5) {$\phi_6$};
        \node (E8) at (3,-5) {$\phi_5$};
        \node[shape=circle,fill=blue!50,draw=black, scale = 0.5] (V1) at (0.5,-4-0.3) {};
        \node[shape=circle,fill=blue!50,draw=black, scale = 0.5] (V3) at (0.5,-4+0.3) {};

        \draw (E2) -- (V1);
        \draw (E3) -- (V3);
        \draw (E4) -- (V3);
        \draw (E6) -- (V3);
        \draw (E7) -- (V1);
        \draw (E8) -- (V1);
        \path[-, draw] (V1) edge node[left] {$\mathcal{O}$} (V3);

        \draw[implies-implies,double equal sign distance] (4,0) -- (5,0);
        \draw[implies-implies,double equal sign distance] (4,-4) -- (5,-4);

        \node (E2) at (6,-1){$\phi_1$};
        \node (E4) at (6,1){$\mathcal{O}_1$};
        \node (E6) at (11,1){$\phi_4$};
        \node (E8) at (11,-1) {$\mathcal{O}_3$};
        \node[shape=circle,fill=blue!50,draw=black, scale = 0.5] (V1) at (7.5,0) {};
        \node[shape=circle,fill=blue!50,draw=black, scale = 0.5] (V3) at (9.5,0) {};

        \draw (E2) -- (V1);

        \draw (E4) -- (V1);

        \draw (E6) -- (V3);

        \draw (E8) -- (V3);
        \path[-, draw] (V1) edge node[above] {$\mathcal{O}$} (V3);

        \def\x{9};
        \def\l{1};
        \node (A1) at (8.5,-1) {};
        \node (A2) at (8.5,-3) {};
        \draw[{Stealth[length=5mm]}-{Stealth[length=5mm]},thick] (A1) -- (A2);

        \node (E2) at (6,-5) {$\phi_1$};
        \node (E4) at (6,-3) {$\mathcal{O}_1$};
        \node (E6) at (11,-3) {$\phi_4$};
        \node (E8) at (11,-5) {$\mathcal{O}_3$};
        \node[shape=circle,fill=blue!50,draw=black, scale = 0.5] (V1) at (8.5,-4.5) {};
        \node[shape=circle,fill=blue!50,draw=black, scale = 0.5] (V3) at (8.5,-3.5) {};

        \draw (E2) -- (V1);

        \draw (E4) -- (V3);

        \draw (E6) -- (V3);

        \draw (E8) -- (V1);
        \path[-, draw] (V1) edge node[right] {$\mathcal{O}$} (V3);
        \node (S) at (4.5,-1) {s-channel};
        \node (C) at (4.5,-2) {to};
        \node (T) at (4.5,-3) {t-channel};
    \end{tikzpicture}
    \caption{Visualising the equivalence between crossing of four-point decompositions of six-point correlators and infinite systems of s- to t-channel four-point crossing equations.}
    \label{fig:SixToFour}
\end{figure}
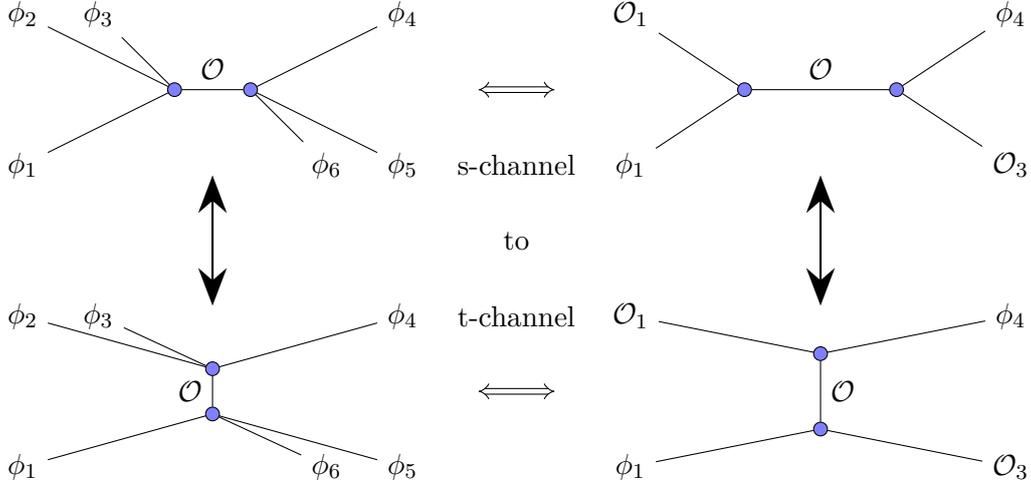

It is reassuring that this perspective now gives us the precise answer to the questions raised at the end of Section~\ref{sec:sixpt-crossing}: We exactly capture all s- to t-channel crossings, involving two external $\phi$ and two arbitrary operators $\mathcal{O}_1,\mathcal{O}_3$ in the $\phi\times\phi$ OPE. But can we do better? How could a multi-point bootstrap access crossing of all possible operators in the $\phi\times\phi$ OPE?

Figure~\ref{fig:SixToFour} already suggests the answer to these questions. We can go to eight-point functions. By the same arguments that led to eq.~\eqref{eq:FourPointMixedAP}, one arrives at the conclusion that imposing the following crossing equation
\begin{align}
\sum\limits_{\mathcal{O}} \sum\limits_{b} \langle \phi_1 \phi_2 \phi_3 \phi_4 \mathcal{O}^{b} \rangle  \langle \mathcal{O}^{b} \phi_5 \phi_6 \phi_7 \phi_8\rangle &=\sum\limits_{\mathcal{O}} \sum\limits_{b} \langle \phi_1 \phi_2 \phi_7 \phi_8 \mathcal{O}^{b} \rangle  \langle \mathcal{O}^{b} \phi_3 \phi_4 \phi_5 \phi_6\rangle \label{crossing:st}.
\end{align}
is directly related to imposing s- to t-channel crossing for the entirety of all operators in the $\phi\times\phi$ OPE.  This is visualised in the figure below.

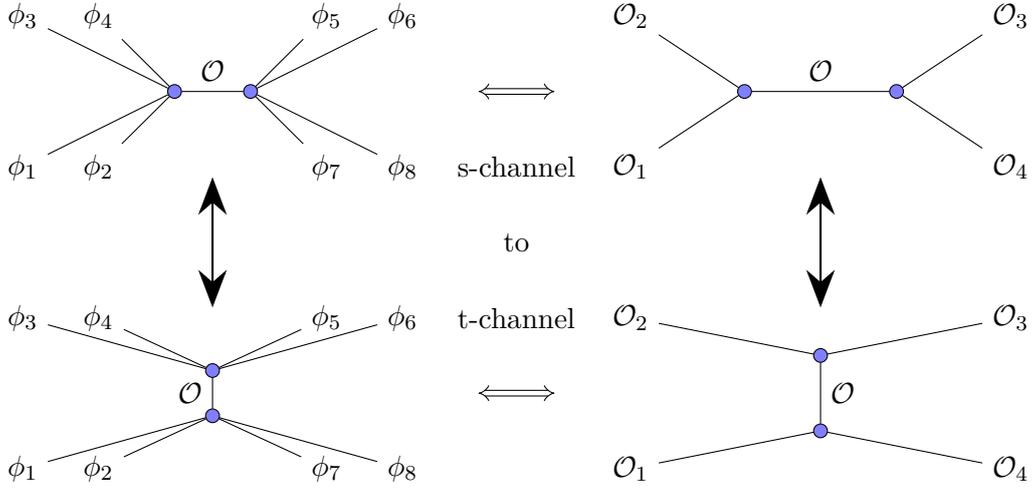
\begin{figure}[ht]
    \centering
    \begin{tikzpicture}
        \node (E1) at (-1,-1) {$\phi_2$};
        \node (E2) at (-2,-1) {$\phi_1$};
        \node (E3) at (-1,1) {$\phi_4$};
        \node (E4) at (-2,1) {$\phi_3$};
        \node (E5) at (2,1) {$\phi_5$};
        \node (E6) at (3,1) {$\phi_6$};
        \node (E7) at (2,-1) {$\phi_7$};
        \node (E8) at (3,-1) {$\phi_8$};
        \node[shape=circle,fill=blue!50,draw=black, scale = 0.5] (V1) at (0,0) {};
        \node[shape=circle,fill=blue!50,draw=black, scale = 0.5] (V3) at (1,0) {};

        \draw (E1) -- (V1);
        \draw (E2) -- (V1);
        \draw (E3) -- (V1);
        \draw (E4) -- (V1);
        \draw (E5) -- (V3);
        \draw (E6) -- (V3);
        \draw (E7) -- (V3);
        \draw (E8) -- (V3);
        \path[-, draw] (V1) edge node[above] {$\mathcal{O}$} (V3);

        \def\x{9};
        \def\l{1};
        \node (A1) at (0.5,-1) {};
        \node (A2) at (0.5,-3) {};
        \draw[{Stealth[length=5mm]}-{Stealth[length=5mm]},thick] (A1) -- (A2);

        \node (E1) at (-1,-5) {$\phi_2$};
        \node (E2) at (-2,-5) {$\phi_1$};
        \node (E3) at (-1,-3) {$\phi_4$};
        \node (E4) at (-2,-3) {$\phi_3$};
        \node (E5) at (2,-3) {$\phi_5$};
        \node (E6) at (3,-3) {$\phi_6$};
        \node (E7) at (2,-5) {$\phi_7$};
        \node (E8) at (3,-5) {$\phi_8$};
        \node[shape=circle,fill=blue!50,draw=black, scale = 0.5] (V1) at (0.5,-4-0.3) {};
        \node[shape=circle,fill=blue!50,draw=black, scale = 0.5] (V3) at (0.5,-4+0.3) {};

        \draw (E1) -- (V1);
        \draw (E2) -- (V1);
        \draw (E3) -- (V3);
        \draw (E4) -- (V3);
        \draw (E5) -- (V3);
        \draw (E6) -- (V3);
        \draw (E7) -- (V1);
        \draw (E8) -- (V1);
        \path[-, draw] (V1) edge node[left] {$\mathcal{O}$} (V3);

        \draw[implies-implies,double equal sign distance] (4,0) -- (5,0);
        \draw[implies-implies,double equal sign distance] (4,-4) -- (5,-4);

        \node (E2) at (6,-1) {$\mathcal{O}_1$};
        \node (E4) at (6,1) {$\mathcal{O}_2$};
        \node (E6) at (11,1) {$\mathcal{O}_3$};
        \node (E8) at (11,-1) {$\mathcal{O}_4$};
        \node[shape=circle,fill=blue!50,draw=black, scale = 0.5] (V1) at (7.5,0) {};
        \node[shape=circle,fill=blue!50,draw=black, scale = 0.5] (V3) at (9.5,0) {};

        \draw (E2) -- (V1);

        \draw (E4) -- (V1);

        \draw (E6) -- (V3);

        \draw (E8) -- (V3);
        \path[-, draw] (V1) edge node[above] {$\mathcal{O}$} (V3);

        \def\x{9};
        \def\l{1};
        \node (A1) at (8.5,-1) {};
        \node (A2) at (8.5,-3) {};
        \draw[{Stealth[length=5mm]}-{Stealth[length=5mm]},thick] (A1) -- (A2);

        \node (E2) at (6,-5) {$\mathcal{O}_1$};
        \node (E4) at (6,-3) {$\mathcal{O}_2$};
        \node (E6) at (11,-3) {$\mathcal{O}_3$};
        \node (E8) at (11,-5) {$\mathcal{O}_4$};
        \node[shape=circle,fill=blue!50,draw=black, scale = 0.5] (V1) at (8.5,-4.5) {};
        \node[shape=circle,fill=blue!50,draw=black, scale = 0.5] (V3) at (8.5,-3.5) {};

        \draw (E2) -- (V1);

        \draw (E4) -- (V3);

        \draw (E6) -- (V3);

        \draw (E8) -- (V1);
        \path[-, draw] (V1) edge node[right] {$\mathcal{O}$} (V3);
        \node (S) at (4.5,-1) {s-channel};
        \node (C) at (4.5,-2) {to};
        \node (T) at (4.5,-3) {t-channel};
\end{tikzpicture}
    \caption{Equivalence between eight-point crossing and the full system of all four-point crossing equations, s- and t-channel.}
    \label{fig:eightpointst}
\end{figure}

\section{Semi-definite programming implementation: polynomial SOS}
\label{app:SDP}
\label{sec:Polynomial SOS}
The notebook \texttt{polynomial\_sos.nb} attached to the arXiv submission contains the algorithm that we use to recast the problem of constructing SOS linear combinations of two variable polynomials as an SDP. At the current stage of the numerical six-point bootstrap, we found that solving this SDP is not the bottleneck holding us back from higher precision results and therefore decided to stick to the minimalistic implementation described below. Nevertheless, due to the notoriously bad scaling of SDP formulations of polynomial SOS optimisation problems with the degree of the polynomials, it might become necessary at high derivative order to implement some more sophisticated algorithm. For a short review of directions for optimisation see \cite{SOSReview}.

The input of the notebook \texttt{polynomial\_sos.nb} is a list $\texttt{P}$ of polynomials of the form
\begin{align}
    \sum\limits_{i=0}^{2n} \sum\limits_{j=0}^{2m} c_{ij} x^i y^j\,,
\end{align}
for some reals coefficients $c$. Corresponding to these polynomials, we consider a list
\begin{align}
    \texttt{v} = \{x^i y^j | 0\leq i \leq n, 0 \leq j \leq m\}
\end{align}
of monomials. Now, for each polynomial $\texttt{p} \in \texttt{P}$, we construct a $(n+1)(m+1) \times (n+1)(m+1)$ matrix $\texttt{Q}[\texttt{p}]$ such that
\begin{align}
    \texttt{v}^T \texttt{Q}[\texttt{p}] \texttt{v} = \texttt{p}\,.
\end{align}
Next, we construct a basis of the kernel of the conjugation by \texttt{v} i.e.~a basis
\texttt{K} of matrices $\texttt{K}[\![j]\!]$ s.t.
\begin{align}\label{eq:vker}
     \texttt{v}^T \texttt{K}[\![j]\!] \texttt{v} = 0.
\end{align}
We then input the problem of finding positive semi-definite linear combinations of the matrices $\texttt{Q}[\texttt{P}[\![i]\!]]$ and the kernel matrices $\texttt{K}$ into \texttt{SDPA}. If we manage to find such a positive semi-definite linear combination
\begin{align}
    \texttt{M}\equiv \sum\limits_i a_i \texttt{Q}[\texttt{P}[\![i]\!]]+\sum\limits_j b_j \texttt{K}[\![j]\!]
\end{align}
with some real coefficients $a$ and $b$, then the polynomial
\begin{align}
    \texttt{v}^T \texttt{M}  \texttt{v} = \sum\limits_i a_i \texttt{v}^T  \texttt{Q}[\texttt{P}[\![i]\!]]\texttt{v} = \sum\limits_i a_i \texttt{P}[\![i]\!]
\end{align}
is manifestly a sum of squares.

\section{Conformal block expansions of selected six-point correlators}
\label{app:blockexpansion}
In this appendix, we provide explicit expressions for the conformal block expansions of some of the correlators which appeared in the main text. In practice, we consider a given correlator and observe its leading power law dependence on the cross-ratios $\chi_i$. We then write an Ansatz with a finite sum of conformal blocks \eqref{eq:combblock} which we can be Taylor expanded and matched to the correlator's Taylor expansion, up to an order determined by the number of blocks included. In the cases we consider, this is simple to do since the scaling dimensions are of the form $\Delta_i= a_i \Delta_\phi +n_i$, for some integer or rational numbers $a_i$ and some integers $n_i$. The Ansatz can then be written as a sum over the $n_i$ which are chosen consistently with the order at which we truncate the Taylor expansion.
\paragraph{No-identity correlator}
Let us first consider the ``no-identity'' correlator \eqref{eq:noidentcorr}, repeated here for convenience:
\begin{equation}
\small
    \mathcal{G}= \frac{\chi_1^{8 \Delta_\phi}\chi_2^{9 \Delta_\phi}\chi_3^{8 \Delta_\phi}}{\big((1-\chi_1)(1-\chi_2)(1-\chi_3)(1-\chi_1-\chi_2)(1-\chi_2-\chi_3)(1-\chi_1-\chi_2-\chi_3+\chi_1 \chi_3)\big)^{2\Delta_\phi}}\,.
\end{equation}
By Taylor expanding it becomes clear that the dimensions of the exchanged operators have to be of the form $\Delta_1=8\Delta_\phi +n_1$ and $\Delta_2=9\Delta_\phi +n_2$.  We can then propose an Ansatz
\begin{equation}
\mathcal{G}= \sum_{n_1,n_2,n_3=0}^\infty P_{n_1,n_2,n_3} G_{8\Delta_\phi +n_1,9\Delta_\phi +n_2,8\Delta_\phi +n_3}
\end{equation}
where $P_{n_1,n_2,n_3}=P_{n_3,n_2,n_1}$, and we recall that the external operator has dimension $5\Delta_\phi$ in these conventions. Truncating the Taylor expansion to the appropriate order we can read off the first few non-vanishing OPE coefficients
\begin{align}
P_{000}&=1\,, \quad P_{002}=\frac{6\Delta_\phi^2}{1+16\Delta_\phi}\,, \quad P_{004}= \frac{3 \Delta _{\phi }^2 \left(2 \Delta _{\phi }+1\right) \left(6 \Delta _{\phi }+1\right)}{2 \left(16 \Delta _{\phi }+3\right) \left(16 \Delta _{\phi }+5\right)}\,,\\
P_{020}&=\frac{8\Delta_\phi^2}{1+18\Delta_\phi}\,, \quad P_{022}= \frac{4 \Delta _{\phi }^2 \left(4 \Delta _{\phi } \left(3 \Delta _{\phi }+7\right)+1\right)}{\left(16 \Delta _{\phi }+1\right) \left(18 \Delta _{\phi }+1\right)}\,, \quad P_{030}= \frac{8 \Delta _{\phi }^3}{3 \left(9 \Delta _{\phi }+1\right) \left(9 \Delta _{\phi }+2\right)} \,,\nonumber\\
P_{202}&= \frac{36 \Delta _{\phi }^4}{\left(16 \Delta _{\phi }+1\right){}^2}\,, \quad P_{040}=\frac{2 \Delta _{\phi }^2 \left(2 \Delta _{\phi }+1\right) \left(4 \Delta _{\phi }+1\right)}{3 \left(3 \Delta _{\phi }+1\right) \left(18 \Delta _{\phi }+5\right)} \,,\quad\dots \qquad\qquad \qquad \qquad \qquad  .\nonumber
\end{align}
We note that $n_1$ and $n_3$ can only take even values, while $n_2$ can take all even and odd values except 1. This is consistent with what one would naively expect from parity and a counting of primaries vs.~descendants. Unitarity is also easy to check: For fixed $n_2$, we found that the matrices $(P_{n,n_2,m})_{nm}$ that we have computed are positive semi-definite.
\paragraph{Generalised free fermion}
We now consider the non-trivial piece of the GFF six-point function \eqref{eq:GFF6pt}, repeated below
\begin{align}
   &(\mathcal{G}^{\textrm{GFF}}-\mathcal{G}_{\textrm{4-pt}}^{\textrm{GFF}})/\chi_{2}^{\Delta_\psi}=\left(\frac{\chi_1 \chi_2 \chi_3}{(1-\chi_1-\chi_2)(1-\chi_3)} \right)^{2\Delta_\psi} +\left(\frac{\chi_1 \chi_2 \chi_3}{(1-\chi_3-\chi_2)(1-\chi_1)} \right)^{2\Delta_\psi} \nonumber \\
  & - \left(\frac{\chi_1 \chi_2 \chi_3}{(1-\chi_1)(1-\chi_2)(1-\chi_3)} \right)^{2\Delta_\psi} -\left(\frac{\chi_1 \chi_2 \chi_3}{(1-\chi_1-\chi_2)(1-\chi_2-\chi_3)} \right)^{2\Delta_\psi}\nonumber\\
  & -\left(\frac{\chi_1 \chi_2 \chi_3}{1-\chi_1-\chi_2-\chi_3+ \chi_1 \chi_3} \right)^{2\Delta_\psi} + \left(\frac{\chi_1 \chi_2 \chi_3}{(1-\chi_1-\chi_2-\chi_3+ \chi_1 \chi_3)(1-\chi_2)} \right)^{2\Delta_\psi} \,.
\end{align}
This is the piece of the correlator with triple-twist exchange in the $\mathcal{O}_2$ sector as can be seen from the behaviour as $\chi_2 \to 0$. Using the Ansatz
\begin{equation}
\mathcal{G}_{\textrm{3t}}^{\textrm{GFF}}= \sum_{n_1,n_2,n_3=0}^\infty P_{n_1,n_2,n_3} G_{2\Delta_\psi +n_1,3\Delta_\psi +n_2,2\Delta_\psi +n_3}\,,
\end{equation}
where the subscript $\textrm{3t}$ denotes the triple-twist contribution, allows us to extract the first non-vanishing OPE coefficients
\begin{align}\label{eq:PsforFermion}
P_{131}&= 2\Delta_\psi^2(1+2\Delta_\psi)\,, \quad P_{133}= -\frac{2 \Delta_\psi^2(1+\Delta_\psi)(1+2\Delta_\psi)}{3+4\Delta_\psi}\,,  \\
\quad P_{333}&= \frac{2\Delta_\psi^2(1+\Delta_\psi)^2(1+2\Delta_\psi)}{(3+4\Delta_\psi)^2}\,, \quad P_{151}=\frac{\Delta _{\psi }^2 \left(\Delta _{\psi }+1\right) \left(2 \Delta _{\psi }+1\right){}^2 \left(2 \Delta _{\psi }+3\right)}{3 \left(6 \Delta _{\psi }+7\right)}\,, \nonumber \\
P_{161}& = \frac{2 \Delta _{\psi }^3 \left(\Delta _{\psi }+1\right) \left(\Delta _{\psi }+2\right) \left(2 \Delta _{\psi }+1\right){}^2 \left(2 \Delta _{\psi }+3\right)}{45
   \left(3 \Delta _{\psi }+4\right) \left(3 \Delta _{\psi }+5\right)}\,, \quad \dots \nonumber \qquad \qquad\qquad \qquad  \quad \quad .
\end{align}
The OPE coefficients pass all the unitarity checks mentioned above. Furthermore one observes that $n_1$ and $n_3$ take only odd values as expected from the fermionic double-trace spectrum. One also immediately sees that the leading triple-trace operator has dimension $3\Delta_\psi+3$ as discussed in the main text. Additionally we see that there is no state with $n_2=4$ which is compatible with a counting of triple-twist primaries in GFF discussed in Appendix~\ref{app:countmulti}.

Using the double trace OPE coefficients
\begin{equation}\label{eq:OPEpsi}
C_{\psi\psi \Delta_n}^2= \frac{2 \Gamma^2(\Delta_n)\Gamma (\Delta_n +2\Delta_\psi-1)}{\Gamma^2(2\Delta_\psi)\Gamma(2\Delta_n-1)\Gamma(\Delta_n-2\Delta_\psi+1)}
\end{equation}
one can extract squares of triple-trace OPE coefficients from the coefficients $P_{n_1,n_2,n_3}$ given in Eq.~\eqref{eq:PsforFermion}
\paragraph{Generalised free boson}
Finally we take the triple-twist piece of the GFB correlator \eqref{eq:GFB6pt} given by
\begin{align}
   &(\mathcal{G}^{\textrm{GFB}}-\mathcal{G}_{\textrm{4-pt}}^{\textrm{GFB}})/\chi_{2}^{\Delta_\phi}= \left(\frac{\chi_1 \chi_2 \chi_3}{(1-\chi_1-\chi_2)(1-\chi_3)} \right)^{2\Delta_\phi} +\left(\frac{\chi_1 \chi_2 \chi_3}{(1-\chi_3-\chi_2)(1-\chi_1)} \right)^{2\Delta_\phi} \nonumber \\
  & +\left(\frac{\chi_1 \chi_2 \chi_3}{(1-\chi_1)(1-\chi_2)(1-\chi_3)} \right)^{2\Delta_\phi} +\left(\frac{\chi_1 \chi_2 \chi_3}{(1-\chi_1-\chi_2)(1-\chi_2-\chi_3)} \right)^{2\Delta_\phi}\nonumber\\
  & +\left(\frac{\chi_1 \chi_2 \chi_3}{1-\chi_1-\chi_2-\chi_3+ \chi_1 \chi_3} \right)^{2\Delta_\phi} + \left(\frac{\chi_1 \chi_2 \chi_3}{(1-\chi_1-\chi_2-\chi_3+ \chi_1 \chi_3)(1-\chi_2)} \right)^{2\Delta_\phi} \,.
\end{align}
Once again we can use the Ansatz
\begin{equation}
\mathcal{G}_{\textrm{3t}}^{\textrm{GFB}}= \sum_{n_1,n_2,n_3=0}^\infty P_{n_1,n_2,n_3} G_{2\Delta_\phi +n_1,3\Delta_\phi +n_2,2\Delta_\phi +n_3}\,,
\end{equation}
which allows us to extract the leading OPE coefficients
\begin{align}
    P_{0,0,0}&=6\,,\quad P_{0,2,0}= \frac{8 \Delta _{\phi }^2 \left(2 \Delta _{\phi }+1\right)}{6 \Delta _{\phi }+1}\,, \quad P_{0,2,2}= \frac{4 \Delta _{\phi }^2 \left(2 \Delta _{\phi }+1\right)}{4 \Delta _{\phi }+1} \,,  \\
    P_{0,3,0}&= \frac{8 \Delta _{\phi }^3 \left(\Delta _{\phi }+1\right) \left(2 \Delta _{\phi }+1\right)}{3 \left(3 \Delta _{\phi }+1\right) \left(3 \Delta _{\phi }+2\right)}\,, \quad P_{0,3,2}= -\frac{8 \Delta _{\phi }^3 \left(\Delta _{\phi }+1\right) \left(2 \Delta _{\phi }+1\right)}{\left(3 \Delta _{\phi }+2\right) \left(4 \Delta _{\phi }+1\right)}\,, \nonumber\\
    P_{0,4,0}&= \frac{2 \Delta _{\phi }^2 \left(2 \Delta _{\phi }+1\right){}^2 \left(2 \Delta _{\phi }+3\right)}{9 \left(6 \Delta _{\phi }+5\right)} \,, \quad P_{0,4,2}= \frac{4 \Delta _{\phi }^3 \left(2 \Delta _{\phi }+1\right){}^2 \left(2 \Delta _{\phi }+3\right)}{3 \left(4 \Delta _{\phi }+1\right) \left(6 \Delta _{\phi }+5\right)}\,, \quad \dots \qquad , \nonumber
\end{align}
which yet again passes all the unitarity checks, is consistent with a double-trace spectrum with only even integer shifts and with the counting of triple-twist primary operators.

Finally note that these expansion coefficients fit nicely with the closed OPE-coefficient formulas: The OPE coefficient between two $\phi$ and a double trace operator $[\phi\phi]_n$ of dimension $\Delta_n^{\text{B}}$ is
\begin{align}
\label{eq:OPEboson}
    C_{[\phi \phi]_{n}\phi\phi}^2 =
\frac{2 (2 \Delta_\phi)_{2n}^2}{(4 \Delta_\phi + 2n -1)_n {(2n)}!}\,.
\end{align}
The squared OPE coefficient $C_{\phi [\phi \phi]_n [\phi \phi \phi]_m}^2$ between $\phi$ the double twist $[\phi \phi]_n$ and the sum of all $m$-derivative triple-twist operators $[\phi\phi\phi]_m$ vanishes if $m<2n$ and is given by
\begin{align}
 &C_{\phi [\phi \phi]_n [\phi \phi \phi]_m}^2 = \frac{(2 \Delta )_{m-2 n} (4 n+4 \Delta )_{m-2 n}}{(m-2 n)! (m+2 n+6 \Delta -1)_{m-2 n}} \biggl(1 +  \\
 &\frac{2 (-1)^m (4 \Delta +4 n-1) (2 \Delta )_m}{(2 n)! (2 n+4 \Delta -1)_{m-2 n+1}}\,  _4F_3\left[\begin{matrix}-2 n,1-2\Delta-2 n,2 \Delta +2 n,4 \Delta+2 n -1\\1-2 \Delta-m,2 \Delta ,4 \Delta+m\end{matrix};1\right]
\biggr)\,, \nonumber
\end{align}
for $m\ge 2n$. In terms of these OPE coefficients, we have
\begin{align}
    P_{2n_1,n_2,2n_3}^2=C^2_{[\phi\phi]_{n_1}\phi\phi}C^2_{[\phi\phi\phi]_{n_2} [\phi \phi]_{n_1} \phi} C^2_{[\phi\phi\phi]_{n_2} [\phi \phi]_{n_3} \phi}C^2_{[\phi\phi]_{n_3}\phi\phi} .
\end{align}
\section{Counting multi-twist operators in GFF Theory}
\label{app:countmulti}
A straightforward way to obtain the spectrum of multi-twist operators in GFF is to use its Fock space description and compute the thermal partition function \cite{Kraus:2020nga}.

We define the partition function
\begin{equation}
Z(\beta)= \textrm{Tr} e^{-\beta\Delta}\,,
\end{equation}
which is guaranteed by conformal representation theory to admit the decomposition
\begin{equation}
    Z(\beta)= 1+ \sum_{\Delta} N_{\Delta} \chi_\Delta(\beta)\,,
\end{equation}
where we have the conformal character $\chi_\Delta$\footnote{Hopefully, there is no confusion with the cross-ratios $\chi_i$}
\begin{equation}
    \chi_{\Delta}(\beta)= \frac{q^\Delta}{1-q}\,, \quad q\equiv e^{-\beta}\,,
\end{equation}
and the degeneracies $N_\Delta$.
We can use the fermionic Fock space to directly compute $Z(\beta)$. For that we define the partial partition function $Z_n$ built out of constituents with exactly $n$ insertions of the momentum generator $P$. The Pauli exclusion principle leads to
\begin{equation}
Z_n=1+q^{\Delta_\psi+n}\,.
\end{equation}
The full partition function is then easily computed through
\begin{equation}
    Z(\beta)= \prod_{n=0}^\infty Z_n = (-q^{\Delta_\psi},q)_\infty\,,
\end{equation}
where we used the $q-$ deformed Pochhammer symbol
\begin{equation}
    (a,q)_n=\prod_{k=0}^{n-1}(1-aq^k)\,.
\end{equation}
It is straightforward to expand the explicit partition function in powers of the form $q^{n\Delta_\psi+m}$ and match this to the character expansion, reading off the degeneracies
\begin{equation}
    N_{2\Delta_\psi+n}= \frac{1+ (-1)^{n+1}}{2}\,,
\end{equation}
which correspond to the usual non-degenerate fermionic double-twists with an odd number of derivatives. For the triple-twists we find in terms of the floor function
\begin{equation}
     N_{3\Delta_\psi+n}= \lfloor (n-1)/2\rfloor - \lfloor (n-1)/3 \rfloor\,.
\end{equation}
In particular, we see that the first non-vanishing primary has dimension $3\Delta_\psi+3$, and that there is no primary of dimension $3\Delta_\psi+4$, in agreement with the explicit conformal block expansion.
For completeness we add the quadruple-twist degeneracies
\begin{equation}
   N_{4\Delta_\psi+n}= \lfloor ((n-3)^2+6) /12\rfloor - \lfloor(n-3)/4 \rfloor\lfloor(n-1)/4 \rfloor \,.
\end{equation}

\bibliographystyle{JHEP}
\bibliography{draft.bib}
\end{document}